\documentclass{article}

\usepackage[square, numbers, sort&compress]{natbib}
    \usepackage[final]{neurips_2025}

\usepackage[utf8]{inputenc} % allow utf-8 input
\usepackage[T1]{fontenc}    % use 8-bit T1 fonts
\usepackage{hyperref}       % hyperlinks
\usepackage{url}            % simple URL typesetting
\usepackage{booktabs}       % professional-quality tables
\usepackage{amsfonts}       % blackboard math symbols
\usepackage{nicefrac}       % compact symbols for 1/2, etc.
\usepackage{microtype}      % microtypography
\usepackage{xcolor}         % colors
\usepackage{graphicx}
\usepackage{amsmath}
\usepackage{amsthm, amssymb, bm}
\usepackage{adjustbox}
\usepackage{comment}
\usepackage{enumitem}
\usepackage{hyperref}

\newtheorem{proposition}{Proposition}

\setlist[itemize]{noitemsep, topsep=0pt}

\newcommand{\insettitle}[1]{\textbf{#1}\hspace{0.5em}}

\newcommand{\beginsupplement}{%
        \setcounter{table}{0}
        \renewcommand{\thetable}{S\arabic{table}}%
        \renewcommand{\theHtable}{S\arabic{table}}%
        \setcounter{figure}{0}
        \renewcommand{\thefigure}{S\arabic{figure}}%
        \renewcommand{\theHfigure}{S\arabic{figure}}  % <-- THIS IS THE FIX
     }

\title{Characterizing control between interacting subsystems \\ with deep Jacobian estimation}

\author{%
  Adam J. Eisen \\
  Brain and Cognitive Sciences\\
  MIT\\
  Cambridge, MA 02139\\
  \texttt{eisenaj@mit.edu} \\
  \And
  Mitchell Ostrow \\
  Brain and Cognitive Sciences\\
  MIT\\
  Cambridge, MA 02139\\
  \texttt{ostrow@mit.edu} \\
  \And
  Sarthak Chandra\thanks{This author is now at The International Centre for Theoretical Sciences.}\\
  Brain and Cognitive Sciences\\
  MIT\\
  Cambridge, MA 02139\\
  \texttt{sarthak.chandra@icts.res.in}\\
  \And
  Leo Kozachkov\thanks{This author is now at Brown University.}\\
  IBM Thomas J. Watson Research Center\\
  IBM Research\\
  Yorktown Heights, NY 10598\\
  \texttt{leokoz8@brown.edu} \\
  \And
  Earl K. Miller\\
  Brain and Cognitive Sciences\\
  MIT\\
  Cambridge, MA 02139\\
  \texttt{ekmiller@mit.edu}\\
  \And
  Ila R. Fiete\\
  Brain and Cognitive Sciences\\
  MIT\\
  Cambridge, MA 02139\\
  \texttt{fiete@mit.edu}\\
}

\begin{document}

\maketitle

\begin{abstract}

Biological function arises through the dynamical interactions of multiple subsystems, including those between brain areas, within gene regulatory networks, and more. A common approach to understanding these systems is to model the dynamics of each subsystem and characterize communication between them. An alternative approach is through the lens of control theory: how the subsystems control one another. This approach involves inferring the directionality, strength, and contextual modulation of control between subsystems. However, methods for understanding subsystem control are typically linear and cannot adequately describe the rich contextual effects enabled by nonlinear complex systems. To bridge this gap, we devise a data-driven nonlinear control-theoretic framework to characterize subsystem interactions via the Jacobian of the dynamics. We address the challenge of learning Jacobians from time-series data by proposing the JacobianODE, a deep learning method that leverages properties of the Jacobian to directly estimate it for arbitrary dynamical systems from data alone. We show that JacobianODE models outperform existing Jacobian estimation methods on challenging systems, including high-dimensional chaos. Applying our approach to a multi-area recurrent neural network (RNN) trained on a working memory selection task, we show that the “sensory” area gains greater control over the “cognitive” area over learning. Furthermore, we leverage the JacobianODE to directly control the trained RNN, enabling precise manipulation of its behavior. Our work lays the foundation for a theoretically grounded and data-driven understanding of interactions among biological subsystems.

\end{abstract}

% \vspace{-4.5mm}
\section{Introduction}
\label{sec:introduction}
% \vspace{-2mm}
Complex systems are ubiquitous in nature. These systems exhibit a wide range of behavior and function, in large part through the dynamic interaction of multiple component subsystems within them. One approach to understanding such complex systems is to build detailed models of their underlying dynamics. An alternative and simpler yet powerful approach is offered by control theory, focusing instead on how subsystems influence and regulate one another, and how they can be controlled. 

\begin{figure}[!htbp] % 'h' means "here" (try placing it near the text)
    \centering
    % \vspace{-2mm}
    \includegraphics[width=0.8\linewidth]{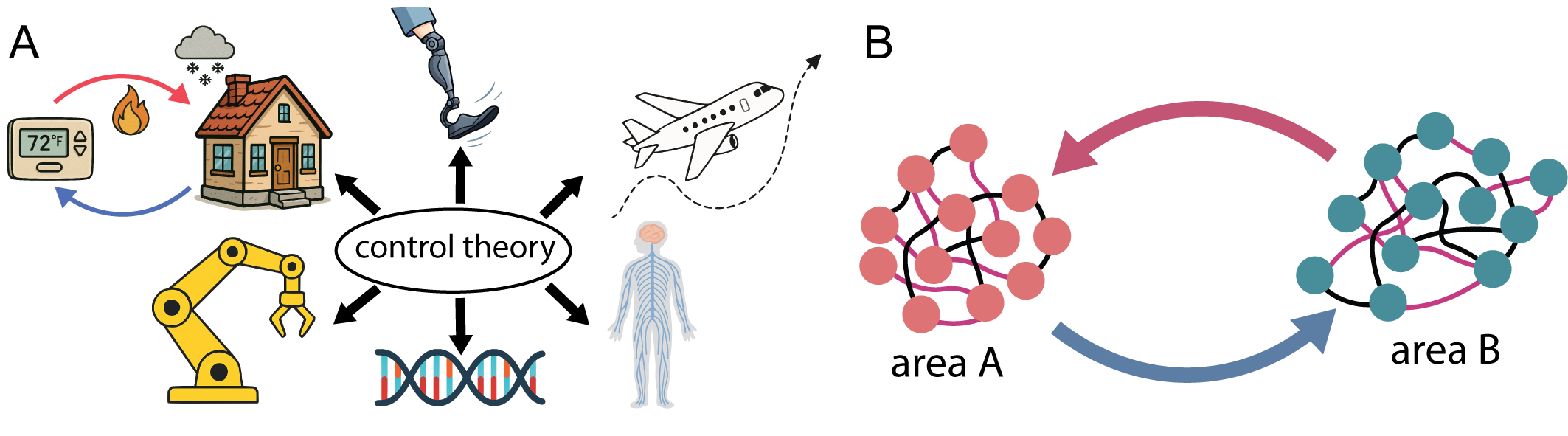} % Adjust width as needed
    \caption{\textbf{Schematic overview of control-theoretic framework applied to neural interactions.} (A) Control theory generalizes across diverse systems. (B) Illustration of interareal control, highlighting how neural activity in one area directly influences dynamics in another.}
    \label{fig:introduction}
\end{figure}

Control theory thus offers a complementary approach to both understanding and manipulating biological systems. The theory describes how inputs must be coordinated with system dynamics to achieve desired behaviors, and can be applied across domains ranging from robotics to biology (Figure \ref{fig:introduction}A). The brain coordinates neural activity across multiple interconnected brain areas, dynamically modulating which regions receive information from which others depending on need and context \cite{colgin_frequency_2009, ni2024distributeddynamic}. Interareal interactions play central roles in cognition and consciousness \citep{crick1990consc, engel2016consc, thompson2001consc, ward2003cog, siegel2012cog}, in selective attention \citep{palva2007osc, buschman2007attention, gregoriou2009attention, gregoriou2012attention, salinas2001correlated, fries2001attention, siegel2008attention, vankempen2021attention, gray1989osc}, decision making \citep{siegel2011decision, siegel2015decision}, working memory \cite{brincat2021wm, salazar2012wm}, feature binding \cite{engel2001fb, singer1995fb}, motor control \cite{murthy1996mc,logiaco2021mc,arcemcshane2016mc,perich2018mc,kaufman2014mc}, and learning and memory \cite{bricat2015learn,jones2005learn,siapas1998learn,fernandezruiz2021learn,antzoulatos2014learn}.

A common approach to characterizing interareal interactions is to quantify {\em communication} between them, using methods such as reduced-rank regression to define ``communication subspaces''. These subspaces determine low-dimensional projections that maximally align high-dimensional states of the input area with high-dimensional states of the target area \cite{semedo2019commsub, semedo2020interaction, macdowell2023multiplex}. However, effective {\em control} not only involves alignment of high-variance input states with high-variance target states, but also appropriate alignment of the inputs with the {\em dynamics} of the target area. Given connected subsystems A and B, an identical signal from B will have dramatically different control effects on A, depending on whether the signal aligns with stable or unstable directions of A’s dynamics: projections onto more unstable eigenvectors can much more readily drive the system to novel states. (For more detail see Appendix \ref{supp:control-v-comm}.) 
 %To illustrate this, consider two linear time invariant systems, A and B, that are connected linearly as well from B to A. For the same signal communicated from B to A, projections onto a very stable eigenvector and unstable eigenvector of A's dynamics will communicate "messages" with identical magnitudes. However, the signal projected on to the unstable eigenvector will much more easily be able to drive A to novel states. For more detail, see Supplementary Material. 
% I like this^^!
 
 Accordingly, recent work in neuroscience has espoused control-theoretic perspectives on interareal interactions (Figure \ref{fig:introduction}B) \cite{gu2015netctrl,kao2019ctrl,bassett2017networkneuroscience, kao2021ctrl, logiaco2021mc, kozachkov2022rnns,schimel2021ctrl,muldoon2016}. The dominant approach has involved linear control \cite{bouchard2024ctrl,braun2021netctrl,zhou2023netctrl,zoller2021netctrl,he2022ctrlenergy,singleton2022netctrl,medaglia2018netctrl,wu2024brainnct,cai2021dynamiccausal,amani2024structfunc,li2023funcnet}. However, the inherently nonlinear dynamics of the brain enable richer contextual control than possible to fully model with linear systems, necessitating a nonlinear control approach.

%however the nonlinear brain necessitates a nonlinear control approach. 

One approach to extend control-theoretic analyses to nonlinear systems is by linearizing the nonlinear dynamics through Taylor expansion, which involves the Jacobian. This converts the nonlinear system into a linear state- and time-dependent one \cite{liu2011complexnet, parkes2024nct, tyner2010geojac,kozachkov2022rnns}, allowing for simple control. Jacobian linearization for control is straightforward with access to analytical expressions for the non-linear system, but it becomes non-trivial in purely data-driven scenarios. Estimating the Jacobian involves conjunctively inferring both a function and its derivative, yet good function approximation need not yield good approximations of its derivatives (see Section \ref{sec:jac-est} and Appendix \ref{supp:deriv-v-func}).

%To perform control-theoretic analyses in nonlinear settings (like the brain), one can employ Jacobian-linearized dynamics, which convert a nonlinear system into a linear time-varying one \cite{liu2011complexnet, parkes2024nct, tyner2010geojac}.  Jacobian linearization is trivial when the dynamics are known, but it is highly non-trivial in a purely data-driven setting. Estimating the Jacobian involves conjunctively estimating both a function and its derivative. However, it is in general not the case that approximating a function well will yield a good approximation of its derivative (see section \ref{sec:jac-est}, and Supplementary Material).

This paper introduces several key contributions: \begin{itemize}
    \item \textbf{Robust data-driven Jacobian estimation.} We present JacobianODE, a deep learning-based method for estimating Jacobians from noisy trajectories in high-dimensional dynamical systems. We demonstrate the validity of this approach in several sample systems that include high-dimensional chaos.
   % \item \textbf{Robust data-driven Jacobian estimation.} We construct a deep learning method, called JacobianODE, for learning the Jacobian of high-dimensional dynamical systems from noisy trajectories. We demonstrate the validity of this approach in several sample systems that include high dimensional chaos. 
    %
    \item \textbf{A framework to characterize control between interacting subsystems via data-driven Jacobian estimation.} Harnessing our Jacobian estimation method, we devise a rigorous, data-driven approach for nonlinear control-theoretic analysis of how paired interacting systems, including brain areas, drive and regulate each other across different contexts. 
    \item \textbf{Data-driven inference of control dynamics in trained recurrent neural networks.} We apply our data-driven framework to a recurrent neural network (RNN) trained on a working memory task. We show that, purely from data, we can identify key control-theoretic interactions between the areas, and that these interactions crucially evolve over the course of learning to produce the desired behavior. 
    \item \textbf{Demonstration of accurate  control of rich interacting high-dimensional coupled dynamical subsystems.} We demonstrate high accuracy in a challenging high-dimensional data-driven nonlinear control task, enabling precise control of the behavior of the RNN.
\end{itemize}

Overall, our work lays the foundation for data-driven control-theoretic analyses in complex high-dimensional nonlinear systems, including the brain.

% \vspace{-4.5mm}
\section{Related work}
% \vspace{-3mm}

\insettitle{Interareal communication} A wide range of tools have been developed and harnessed to study interareal communication in neural data \cite{kass2023interaction, semedo2020interaction}. This includes, but is not limited to, methods based on reduced rank regression \cite{semedo2019commsub, macdowell2023multiplex}, recurrent neural network models of neural dynamics \cite{perich2021rnn, perich2020multiregion, andalman2019dynamics, pinto2019dynamics, kleinman2021rnn,kozachkov2022rnns,barbosa2023gating}, Gaussian process factor analysis \cite{gokcen2022dlag, gokcen2023motifs}, canonical correlation analysis \cite{ebrahimi2022interarea, rodu2018regions, semedo2022fffb}, convergent cross mapping \cite{ye2015ccm, tajima2015ccm, sugihara2012ccm}, switching dynamical systems \cite{glaser2020switching, karniol-tambour2022-switching}, granger causality \cite{bressler2011granger, seth2015granger}, dynamic causal mapping \cite{friston2003dcm}, point process models \cite{chen2022pop}, and machine learning methods \cite{calderon2024taci, lu2023attention}. 

\insettitle{Nonlinear controllability} Classical results assess the controllability of nonlinear control systems via the Lie theory \cite{slotine1991appnlctrl, brockett1976nonlineardiff, haynes1970lie, aguilar2014netctrl, whalen2015lie}.  Another approach to nonlinear network controllability is based on attractor strength \cite{wang2016geocont}. \citet{liu2011complexnet} and \citet{parkes2024nct} note that the large literature of linear controllability analyses could be extended locally to nonlinear systems with an appropriate linearization method.

\insettitle{Neural network controllability} Linear structural network controllability has been implicated across a wide range of contexts, tasks, and neuropsychiatric conditions \cite{braun2021netctrl,zhou2023netctrl,zoller2021netctrl,gu2015netctrl,muldoon2016,he2022ctrlenergy,medaglia2018netctrl,wu2024brainnct}. This approach has been extended to functional brain networks \cite{cai2021dynamiccausal, amani2024structfunc, li2023funcnet}. Recent work has characterized the subspaces of neural activity that are most feedback controllable as opposed to feedforward controllable using linear methods \cite{bouchard2024ctrl}. Other approaches to neural control analyses consider the identification of structural driver nodes \cite{tang2012driver}, and input novelty \cite{kumar2014inputnovelty}. 

\insettitle{Data-driven Jacobian estimation} Besides approaches estimating Jacobians through weighted linear regressions~\cite{deyle2016tracking, ahamed2020continuouscomplexity}, some methods have used direct parameterization via neural networks to learn Jacobians of general functions~\cite{latremoliere2022jacest, lorraine2024jacnet}. These approaches inform our method but do not explicitly address dynamical systems. Applying path-integral-based Jacobian estimation to dynamical systems is challenging, as the target function (the system’s time derivative) is typically unobserved. \citet{beik-mohammadi2024-ncds} utilized this idea in dynamical systems to learn contracting latent dynamics from demonstrations. 

\section{JacobianODE: learning Jacobians from data}
% \vspace{-2mm}
\label{sec:jacobianode}

\subsection{Jacobian linearization}
\begin{figure}[ht] % 'h' means "here" (try placing it near the text)
    % \vspace{-2mm}
    \centering
    \includegraphics[width=0.9\linewidth]{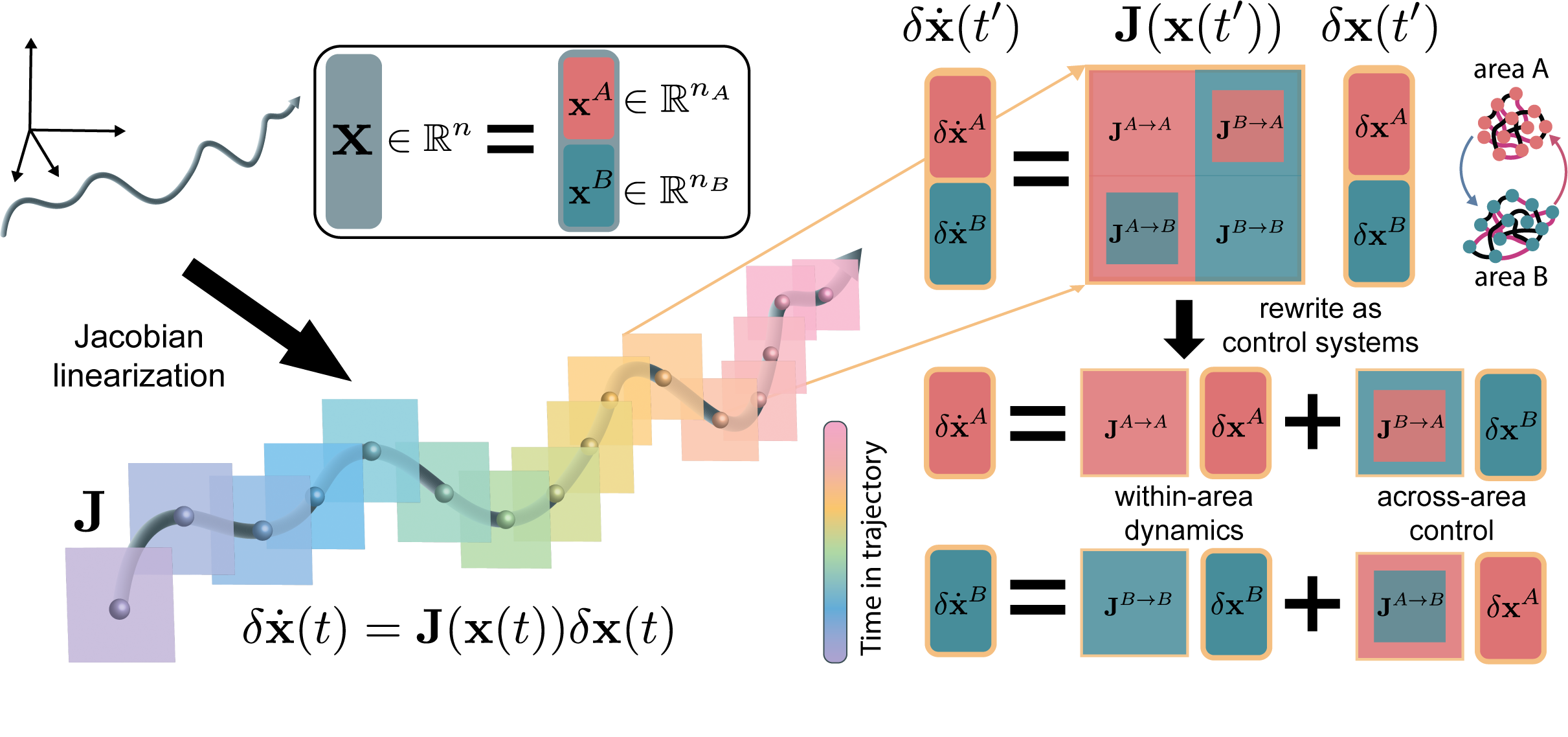} % Adjust width as needed
    \caption{\textbf{Analytical framework for pairwise interacting subsystem control.} Trajectory dynamics (left, top)  are locally linearized via Jacobians (left, bottom), explicitly separating within-area (diagonal blocks) and interareal (off-diagonal blocks) dynamics (right). These separated dynamics can be used to construct interaction-specific control systems.}%(B) Reachability measures how readily trajectories can be driven to other parts of state space. (C) Conceptual illustration of conservative vector fields. Integrated conservative vector fields such as the work done by gravity are independent of path choice.}
    \label{fig:methods-intro}
    % \vspace{-1mm}
\end{figure}

We consider nonlinear dynamical systems in $\mathbb{R}^n$,  defined by
$\dot{\mathbf{x}}(t) = \mathbf{f}(\mathbf{x}(t))$. The Jacobian of the dynamics is a matrix-valued function $\mathbf{J}_\mathbf{f}: \mathbb{R}^n \to \mathbb{R}^{n \times n}$ (henceforth, $\mathbf{J}$) given by

\begin{equation}
\mathbf{J}(\mathbf{x}(t)) = \frac{\partial }{\partial \mathbf{x}}\mathbf{f}(\mathbf{x}(t)) = 
\begin{bmatrix}
\frac{\partial f_1 }{\partial x_1}& \frac{\partial f_1}{\partial x_2} & \dots & \frac{\partial f_1}{\partial x_n} \\
\frac{\partial f_2 }{\partial x_1}& \frac{\partial f_2}{\partial x_2} & \dots & \frac{\partial f_2}{\partial x_n} \\
\vdots & \vdots & \ddots & \vdots \\
\frac{\partial f_n }{\partial x_1}& \frac{\partial f_n}{\partial x_2} & \dots & \frac{\partial f_n}{\partial x_n} \\
\end{bmatrix}.
\end{equation} 
At each time $t$, the Jacobian defines a linear subspace relating input and output changes, capturing how perturbations to the system will propagate. This recasts nonlinear dynamics as linear time-varying dynamics in the tangent space locally along trajectories (formally, $\delta\dot{\mathbf{x}}(t) = \mathbf{J}_\mathbf{f}(\mathbf{x}(t)) \delta \mathbf{x}(t)$, see Figure \ref{fig:methods-intro} left, also see \citet{lohmiller1998contraction} for a discussion in the context of virtual displacements).

%This reframing enables the dynamics in the original space to be recast as linear time-varying up to a constant, which we exploit further below. 

\begin{figure}[ht] % 'h' means "here" (try placing it near the text)
    \centering
    % \vspace{-2mm}
    \includegraphics[width=1.0\linewidth]{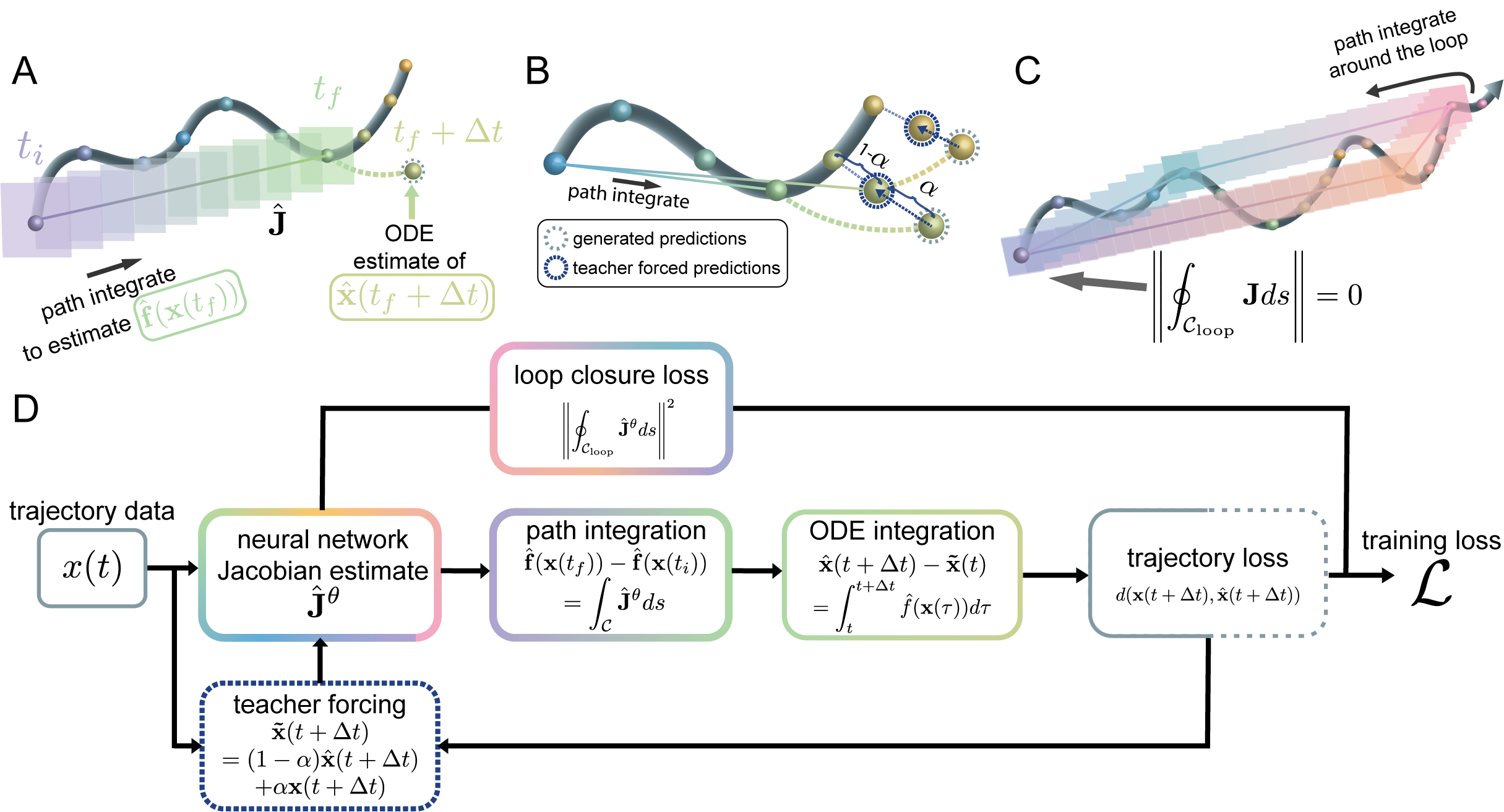} % Adjust width as needed
    \caption{\textbf{Jacobian estimation with JacobianODE models.} (A) Path integration of the Jacobian predicts future states. (B) Generalized teacher forcing stabilizes trajectory predictions during training. (C) Loop-closure constraints enforce consistency of Jacobian estimates. (D) Training pipeline, combining neural Jacobian estimation, path integration, teacher forcing, and self-supervised loop-closure loss.}
    \label{fig:methods-training}
\end{figure}

\subsection{Parameterizing differential equations via the Jacobian}
\label{sec:param-ode-jac}
% \vspace{-2mm}
We now turn to the problem of how to estimate the Jacobian $\mathbf{J}$ from data. We assume that we have access only to observed trajectories of the system, of the form $\mathbf{x}^{(j)}(t_0^{(j)} + k\Delta t), \ \ k=1,2,..., \ \ j=1,2,...$where $\Delta t$ is a fixed sampling interval, $j$ indexes the trajectory, and $t_0^{(j)}$ is a trajectory-specific start time. Crucially, we do not assume access to the function $\mathbf{f}$. Our method estimates the Jacobian directly via a neural network. To do this,  we parameterize a neural network function $\hat{\mathbf{J}}^{\theta}$ with learnable parameters $\theta$ that is then trained to approximate $\mathbf{J}$. 

 % To enable the utility of this approach for arbitrary dynamical systems (which present unique challenges in comparison with ordinary functions), we developed a novel parameterization of ordinary differential equations (ODEs) via the Jacobian, and introduced dynamics-specific loss functions.

\insettitle{Path integration} Following \citet{lorraine2024jacnet} and \citet{beik-mohammadi2024-ncds},  we exploit the fact that the path integral of the Jacobian is path independent in the construction of the loss function. This is because the rows of the Jacobian are conservative vector fields (i.e., they are the gradients of scalar functions). For an intuitive picture, consider the work done by the force of gravity as you climb a mountain. Regardless of the path you take to climb, the resulting work is dependent only on the start and end points of the path. Formally, for the time derivative function $\mathbf{f}$ we have that \begin{equation}
\mathbf{f}(\mathbf{x}(t_f)) - \mathbf{f}(\mathbf{x}(t_i)) \ \ = \ \ \int_{\mathcal{C}}\mathbf{J}\ ds \ \ = \ \ \int_{t_i}^{t_f} \mathbf{J}(\mathbf{c}(r))\mathbf{c}'(r) \ dr,
\label{eq:pathint}
\end{equation}
where $\mathcal{C}$ is a piecewise smooth curve in $\mathbb{R}^n$ and $\mathbf{c}: [t_i, t_f] \to \mathcal{C}$ is a parameterization of $\mathcal{C}$ with $\mathbf{c}(t_i) = \mathbf{x}(t_i)$  and $\mathbf{c}(t_f) = \mathbf{x}(t_f)$ (Figure \ref{fig:methods-training}A). Given estimates of $\mathbf{f}(\mathbf{x}(t_i))$ and the Jacobian $\mathbf{J}$, we can then use Equation \ref{eq:pathint} to approximate $\mathbf{f}(\mathbf{x}(t_f))$ at any time $t_f$. For these integrals, we use a line between the endpoints as a simple choice of path (Figure \ref{fig:methods-training}A). Then, to generate an estimate $\hat{\mathbf{x}}(t_f + \Delta t)$ of the next step, we can use a standard ordinary differential equation (ODE) integrator (e.g., Euler, fourth-order Runge–Kutta, etc.) to integrate the estimated time derivative $\hat{\mathbf{f}}(\mathbf{x}(t))$ (see Appendix \ref{supp:jacobianODE-technical-detail} for more detail). To avoid the need to represent $\mathbf{f}$ directly (thereby
enabling all gradients to backpropagate through the Jacobian network) we note that we parameterize an estimate of $\mathbf{f}(\mathbf{x}(t_i))$ in Equation \ref{eq:pathint} in terms of the Jacobian (see Appendix \ref{supp:jacparam}).

% \vspace{-2mm}
\subsection{Loss functions}
\label{sec:lossfuncs}

\insettitle{Trajectory reconstruction loss } Given an observed trajectory $\mathbf{x}(t_0 + k\Delta t), k=0,...,T-1$ of length $T$, we can compute the trajectory reconstruction loss, $\mathcal{L}_{\text{traj}}(\theta;\mathbf{x})$, between the true trajectory and the estimated trajectory using an appropriate distance measure $d$ (e.g., mean squared error, see Appendix \ref{supp:path-pred-gen} for more detail on generating predictions and the trajectory prediction loss).

\insettitle{Generalized Teacher Forcing} To avoid trajectory divergence in chaotic or noisy systems, we employ Generalized Teacher Forcing, generating recursive predictions partially guided by true states (Figure \ref{fig:methods-training}B, and Appendix \ref{supp:teacher-forcing}) \cite{hess2023generalizedtf}.

%The Jacobian can be best learned when training predictions are generated recursively (i.e., replacing $\mathbf{x}(t)$ by $\hat{\mathbf{x}}(t)$). However, in chaotic systems, and/or systems with measurement noise (as considered here), this could lead to catastrophic divergence of the predicted trajectory from the true trajectory during training. We therefore employ Generalized Teacher Forcing when training the model \cite{hess2023generalizedtf}. Generalized Teacher Forcing prevents catastrophic divergence by forcing the generated predictions along the line from the prediction to the true state.

\begin{comment}
This generates a state expressed by

\begin{equation}
\tilde{\mathbf{x}}(t + \Delta t) = (1 - \alpha) \mathbf{x}(t + \Delta t) + \alpha \mathbf{x}(t + \Delta t), \ \ \alpha \in [0, 1]
\end{equation}

We thus replace $\mathbf{x}(t)$ by $\tilde{\mathbf{x}}(t)$ in equations (\ref{eq:universal-f}) and (\ref{eq:traj-pred}) during training. We employ Generalized Teacher Forcing using the suggested annealing process for $\alpha$, which guarantees the boundedness of the generated trajectories.

\end{comment}

\insettitle{Loop closure loss} The Jacobian captures how perturbations to the system will propagate along \textit{any} direction in state space. Estimating it purely from dynamics constrains only the direction of the flow, leaving the full solution underdetermined. To address this, we again exploit the fact that each row of the Jacobian is a conservative vector field. Specifically, we note that for any piecewise smooth loop $\mathcal{C}_{\text{loop}}$, we have $\left \Vert \oint_{\mathcal{C}_{\mathrm{loop}}} \mathbf{J} ds \right \Vert_2 = 0$ (see Figure \ref{fig:methods-training}C). Thus, by integrating along loops that contain directions \textit{orthogonal} to the system's dynamics (and penalizing the deviation from zero), we encourage the estimated Jacobians to capture information about other directions in state space (see Appendices \ref{supp:loop-closure} and \ref{supp:loop-training} for full technical details). To ensure broad coverage of tangent space directions, we form loops from concatenations of line integrals between randomly selected data points. This strategy samples diverse directions from the tangent space while remaining easy to compute. The resulting self-supervised loss term, $\mathcal{L}_{\text{loop}}(\theta;\mathbf{x})$,  builds on the loss introduced by \citet{iyer2024velocities}. It constrains $\hat{\mathbf{J}}^{\theta}$ to satisfy both the dynamics and conservativity. This improves Jacobian estimation accuracy significantly (see Appendix \ref{supp:ablation-studies} for ablation studies).

\insettitle{Training loss} We therefore minimize the following loss function with respect to the parameters $\theta$: \begin{equation}
\mathcal{L}(\theta;\mathbf{x}) = \mathcal{L}_{\text{traj}}(\theta;\mathbf{x}) + \lambda_{\text{loop}}\mathcal{L}_{\text{loop}}(\theta;\mathbf{x})
\end{equation} where $\lambda_{\text{loop}}$ controls the relative weighting of the loop closure loss $\mathcal{L}_{\text{loop}}(\theta;\mathbf{x})$ compared to the trajectory prediction, and is a hyperparameter of the learning procedure (Figure \ref{fig:methods-training}D).

% \vspace{-4.5mm}
\section{Jacobian estimation in dynamical systems}
\label{sec:jac-est}
% \vspace{-2mm}
\insettitle{Data} To evaluate the quality of the Jacobian estimation procedure, we apply our approach to several example systems for which the dynamics are known. For this analysis, we used the Van der Pol oscillator \cite{vanderpol1926osc}, Lorenz system \cite{lorenz1963flow}, and the Lorenz 96 system across three different system sizes (12, 32, and 64 dimensional) \cite{lorenz1996pred}. All systems were simulated using the \texttt{dysts} package, which samples dynamical systems with respect to the characteristic timescale $\tau$ of their Fourier spectrum \cite{gilpin2021chaos, gilpin2023modelscale}. All training data consisted of 26 trajectories of 12 periods, sampled at 100 time steps per $\tau$. All models were trained on a 10 time-step prediction task with teacher forcing.

To evaluate the performance of the methods in the presence of noise, we trained the models on data with 1\%, 5\%, and 10\% Gaussian observation noise added i.i.d over time, where the percentage is defined via the ratio of the euclidean norm of the noise to the mean euclidean norm of the data.

\insettitle{JacobianODE model} For the JacobianODE framework, loop closure loss weights were chosen via line search (Appendix \ref{supp:hyperparameter-selection}), where the neural network $\mathbf{J}^{\theta}$ was taken to be a four-layer multilayer perceptron (MLP) with hidden layer sizes 256, 1024, 2048 and 2048. Path integration was performed using the trapezoid method from the \texttt{torchquad} package, with each integral discretized into 20 steps \cite{gomez2021torchquad}. ODE integration was performed using the fourth-order Runge–Kutta (RK4) method from the \texttt{torchdiffeq} package \cite{chen2018neuralode}. The JacobianODE models used 15 observed points to generate the initial estimate of $\mathbf{f}$ (see Section \ref{sec:param-ode-jac} and Appendices \ref{supp:jacparam} and \ref{supp:path-pred-gen}). All models were built in PyTorch \cite{paszke2019pytorch}. Full implementation details are provided in appendices \ref{supp:jacobianODE-technical-detail} and \ref{supp:experimental-details}.

\insettitle{Baselines} We chose two different Jacobian estimation procedures for comparison. The first was a neural ordinary differential equation (NeuralODE) model trained to reproduce the dynamics \cite{chen2018neuralode}. The NeuralODE was implemented as a four-layer MLP with hidden layers of the same size as the one used for the JacobianODE model. Jacobians were computed via automatic differentiation. NeuralODEs were regularized via a penalty on the Frobenius norm of the estimated Jacobians to prevent the model from learning unnecessarily large negative eigenvalues (see Appendix \ref{supp:neuralode-details}) \cite{hoffman2019jacpen, wikner2024reservoirstab, schneider2025timeseries-att}. We also employed a baseline that estimates Jacobian via a weighted linear regression, which computes locally linear models at each point in the space \cite{deyle2016tracking, ahamed2020continuouscomplexity} (see Appendix \ref{supp:weighted-linear-jacobian}).

\begin{table}[ht]
  \centering
  \caption{Mean Frobenius norm error on Jacobian estimation, $\langle\|\mathbf{J} - \hat{\mathbf{J}}\|_F \rangle$, for each system and noise level. Errors are reported as mean $\pm$ standard deviation, with mean and standard deviation computed over five random initializations of the model architectures.}
  \label{tab:fro_metrics}
  \scriptsize
  \setlength{\tabcolsep}{6pt}
  \begin{adjustbox}{width=\textwidth}
    \begin{tabular}{llccc}
      \toprule
      Project & Training noise & JacobianODE & NeuralODE & Weighted Linear \\
      \midrule
      VanDerPol (2 dim) & 1\% & \textbf{0.7 ± 0.1} & 1.0 ± 0.3 & 6.11 \\
       & 5\% & \textbf{0.72 ± 0.05} & \textbf{0.72 ± 0.08} & 6.09 \\
       & 10\% & \textbf{1.35 ± 0.06} & 2.2 ± 0.2 & 6.08 \\
      \midrule
      Lorenz (3 dim) & 1\% & \textbf{3.3 ± 0.2} & 8.7 ± 0.3 & 21.94 \\
       & 5\% & \textbf{5.1 ± 0.9} & 26.0 ± 1.5 & 21.90 \\
       & 10\% & \textbf{6.4 ± 0.1} & 26.7 ± 0.9 & 21.84 \\
      \midrule
      Lorenz 96 (12 dim) & 1\% & \textbf{1.2 ± 0.2} & 4.8 ± 0.2 & 28.67 \\
       & 5\% & \textbf{2.7 ± 0.2} & 5.9 ± 0.2 & 28.64 \\
       & 10\% & \textbf{4.6 ± 0.1} & 6.1 ± 0.1 & 28.56 \\
      \midrule
      Lorenz 96 (32 dim) & 1\% & \textbf{8.7 ± 0.2} & 16.8 ± 0.4 & 47.13 \\
       & 5\% & \textbf{7.8 ± 0.2} & 17.7 ± 0.6 & 46.96 \\
       & 10\% & \textbf{13.45 ± 0.09} & 19.5 ± 0.4 & 47.03 \\
      \midrule
      Lorenz 96 (64 dim) & 1\% & \textbf{30.9 ± 0.5} & 45.5 ± 0.1 & 66.39 \\
       & 5\% & \textbf{34.0 ± 0.2} & 45.7 ± 0.1 & 66.26 \\
       & 10\% & \textbf{36.0 ± 0.2} & 46.0 ± 0.2 & 66.29 \\
      \midrule
      Task-trained RNN & 1\% & \textbf{188.5 ± 7.1} & 294.02 ± 0.03 & 301.46 \\
       & 5\% & \textbf{166.8 ± 3.6} & 294.357 ± 0.003 & 297.63 \\
       & 10\% & \textbf{180.4 ± 0.5} & 294.328 ± 0.004 & 296.63 \\
      \midrule
      \bottomrule
    \end{tabular}
  \end{adjustbox}
\end{table}

\begin{figure}[ht] % 'h' means "here" (try placing it near the text)
\label{fig:jac-est}
    % \vspace{-2mm}
    \centering
    \includegraphics[width=1.0\linewidth]{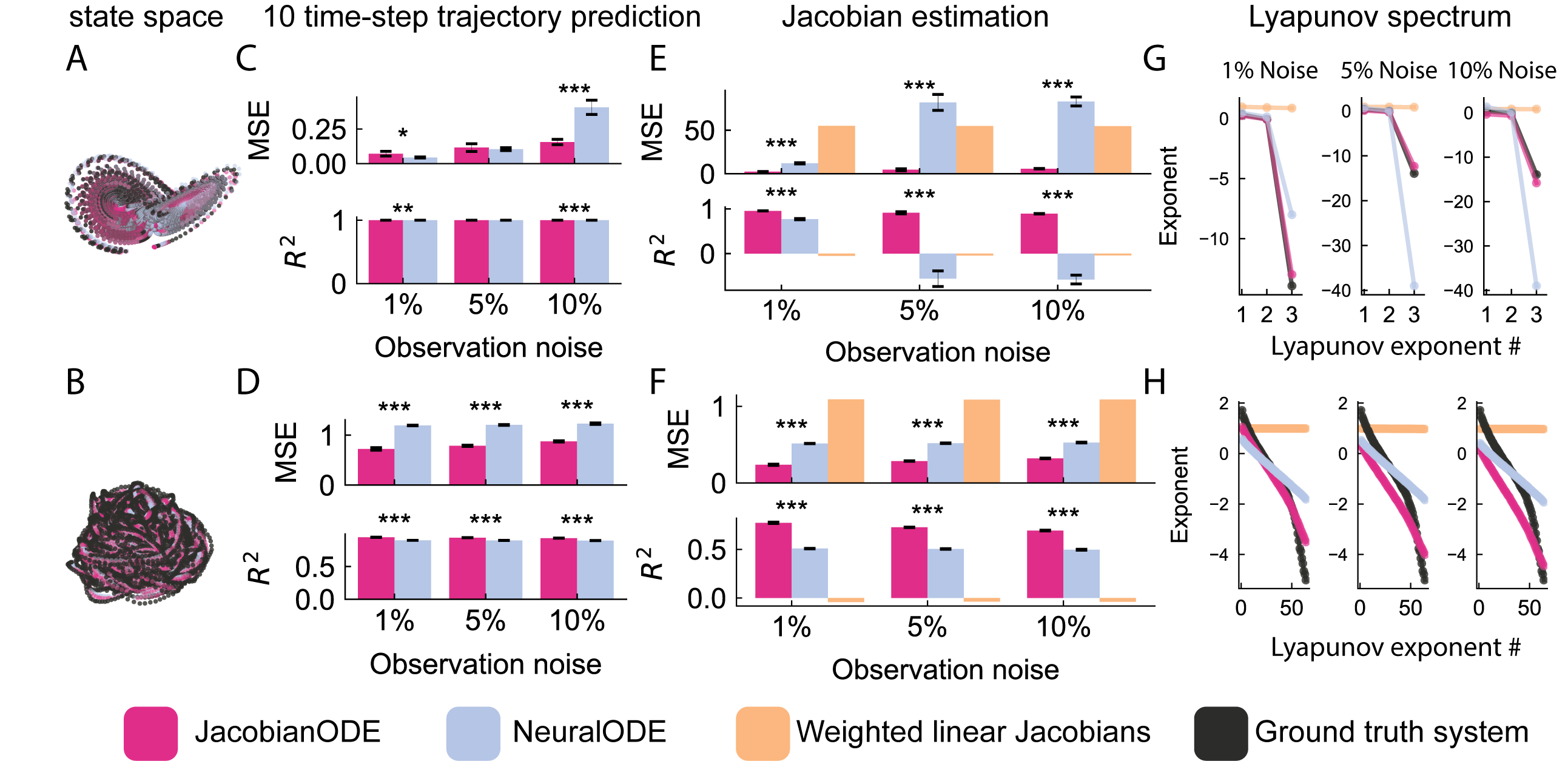} % Adjust width as needed
    \caption{\textbf{JacobianODE surpasses benchmark models on chaotic dynamical systems.} Error bars indicate standard deviation, with statistics computed over five different model initializations. (A,B) State-space trajectories for (A) Lorenz and (B) 64-dimensional Lorenz 96 systems, with 10 time-step predictions. Spectral embedding was used to reduce the Lorenz 96 data to three dimensions. (C,D) Accuracy of 10 time-step trajectory predictions at varying noise levels. (E,F) Comparison of Jacobian estimation accuracy, quantified by mean squared error (MSE) and $R^2$. (G,H) Estimated Lyapunov spectra averaged over initializations.}
    \label{fig:results-sample-dyns}
    % \vspace{-2mm}
\end{figure}
\insettitle{Performance} We tested the approaches on Jacobian estimation on held-out trajectories without noise. The JacobianODE method outperforms the baseline methods in terms of mean Frobenius norm error for virtually all systems (Table \ref{tab:fro_metrics}).  This was also true when considering the spectral matrix 2-norm (see Table \ref{tab:2_norm_metrics} in Appendix \ref{supp:dyn-sys-supp-results}). 

We plot performance in Figure \ref{fig:jac-est}. While both the JacobianODEs and the NeuralODEs reproduce the observed dynamics (Figure \ref{fig:jac-est}A-D), the JacobianODE learns a more accurate estimate of the Jacobian (Figure \ref{fig:jac-est}
E,F). In particular, looking at Lyapunov spectra learned by the models, we note that the JacobianODE exceeds the other methods in estimating the full Lyapunov exponent spectrum, indicating a better overall representation of how perturbations along different directions will evolve (Figure \ref{fig:jac-est}G,H). 

% \vspace{-4.5mm}
\section{Control-theoretic analyses in a task-trained RNN}
% \vspace{-2mm}
\label{sec:wmtask-control-analysis}

\begin{figure}[ht] % 'h' means "here" (try placing it near the text)
    \centering
    % \vspace{-2mm}
    \includegraphics[width=1.0\linewidth]{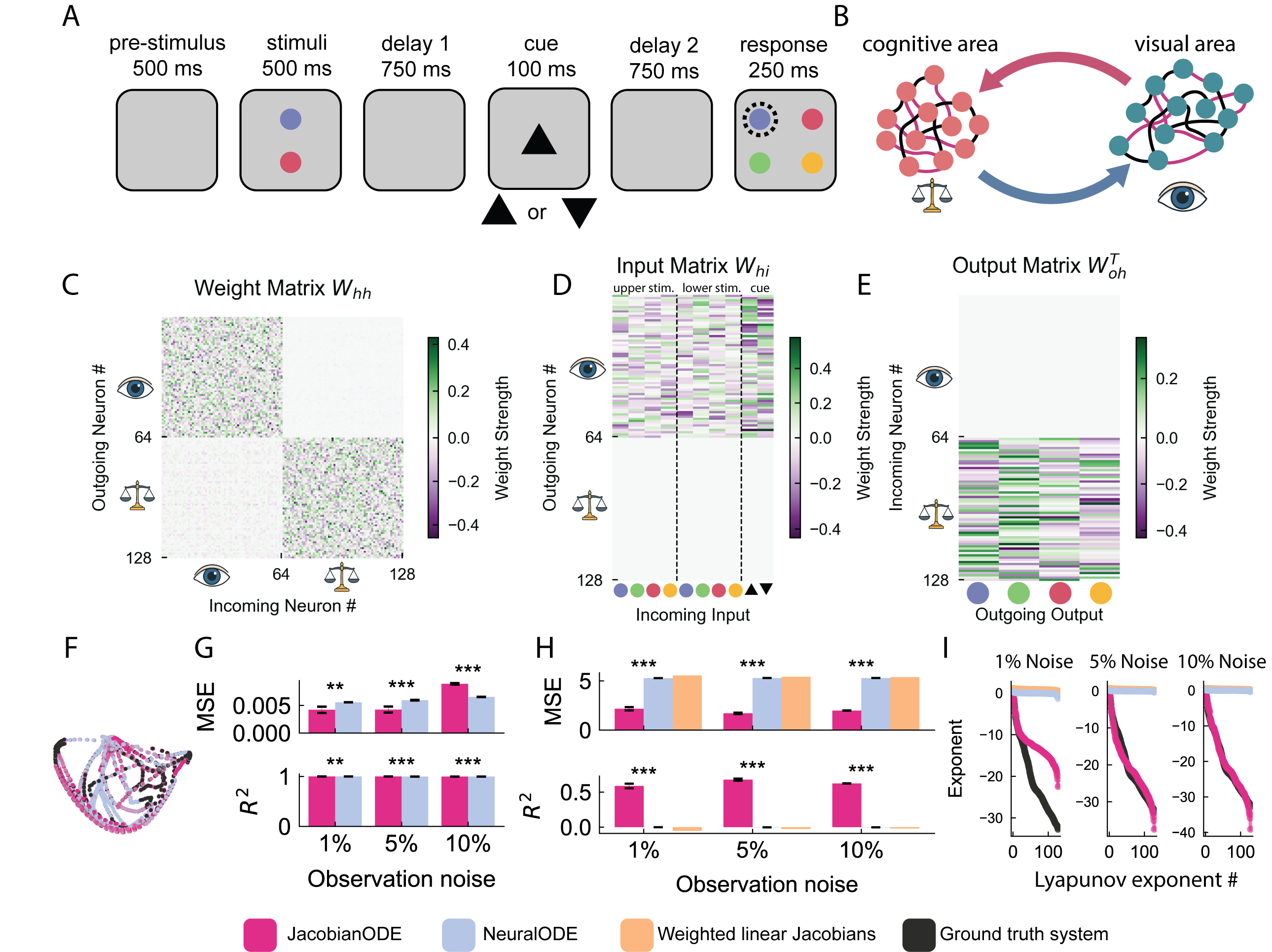} % Adjust width as needed
    % \makebox[\textwidth][c]{\includegraphics[width=1.2\linewidth]{figures/wmtask-1.png}}
    \caption{\textbf{JacobianODE accurately infers trained RNN Jacobians.} Error bars indicate standard deviation, with statistics computed over five different model initializations. (A) Task schematic, involving stimuli presentation, delays, cueing, and response. (B) RNN architecture with distinct visual and cognitive areas interacting. (C–E) Structure of trained weight matrices: (C) recurrent weights within RNN, (D) input connectivity pattern, and (E) output connectivity. (F) State-space trajectories from the trained RNN, with 10 time-step predictions. Spectral embedding was used to reduce the data to three dimensions. (G,H) JacobianODE performance evaluated against other models for trajectory prediction (G) and Jacobian estimation accuracy (H) at multiple noise levels. (I) Estimated Lyapunov spectra averaged over initializations.}
    \label{fig:wmtask-intro}
\end{figure}

\subsection{Characterizing control between subsystems with Jacobian linearization}
\label{sec:framework-v2}
Consider neural data recorded from two areas, A and B. Concatenating their data into a state vector $\mathbf{x} \in \mathbb{R}^n$, composed of $\mathbf{x}^A \in \mathbb{R}^{n_A}$ and $\mathbf{x}^B \in \mathbb{R}^{n_B}$, and assuming dynamics governed by $\mathbf{f}$, we linearize around a reference trajectory ($
\delta\dot{\mathbf{x}}(t) = \mathbf{J}(\mathbf{x}(t))\,\delta \mathbf{x}(t)$). Splitting the Jacobian into block matrices yields:
\begin{equation}
\begin{aligned}
\delta \dot{\mathbf{x}}^A(t) \ &= \ \mathbf{J}^{A \to A}(\mathbf{x}(t))\,\delta \mathbf{x}^A + \mathbf{J}^{B \to A}(\mathbf{x}(t))\,\delta \mathbf{x}^B \\
\delta \dot{\mathbf{x}}^B(t) \ &= \ \mathbf{J}^{A \to B}(\mathbf{x}(t))\,\delta \mathbf{x}^A + \mathbf{J}^{B \to B}(\mathbf{x}(t))\,\delta \mathbf{x}^B
\end{aligned}
\label{eq:area-interaction}
\end{equation}
where diagonal blocks $\mathbf{J}^{A \to A} \in \mathbb{R}^{n_A \times n_A}$ and $\mathbf{J}^{B \to B}\in \mathbb{R}^{n_B \times n_B}$ represent within-area dynamics, and off-diagonal blocks $\mathbf{J}^{B \to A} \in \mathbb{R}^{n_A \times n_B}$, $\mathbf{J}^{A \to B} \in \mathbb{R}^{n_B \times n_A}$ represent interareal interactions (Figure \ref{fig:methods-intro}, right). Explicit separation of each area’s control dynamics quantifies the direct influence each exerts on the other, and readily generalizes beyond two areas (see Appendix \ref{supp:extend-multiple}).

Since Jacobians are time-dependent, we obtain a linear time-varying representation of control dynamics along the trajectory. This enables computation of time-varying reachability ease, capturing how readily each area drives the other toward novel states~\cite{antsaklis2006linear, kawano2021diffgram,lindmark2018ctrl}. Reachability is quantified via the reachability Gramian, a matrix defining a local metric in tangent space. For the above control system capturing the influence of area B on area A, the time-varying reachability Gramian on the interval $[t_0,t_1]$ is defined as
\begin{equation}
\mathbf{W}_r(t_0, t_1) \triangleq \int_{t_0}^{t_1}\mathbf{\Phi}(t_1, \tau)\mathbf{B} (\tau)\mathbf{B}^T(\tau)\mathbf{\Phi}^T(t_1, \tau) d\tau
\end{equation}
where $\mathbf{\Phi}$ (computed from $\mathbf{J}^{A \to A}$) denotes the state-transition matrix of the intrinsic dynamics of subsystem A without any input (i.e., $\delta\mathbf{x}^A(t) = \mathbf{\Phi}(t, t_0)\delta\mathbf{x}^A(t_0)$), and $\mathbf{B}(\tau) = \mathbf{J}^{B \to A}(\mathbf{x}(\tau))$  (see Appendix \ref{supp:gramian-comp}) \cite{antsaklis2006linear}. The Gramian $\mathbf{W}_r(t_0, t_1)$ is symmetric and positive semidefinite for every $t_1 > t_0$ \cite{antsaklis2006linear}. Each eigenvalue of the reachability Gramian quantifies how easily the target system can be driven along its corresponding eigenvector. Thus, the trace of the reachability Gramian reflects average ease of reaching new states, and its minimum eigenvalue reflects the ease along the most challenging direction of control.

\subsection{Estimating the Jacobian of a task-trained RNN}

\insettitle{Task} To demonstrate how JacobianODE models could be used in neuroscience, we performed a control-theoretic analysis of a task-trained RNN. We used a working memory selection task from \citet{panichello2021wmattention} (Figure \ref{fig:wmtask-intro}A). On each trial, the network is presented with two of four possible "colors", denoted by one-hot vectors. After a delay, the network is presented with a cue indicating which of the colors to select. The network is then asked to reach a state of sustained activation that corresponds to the selected color. 

%This task was selected as the neural instantiation of the solution to this task involves subspaces that transform over time, indicating nonlinearity in dynamics \cite{panichello2021wmattention}.

\insettitle{RNN model} To perform this task, we trained a 128-dimensional continuous-time RNN. The RNN had hidden dynamics defined by
\begin{equation}
\begin{aligned}
\tau\dot{\mathbf{h}}(t) &= -\mathbf{h} + \mathbf{W}_{hh}\sigma( \mathbf{h}(t))+ \mathbf{W}_{hi}\mathbf{u}(t) + \mathbf{b} \\
\mathbf{o}(t) &= \mathbf{W}_{oh} \mathbf{h}(t)
\end{aligned}
\end{equation}
where $\mathbf{W}_{hh}$ defines the internal dynamics, $\mathbf{W}_{hi}$ maps the input into the hidden state,  $\mathbf{W}_{oh}$ maps the hidden state to a four-dimensional output $\mathbf{o}(t)$,  $\mathbf{b}$ is a bias term, and $\sigma$ is the exponential linear unit activation. The RNN had two 64-neuron areas: a "visual" area (which received sensory input) and a "cognitive" (which output the RNN's color choice) (Figure \ref{fig:wmtask-intro}B). To encourage multi-area structure, we initialized the within-area weights with greater connectivity strength than the across-area weights (Figure \ref{fig:wmtask-intro}C). Since input comes only to the visual area and output only from the cognitive area, the two areas are forced to interact to solve the task (Figure \ref{fig:wmtask-intro}D,E). The RNN solves this task with 100\% accuracy.

\insettitle{Jacobian reconstruction quality} We trained JacobianODEs on RNN trajectories from the post-cue delay and response portion of the trials. We used the same baselines as in Section \ref{sec:jac-est}. JacobianODEs are not only robust to noise, but can also benefit from it (for multiple systems in Table \ref{tab:fro_metrics}, 5\% training noise improves estimation). Noise encourages the model to explore how perturbations around the observed trajectory evolve, which is crucial for learning accurate Jacobians in high-dimensional systems. Thus (for both the JacobianODE and NeuralODE) we add a small amount of additional noise to the data during learning. Although the models perform similarly on trajectory reconstruction (Figure \ref{fig:wmtask-intro}F,G), on Jacobian estimation, JacobianODEs drastically outperform both baseline models (Figure \ref{fig:wmtask-intro}H,I, and Table \ref{tab:fro_metrics}). 

%The relative difference between the models on the RNN's dynamics is additionally larger than for the previously considered high-dimensional chaotic systems. This is likely because chaotic sensitivity to initial conditions entails modeling perturbation dynamics, whereas this is unnecessary in stable systems.

\begin{figure}[ht] 
    % \vspace{-2mm}
    \centering    \includegraphics[width=1.0\linewidth]{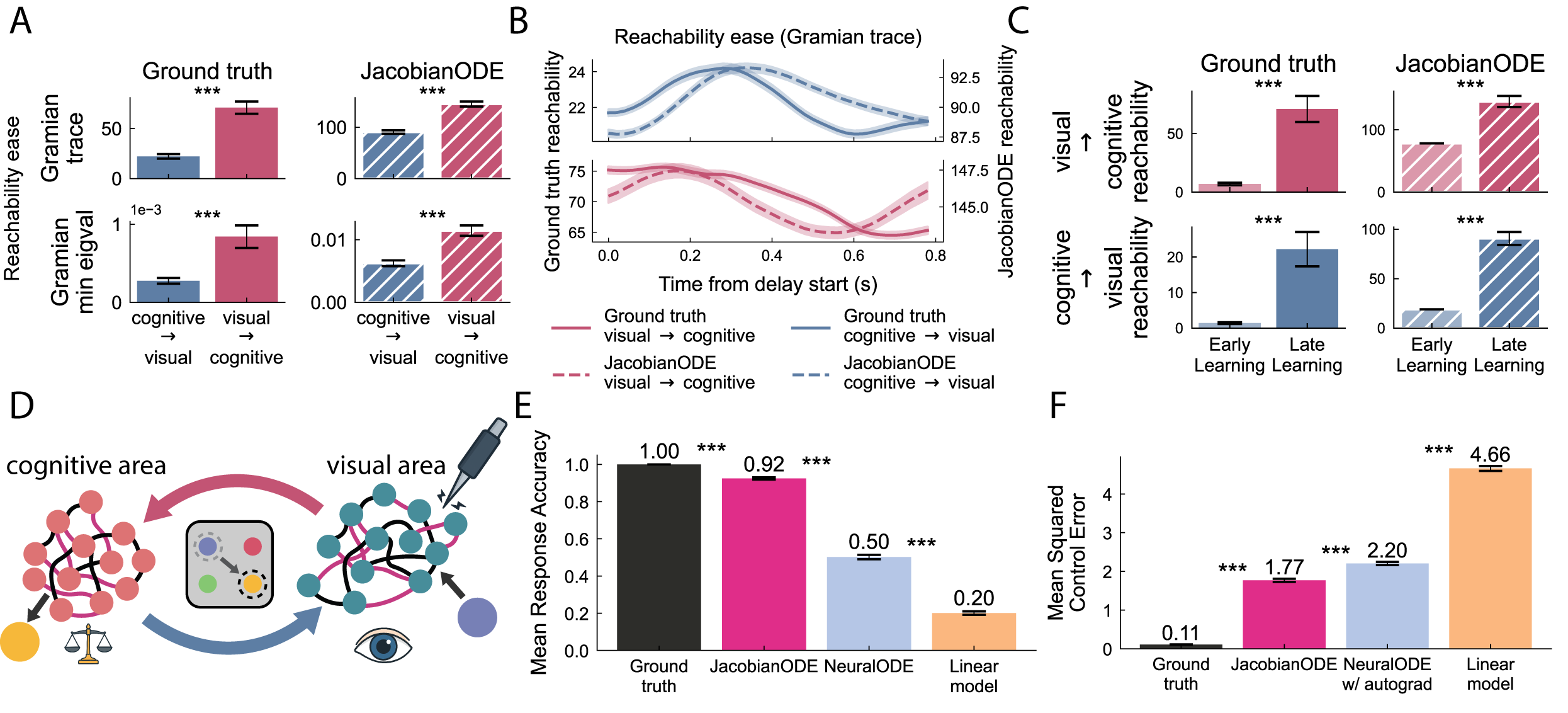} % Adjust width as needed
    \caption{\textbf{JacobianODE reveals differential interareal reachability.} Error bars indicate standard deviation for (A) and (C), and standard error for (B), (E), and (F), with statistics computed over trajectories. (A) Comparison of interareal reachability between ground truth and JacobianODE estimates. (B) Temporal evolution of reachability (Gramian trace) throughout the delay period. (C) Comparison of reachability in early and late learning. (D) Schematic illustrating targeted control of cognitive-area via visual-area stimulation. (E) Mean accuracy of targeted responses using JacobianODE versus alternative methods. (F) MSE for multiple ILQR-based methods.}
    \label{fig:wmtask-control}
    % \vspace{1mm}
\end{figure}

% \vspace{-2mm}
\subsection{Reachability in the task-trained RNN across contexts}
% \vspace{-2mm}
Next, we used the Jacobians learned by the JacobianODE (trained on 5\% noise) to evaluate reachability control in the RNN, and compared it to evaluations using ground truth Jacobians. All analyses were performed using the 10 time-step reachability Gramian. We first found that it was on average easier for the visual area to drive the cognitive area, both when considering overall ease (Gramian trace) and worst-case ease (Gramian minimum eigenvalue) (Figure \ref{fig:wmtask-control}A, larger values indicate greater reachability). We next considered how reachability ease varied throughout the delay period. We found that the visual area could drive the cognitive area more easily at the beginning of the delay, with ease decreasing into the middle of the delay period (Figure \ref{fig:wmtask-control}B, bottom). The cognitive area was able to drive the visual area more easily slightly later in the delay period (Figure \ref{fig:wmtask-control}B, top). Finally, we considered whether reachability changes over the course of learning. We found that both directions of reachability increased after learning (Figure \ref{fig:wmtask-control}C). The JacobianODE reproduced all results accurately (Figure \ref{fig:wmtask-control}A-C, comparison with ground truth). Our analysis reveals that reachability is crucial in the RNN's ability to perform the working memory task, with the visual area's ability to drive the cognitive area shortly after the cue being especially important.

% \vspace{-2mm}
\subsection{Controlling the task-trained RNN}
% \vspace{-1mm}

%To further demonstrate the utility of the learn Jacobians in the context of control, we aimed to control the task-trained RNN using the learned Jacobians from the JacobianODE. Specifically, given a trial in which one of the four colors was cued, we wondered whether we could control the RNN to output a specific incorrect response corresponding to one of the three incorrect colors. Further, we wondered whether this was possible only by providing input to the visual area, making it crucial to have an appropriate representation of the interaction between the visual and cognitive area. 

To further validate our approach, we used JacobianODE-learned Jacobians to control the task-trained RNN (Figure \ref{fig:wmtask-control}D). Given a trial in which the network was cued to respond with a particular color, we tested if we could induce a specific incorrect response by input to the visual area alone. To do so, we implemented Iterative Linear Quadratic Regulator (ILQR) control, which relies on knowledge of the Jacobian \cite{li2004ilqr, tassa2012ilqr}. The controller guided the network towards the mean hidden state of training trials corresponding to the desired incorrect color. We defined accuracy as the percentage of time points during which the RNN output the desired color. We found that the JacobianODE widely outperformed the baseline models on this task, achieving an accuracy nearing that of the ground truth system (Figure \ref{fig:wmtask-control}E). The JacobianODE was furthermore able to achieve the lowest mean squared error on the desired control trajectory (Figure \ref{fig:wmtask-control}F). This illustrates that while both the JacobianODE and the NeuralODE can learn the dynamics, only the JacobianODE learns a representation of the Jacobian that is sufficient for control.

\vspace{-2mm}
\section{Discussion}
\label{sec:discussion}
\vspace{-2mm}
\insettitle{Extended Jacobian estimation} A natural extension of JacobianODE models would add inductive biases for particular classes of dynamics. For example, one could parameterize a negative definite matrix and thus ensure contracting dynamics \cite{beik-mohammadi2024-ncds}. Other extensions could include an L1 penalty to encourage sparsity, as well as the inclusion of known dynamic structure (e.g., hierarchical structure, low-rank structure, etc.). In neuroscience, connectomic constraints could be incorporated to capture interactions within a neural circuit. 

% We chose to enforce the conservation constraint on the rows of the Jacobian with a loop closure loss. However, other choices are possible. Namely, since each row, of the Jaocbian corresponds to the gradient of a scalar function, taking a second gradient should result in a Hessian, which must be symmetric. Thus a regularization enforcing the symmetry of these Hessians would be another possible choice--though back-propagating through this loss could be prohibitively computationally costly.

%Another approach to combatting measurement noise--which was handled here through robust dynamics estimation--could be to simultaneously estimate a Kalman filter. Future work should investigate whether this approach leads to overall better performance on Jacobian learning tasks.

\insettitle{Limitations} A future challenge for JacobianODE models is partially observed dynamics. Recent work has identified that it is possible to learn latent embeddings that approximately recover the true state from partial observation \cite{ouala2020partial,gilpin2020deeprecon,lu2022recon,ozalp2023recon,young2023recon}.  Jacobian-based dynamics learning has been performed in a latent space \cite{beik-mohammadi2024-ncds, jaffe2024elcd}, however it is unclear whether this translates to accurate Jacobian estimation, given the challenges related to automatic differentiation and Jacobian estimation presented here. We also note that reachability estimates depend sensitively on several factors: the alignment of cross-subsystem interactions and within-subsystem dynamics, the eigenvectors and eigenvalues of within-subsystem Jacobians, and the way activity propagates within each subsystem (see Appendix \ref{supp:control-v-comm}). While our method reliably captures broad trends in reachability over time, fine-grained, time-point-specific comparisons should be interpreted with caution. Finally, although JacobianODE models scale well to moderately high-dimensional systems, their performance in systems that are orders of magnitude larger than those considered here (e.g., recordings of thousands of neurons via calcium imaging) remains to be tested. In many practical settings, this may not be a limitation: neural representations during tasks often exhibit intrinsic dimensionalities comparable to those studied here, enabling JacobianODE models to operate in reduced dimensionality (see Appendix \ref{supp:dimensionality}). Notably, most models converged in under 45 minutes on a single GPU, suggesting favorable scaling (see Table \ref{tab:hyperparameters}). Future work should explore the viability of the method in higher dimensions, and assess whether dimensionality reduction strategies (such as latent state models or low-rank Jacobian approximations) can further improve scalability and accuracy.

\insettitle{Broader impact on dynamical systems} We demonstrated in this work that it is possible to learn high accuracy Jacobians of arbitrary dynamical systems from data. While in this work we focus on control-theoretic analyses, we emphasize that JacobianODE models can be trained in any dynamical setting. Jacobians are widely applicable, and are essential in stability analysis \cite{strogatz2014dyn,sydel2009stab,slotine1991appnlctrl,dieci1997lyap,christiansen1997lyap,duchaine2008robotstab}, contraction theory \cite{lohmiller1998contraction,slotine1991appnlctrl,manchester2017controlcontraction,wensing2020contractgrad,kozachkov2020stable,kozachkov2022rnns}, and nonlinear control \cite{li2004ilqr, tassa2012ilqr,tyner2010geojac,hootsmans2002transjac,moosavian2007transjac,tchon2015lagrangejac}, among many other settings.

\insettitle{Broader impact on control-theoretic analyses} The control-theoretic analysis presented here is performed with respect to a particular reference trajectory. It therefore describes the local controllability properties along the reference trajectory, rather than global properties for the overall system \cite{tyner2010geojac}. This locality is desirable in biological settings due to the context-sensitive nature of the relevant processes. Applications of this approach could involve a comparison of the functioning of the autonomic nervous system, or gene regulatory networks, between control and disease states, given the association of these processes with disease \cite{brudey2015autonomicptsd,alvares2016autonomic,goldberger2019autonomic,xiong2019autonomic, barabasi2011genereg,maurano2012genereg,madhamshettiwar2012genereg,lee2013genereg,unger-avila2024genereg}. Future work should investigate whether a more global Jacobian-based control analysis (e.g., contraction theory) is achievable in a data-driven manner via JacobianODE models. This would be useful especially in engineering settings where global controllability is desirable, such as robotics and prosthetic limbs.

% The analysis described here While in certain settings (e.g. engineering) this is less desireable, in neuroscience and biology we may often wish to characterize controllability and reachability in different external and internal contexts (i.e., along different reference trajectories). 

\begin{ack}
The authors would like to thank Andrew Kirjner, Laureline Logiaco, Federico Claudi, Lakshmi Narasimhan Govindarajan, Jaedong Hwang, and Akhilan Boopathy for providing stimulating discussion about this work.  The authors also acknowledge the MIT Office of Research Computing and Data for providing high-performance computing resources that have contributed to the research results reported within this paper. I.R.F. acknowledges funding from The National Science Foundation Computer and Information Science and Engineering Directorate, The Simons Collaboration on the Global Brain, and The McGovern Institute at MIT. E.K.M. acknowledges funding from ONR MURI N00014-23-1-2768, NIMH 1R01MH131715-01, NIMH R01MH11559, The Simons Center for the Social Brain, The Freedom Together Foundation, and The Picower Institute for Learning and Memory. L.K. acknowledges funding from the Goldstine Postdoctoral Fellowship at IBM Research. M.O. acknowledges funding from NSF GRFP 2141064.

\begin{comment}Use unnumbered first level headings for the acknowledgments. All acknowledgments
go at the end of the paper before the list of references. Moreover, you are required to declare
funding (financial activities supporting the submitted work) and competing interests (related financial activities outside the submitted work).
More information about this disclosure can be found at: \url{https://neurips.cc/Conferences/2025/PaperInformation/FundingDisclosure}.

Do {\bf not} include this section in the anonymized submission, only in the final paper. You can use the \texttt{ack} environment provided in the style file to automatically hide this section in the anonymized submission.
\end{comment}
\end{ack}

\bibliographystyle{unsrtnat}
\bibliography{biblio}

@article{lorraine2024jacnet,
  title={Jacnet: Learning functions with structured jacobians},
  author={Lorraine, Jonathan and Hossain, Safwan},
  journal={arXiv preprint arXiv:2408.13237},
  year={2024}
}

@ARTICLE{latremoliere2022jacest,
  title         = "Estimating the Jacobian matrix of an unknown multivariate
                   function from sample values by means of a neural network",
  author        = "Latrémolière, Frédéric and Narayanappa, Sadananda and
                   Vojtěchovský, Petr",
  journal       = "arXiv [cs.LG]",
  month         =  apr,
  year          =  2022,
  archivePrefix = "arXiv",
  primaryClass  = "cs.LG"
}

@ARTICLE{crick1990consc,
  title   = "Towards a neurobiological theory of consciousness",
  author  = "Crick, F and Koch, C",
  journal = "Seminars in the Neurosciences",
  volume  =  2,
  pages   = "263--275",
  year    =  1990
}

@INCOLLECTION{engel2016consc,
  title     = "Chapter 3 - Neuronal Oscillations, Coherence, and Consciousness",
  author    = "Engel, Andreas K and Fries, Pascal",
  editor    = "Laureys, Steven and Gosseries, Olivia and Tononi, Giulio",
  booktitle = "The Neurology of Conciousness (Second Edition)",
  publisher = "Academic Press",
  address   = "San Diego",
  pages     = "49--60",
  month     =  "1~" # jan,
  year      =  2016
}

@ARTICLE{thompson2001consc,
  title    = "Radical embodiment: neural dynamics and consciousness",
  author   = "Thompson, Evan and Varela, Francisco J",
  journal  = "Trends Cogn. Sci.",
  volume   =  5,
  number   =  10,
  pages    = "418--425",
  month    =  "1~" # oct,
  year     =  2001,
  language = "en"
}

@ARTICLE{ward2003cog,
  title    = "Synchronous neural oscillations and cognitive processes",
  author   = "Ward, Lawrence M",
  journal  = "Trends Cogn. Sci.",
  volume   =  7,
  number   =  12,
  pages    = "553--559",
  month    =  dec,
  year     =  2003,
  language = "en"
}

@ARTICLE{palva2007osc,
  title    = "New vistas for alpha-frequency band oscillations",
  author   = "Palva, Satu and Palva, J Matias",
  journal  = "Trends Neurosci.",
  volume   =  30,
  number   =  4,
  pages    = "150--158",
  month    =  apr,
  year     =  2007,
  language = "en"
}

@ARTICLE{buschman2007attention,
  title    = "Top-down versus bottom-up control of attention in the prefrontal
              and posterior parietal cortices",
  author   = "Buschman, Timothy J and Miller, Earl K",
  journal  = "Science",
  volume   =  315,
  number   =  5820,
  pages    = "1860--1862",
  month    =  "30~" # mar,
  year     =  2007,
  language = "en"
}

@ARTICLE{gregoriou2009attention,
  title    = "High-frequency, long-range coupling between prefrontal and visual
              cortex during attention",
  author   = "Gregoriou, Georgia G and Gotts, Stephen J and Zhou, Huihui and
              Desimone, Robert",
  journal  = "Science",
  volume   =  324,
  number   =  5931,
  pages    = "1207--1210",
  month    =  "29~" # may,
  year     =  2009,
  language = "en"
}

@ARTICLE{gregoriou2012attention,
  title    = "Cell-type-specific synchronization of neural activity in {FEF}
              with {V4} during attention",
  author   = "Gregoriou, Georgia G and Gotts, Stephen J and Desimone, Robert",
  journal  = "Neuron",
  volume   =  73,
  number   =  3,
  pages    = "581--594",
  month    =  "9~" # feb,
  year     =  2012,
  language = "en"
}

@ARTICLE{salinas2001correlated,
  title    = "Correlated neuronal activity and the flow of neural information",
  author   = "Salinas, E and Sejnowski, T J",
  journal  = "Nat. Rev. Neurosci.",
  volume   =  2,
  number   =  8,
  pages    = "539--550",
  month    =  aug,
  year     =  2001,
  language = "en"
}

@ARTICLE{fries2001attention,
  title    = "Modulation of oscillatory neuronal synchronization by selective
              visual attention",
  author   = "Fries, P and Reynolds, J H and Rorie, A E and Desimone, R",
  journal  = "Science",
  volume   =  291,
  number   =  5508,
  pages    = "1560--1563",
  month    =  "23~" # feb,
  year     =  2001,
  language = "en"
}

@ARTICLE{siegel2008attention,
  title    = "Neuronal synchronization along the dorsal visual pathway reflects
              the focus of spatial attention",
  author   = "Siegel, Markus and Donner, Tobias H and Oostenveld, Robert and
              Fries, Pascal and Engel, Andreas K",
  journal  = "Neuron",
  volume   =  60,
  number   =  4,
  pages    = "709--719",
  month    =  "26~" # nov,
  year     =  2008,
  language = "en"
}

@ARTICLE{vankempen2021attention,
  title    = "Top-down coordination of local cortical state during selective
              attention",
  author   = "van Kempen, Jochem and Gieselmann, Marc A and Boyd, Michael and
              Steinmetz, Nicholas A and Moore, Tirin and Engel, Tatiana A and
              Thiele, Alexander",
  journal  = "Neuron",
  volume   =  109,
  number   =  5,
  pages    = "894--904.e8",
  month    =  "3~" # mar,
  year     =  2021,
  language = "en"
}

@ARTICLE{gray1989osc,
  title    = "Oscillatory responses in cat visual cortex exhibit inter-columnar
              synchronization which reflects global stimulus properties",
  author   = "Gray, C M and König, P and Engel, A K and Singer, W",
  journal  = "Nature",
  volume   =  338,
  number   =  6213,
  pages    = "334--337",
  month    =  "23~" # mar,
  year     =  1989,
  language = "en"
}

@ARTICLE{siegel2015decision,
  title    = "Cortical information flow during flexible sensorimotor decisions",
  author   = "Siegel, Markus and Buschman, Timothy J and Miller, Earl K",
  journal  = "Science",
  volume   =  348,
  number   =  6241,
  pages    = "1352--1355",
  month    =  "19~" # jun,
  year     =  2015,
  language = "en"
}

@ARTICLE{siegel2011decision,
  title    = "Cortical network dynamics of perceptual decision-making in the
              human brain",
  author   = "Siegel, Markus and Engel, Andreas K and Donner, Tobias H",
  journal  = "Front. Hum. Neurosci.",
  volume   =  5,
  pages    =  21,
  month    =  "28~" # feb,
  year     =  2011,
  language = "en"
}

@ARTICLE{siegel2012cog,
  title    = "Spectral fingerprints of large-scale neuronal interactions",
  author   = "Siegel, Markus and Donner, Tobias H and Engel, Andreas K",
  journal  = "Nat. Rev. Neurosci.",
  volume   =  13,
  number   =  2,
  pages    = "121--134",
  month    =  "11~" # jan,
  year     =  2012,
  language = "en"
}

@ARTICLE{brincat2021wm,
  title    = "Interhemispheric transfer of working memories",
  author   = "Brincat, Scott L and Donoghue, Jacob A and Mahnke, Meredith K and
              Kornblith, Simon and Lundqvist, Mikael and Miller, Earl K",
  journal  = "Neuron",
  volume   =  109,
  number   =  6,
  pages    = "1055--1066.e4",
  month    =  "17~" # mar,
  year     =  2021,
  language = "en"
}

@ARTICLE{salazar2012wm,
  title    = "Content-specific fronto-parietal synchronization during visual
              working memory",
  author   = "Salazar, R F and Dotson, N M and Bressler, S L and Gray, C M",
  journal  = "Science",
  volume   =  338,
  number   =  6110,
  pages    = "1097--1100",
  month    =  "23~" # nov,
  year     =  2012,
  language = "en"
}

@ARTICLE{engel2001fb,
  title    = "Temporal binding and the neural correlates of sensory awareness",
  author   = "Engel, A K and Singer, W",
  journal  = "Trends Cogn. Sci.",
  volume   =  5,
  number   =  1,
  pages    = "16--25",
  month    =  "1~" # jan,
  year     =  2001,
  language = "en"
}

@ARTICLE{singer1995fb,
  title    = "Visual feature integration and the temporal correlation hypothesis",
  author   = "Singer, W and Gray, C M",
  journal  = "Annu. Rev. Neurosci.",
  volume   =  18,
  pages    = "555--586",
  year     =  1995,
  language = "en"
}

@ARTICLE{murthy1996mc,
  title    = "Oscillatory activity in sensorimotor cortex of awake monkeys:
              synchronization of local field potentials and relation to behavior",
  author   = "Murthy, V N and Fetz, E E",
  journal  = "J. Neurophysiol.",
  volume   =  76,
  number   =  6,
  pages    = "3949--3967",
  month    =  dec,
  year     =  1996,
  language = "en"
}

@ARTICLE{logiaco2021mc,
  title    = "Thalamic control of cortical dynamics in a model of flexible motor
              sequencing",
  author   = "Logiaco, Laureline and Abbott, L F and Escola, Sean",
  journal  = "Cell Rep.",
  volume   =  35,
  number   =  9,
  pages    =  109090,
  month    =  "1~" # jun,
  year     =  2021,
  language = "en"
}

@ARTICLE{arcemcshane2016mc,
  title    = "Primary motor and sensory cortical areas communicate via
              spatiotemporally coordinated networks at multiple frequencies",
  author   = "Arce-McShane, Fritzie I and Ross, Callum F and Takahashi, Kazutaka
              and Sessle, Barry J and Hatsopoulos, Nicholas G",
  journal  = "Proc. Natl. Acad. Sci. U. S. A.",
  volume   =  113,
  number   =  18,
  pages    = "5083--5088",
  month    =  "3~" # may,
  year     =  2016,
  language = "en"
}

@ARTICLE{perich2018mc,
  title    = "A Neural Population Mechanism for Rapid Learning",
  author   = "Perich, Matthew G and Gallego, Juan A and Miller, Lee E",
  journal  = "Neuron",
  volume   =  100,
  number   =  4,
  pages    = "964--976.e7",
  month    =  "21~" # nov,
  year     =  2018,
  language = "en"
}

@ARTICLE{kaufman2014mc,
  title    = "Cortical activity in the null space: permitting preparation
              without movement",
  author   = "Kaufman, Matthew T and Churchland, Mark M and Ryu, Stephen I and
              Shenoy, Krishna V",
  journal  = "Nat. Neurosci.",
  volume   =  17,
  number   =  3,
  pages    = "440--448",
  month    =  mar,
  year     =  2014,
  language = "en"
}

@ARTICLE{bricat2015learn,
  title    = "Frequency-specific hippocampal-prefrontal interactions during
              associative learning",
  author   = "Brincat, Scott L and Miller, Earl K",
  journal  = "Nat. Neurosci.",
  volume   =  18,
  number   =  4,
  pages    = "576--581",
  month    =  apr,
  year     =  2015,
  language = "en"
}

@ARTICLE{jones2005learn,
  title    = "Theta rhythms coordinate hippocampal-prefrontal interactions in a
              spatial memory task",
  author   = "Jones, Matthew W and Wilson, Matthew A",
  journal  = "PLoS Biol.",
  volume   =  3,
  number   =  12,
  pages    = "e402",
  month    =  dec,
  year     =  2005,
  language = "en"
}

@ARTICLE{siapas1998learn,
  title    = "Coordinated interactions between hippocampal ripples and cortical
              spindles during slow-wave sleep",
  author   = "Siapas, A G and Wilson, M A",
  journal  = "Neuron",
  volume   =  21,
  number   =  5,
  pages    = "1123--1128",
  month    =  nov,
  year     =  1998,
  language = "en"
}

@ARTICLE{fernandezruiz2021learn,
  title     = "Gamma rhythm communication between entorhinal cortex and dentate
               gyrus neuronal assemblies",
  author    = "Fernández-Ruiz, Antonio and Oliva, Azahara and Soula, Marisol and
               Rocha-Almeida, Florbela and Nagy, Gergo A and Martin-Vazquez,
               Gonzalo and Buzsáki, György",
  journal   = "Science",
  publisher = "American Association for the Advancement of Science",
  volume    =  372,
  number    =  6537,
  month     =  "2~" # apr,
  year      =  2021,
  language  = "en"
}

@ARTICLE{kao2019ctrl,
  title    = "Neuroscience out of control: control-theoretic perspectives on
              neural circuit dynamics",
  author   = "Kao, Ta-Chu and Hennequin, Guillaume",
  journal  = "Curr. Opin. Neurobiol.",
  volume   =  58,
  pages    = "122--129",
  month    =  oct,
  year     =  2019,
  language = "en"
}

@ARTICLE{kao2021ctrl,
  title    = "Optimal anticipatory control as a theory of motor preparation: A
              thalamo-cortical circuit model",
  author   = "Kao, Ta-Chu and Sadabadi, Mahdieh S and Hennequin, Guillaume",
  journal  = "Neuron",
  volume   =  109,
  number   =  9,
  pages    = "1567--1581.e12",
  month    =  "5~" # may,
  year     =  2021,
  language = "en"
}

@ARTICLE{bouchard2024ctrl,
  title   = "Feedback Controllability is a Normative Theory of Neural Population
             Dynamics",
  author  = "Bouchard, Kristofer and Kumar, Ankit",
  journal = "Research Square",
  month   =  "29~" # mar,
  year    =  2024
}

@INPROCEEDINGS{schimel2021ctrl,
  title     = "{iLQR}-{VAE} : control-based learning of input-driven dynamics
               with applications to neural data",
  author    = "Schimel, Marine and Kao, Ta-Chu and Jensen, Kristopher T and
               Hennequin, Guillaume",
  booktitle = "International Conference on Learning Representations",
  month     =  "6~" # oct,
  year      =  2021
}

@ARTICLE{braun2021netctrl,
  title    = "Brain network dynamics during working memory are modulated by
              dopamine and diminished in schizophrenia",
  author   = "Braun, Urs and Harneit, Anais and Pergola, Giulio and Menara,
              Tommaso and Schäfer, Axel and Betzel, Richard F and Zang,
              Zhenxiang and Schweiger, Janina I and Zhang, Xiaolong and Schwarz,
              Kristina and Chen, Junfang and Blasi, Giuseppe and Bertolino,
              Alessandro and Durstewitz, Daniel and Pasqualetti, Fabio and
              Schwarz, Emanuel and Meyer-Lindenberg, Andreas and Bassett,
              Danielle S and Tost, Heike",
  journal  = "Nat. Commun.",
  volume   =  12,
  number   =  1,
  pages    =  3478,
  month    =  "9~" # jun,
  year     =  2021,
  language = "en"
}

@ARTICLE{zhou2023netctrl,
  title    = "Mindful attention promotes control of brain network dynamics for
              self-regulation and discontinues the past from the present",
  author   = "Zhou, Dale and Kang, Yoona and Cosme, Danielle and Jovanova, Mia
              and He, Xiaosong and Mahadevan, Arun and Ahn, Jeesung and Stanoi,
              Ovidia and Brynildsen, Julia K and Cooper, Nicole and Cornblath,
              Eli J and Parkes, Linden and Mucha, Peter J and Ochsner, Kevin N
              and Lydon-Staley, David M and Falk, Emily B and Bassett, Dani S",
  journal  = "Proc. Natl. Acad. Sci. U. S. A.",
  volume   =  120,
  number   =  2,
  pages    = "e2201074119",
  month    =  "10~" # jan,
  year     =  2023,
  language = "en"
}

@ARTICLE{zoller2021netctrl,
  title    = "Structural control energy of resting-state functional brain states
              reveals less cost-effective brain dynamics in psychosis
              vulnerability",
  author   = "Zöller, Daniela and Sandini, Corrado and Schaer, Marie and Eliez,
              Stephan and Bassett, Danielle S and Van De Ville, Dimitri",
  journal  = "Hum. Brain Mapp.",
  volume   =  42,
  number   =  7,
  pages    = "2181--2200",
  month    =  may,
  year     =  2021,
  language = "en"
}

@ARTICLE{gu2015netctrl,
  title    = "Controllability of structural brain networks",
  author   = "Gu, Shi and Pasqualetti, Fabio and Cieslak, Matthew and Telesford,
              Qawi K and Yu, Alfred B and Kahn, Ari E and Medaglia, John D and
              Vettel, Jean M and Miller, Michael B and Grafton, Scott T and
              Bassett, Danielle S",
  journal  = "Nat. Commun.",
  volume   =  6,
  pages    =  8414,
  month    =  "1~" # oct,
  year     =  2015,
  language = "en"
}

@ARTICLE{singleton2022netctrl,
  title     = "Receptor-informed network control theory links {LSD} and
               psilocybin to a flattening of the brain’s control energy
               landscape",
  author    = "Singleton, S Parker and Luppi, Andrea I and Carhart-Harris, Robin
               L and Cruzat, Josephine and Roseman, Leor and Nutt, David J and
               Deco, Gustavo and Kringelbach, Morten L and Stamatakis, Emmanuel
               A and Kuceyeski, Amy",
  journal   = "Nat. Commun.",
  publisher = "Nature Publishing Group",
  volume    =  13,
  number    =  1,
  pages     = "1--13",
  month     =  "3~" # oct,
  year      =  2022,
  language  = "en"
}

@ARTICLE{beik-mohammadi2024-ncds,
  title         = "Neural Contractive Dynamical Systems",
  author        = "Beik-Mohammadi, Hadi and Hauberg, Søren and Arvanitidis,
                   Georgios and Figueroa, Nadia and Neumann, Gerhard and Rozo,
                   Leonel",
  journal       = "arXiv [cs.RO]",
  month         =  "17~" # jan,
  year          =  2024,
  archivePrefix = "arXiv",
  primaryClass  = "cs.RO"
}

@ARTICLE{iyer2024velocities,
  title     = "Flexible mapping of abstract domains by grid cells via
               self-supervised extraction and projection of generalized velocity
               signals",
  author    = "Iyer, Abhiram and Chandra, Sarthak and Sharma, Sugandha and
               Fiete, I",
  editor    = "Globerson, A and Mackey, L and Belgrave, D and Fan, A and Paquet,
               U and Tomczak, J and Zhang, C",
  journal   = "Neural Inf Process Syst",
  publisher = "Curran Associates, Inc.",
  volume    =  37,
  pages     = "85441--85466",
  year      =  2024
}

@ARTICLE{hess2023generalizedtf,
  title         = "Generalized teacher forcing for learning chaotic dynamics",
  author        = "Hess, Florian and Monfared, Zahra and Brenner, Manuel and
                   Durstewitz, Daniel",
  journal       = "arXiv [cs.LG]",
  month         =  "7~" # jun,
  year          =  2023,
  archivePrefix = "arXiv",
  primaryClass  = "cs.LG"
}

@ARTICLE{cai2021dynamiccausal,
  title     = "Dynamic causal brain circuits during working memory and their
               functional controllability",
  author    = "Cai, Weidong and Ryali, Srikanth and Pasumarthy, Ramkrishna and
               Talasila, Viswanath and Menon, Vinod",
  journal   = "Nat. Commun.",
  publisher = "Springer Science and Business Media LLC",
  volume    =  12,
  number    =  1,
  pages     =  3314,
  month     =  "29~" # jun,
  year      =  2021,
  language  = "en"
}

@ARTICLE{parkes2024nct,
  title     = "A network control theory pipeline for studying the dynamics of
               the structural connectome",
  author    = "Parkes, Linden and Kim, Jason Z and Stiso, Jennifer and
               Brynildsen, Julia K and Cieslak, Matthew and Covitz, Sydney and
               Gur, Raquel E and Gur, Ruben C and Pasqualetti, Fabio and
               Shinohara, Russell T and Zhou, Dale and Satterthwaite, Theodore D
               and Bassett, Dani S",
  journal   = "Nat. Protoc.",
  publisher = "Springer Science and Business Media LLC",
  volume    =  19,
  number    =  12,
  pages     = "3721--3749",
  month     =  "29~" # dec,
  year      =  2024,
  language  = "en"
}

@BOOK{antsaklis2006linear,
  title     = "Linear Systems",
  author    = "Antsaklis, Panos J and Michel, Anthony N",
  publisher = "Birkhäuser Boston",
  year      =  2006
}

@ARTICLE{kawano2021diffgram,
  title     = "Empirical differential Gramians for nonlinear model reduction",
  author    = "Kawano, Yu and Scherpen, Jacquelien M A",
  journal   = "Automatica (Oxf.)",
  publisher = "Elsevier BV",
  volume    =  127,
  number    =  109534,
  pages     =  109534,
  month     =  may,
  year      =  2021,
  language  = "en"
}

@ARTICLE{tyner2010geojac,
  title     = "Geometric Jacobian linearization and {LQR} theory",
  author    = "R. Tyner, David and D. Lewis, Andrew",
  journal   = "J. Geom. Mech.",
  publisher = "American Institute of Mathematical Sciences (AIMS)",
  volume    =  2,
  number    =  4,
  pages     = "397--440",
  year      =  2010,
  language  = "en"
}

@ARTICLE{wang2016geocont,
  title    = "A geometrical approach to control and controllability of nonlinear
              dynamical networks",
  author   = "Wang, Le-Zhi and Su, Ri-Qi and Huang, Zi-Gang and Wang, Xiao and
              Wang, Wen-Xu and Grebogi, Celso and Lai, Ying-Cheng",
  journal  = "Nat. Commun.",
  volume   =  7,
  pages    =  11323,
  month    =  "14~" # apr,
  year     =  2016,
  language = "en"
}

@ARTICLE{wu2024brainnct,
  title     = "Development of the brain network control theory and its
               implications",
  author    = "Wu, Zhoukang and Huang, Liangjiecheng and Wang, Min and He,
               Xiaosong",
  journal   = "Psychoradiology",
  publisher = "Oxford University Press (OUP)",
  volume    =  4,
  pages     = "kkae028",
  month     =  "14~" # dec,
  year      =  2024,
  language  = "en"
}

@ARTICLE{he2022ctrlenergy,
  title     = "Uncovering the biological basis of control energy: Structural and
               metabolic correlates of energy inefficiency in temporal lobe
               epilepsy",
  author    = "He, Xiaosong and Caciagli, Lorenzo and Parkes, Linden and Stiso,
               Jennifer and Karrer, Teresa M and Kim, Jason Z and Lu, Zhixin and
               Menara, Tommaso and Pasqualetti, Fabio and Sperling, Michael R
               and Tracy, Joseph I and Bassett, Dani S",
  journal   = "Sci. Adv.",
  publisher = "American Association for the Advancement of Science",
  volume    =  8,
  number    =  45,
  pages     = "eabn2293",
  month     =  "11~" # nov,
  year      =  2022,
  language  = "en"
}

@ARTICLE{medaglia2018netctrl,
  title     = "Network controllability in the inferior frontal gyrus relates to
               controlled language variability and susceptibility to {TMS}",
  author    = "Medaglia, John D and Harvey, Denise Y and White, Nicole and
               Kelkar, Apoorva and Zimmerman, Jared and Bassett, Danielle S and
               Hamilton, Roy H",
  journal   = "J. Neurosci.",
  publisher = "Society for Neuroscience",
  volume    =  38,
  number    =  28,
  pages     = "6399--6410",
  month     =  "11~" # jul,
  year      =  2018,
  language  = "en"
}

@ARTICLE{amani2024structfunc,
  title     = "Controllability of functional and structural brain networks",
  author    = "Moradi Amani, Ali and Tahmassebi, Amirhessam and Stadlbauer,
               Andreas and Meyer-Baese, Uwe and Noblet, Vincent and Blanc,
               Frederic and Malberg, Hagen and Meyer-Baese, Anke",
  journal   = "Complexity",
  publisher = "Wiley",
  volume    =  2024,
  number    =  1,
  pages     =  7402894,
  month     =  "1~" # jan,
  year      =  2024,
  language  = "en"
}

@ARTICLE{li2023funcnet,
  title     = "Controllability of functional brain networks and its clinical
               significance in first-episode schizophrenia",
  author    = "Li, Qian and Yao, Li and You, Wanfang and Liu, Jiang and Deng,
               Shikuang and Li, Bin and Luo, Lekai and Zhao, Youjin and Wang,
               Yuxia and Wang, Yaxuan and Zhang, Qian and Long, Fenghua and
               Sweeney, John A and Gu, Shi and Li, Fei and Gong, Qiyong",
  journal   = "Schizophr. Bull.",
  publisher = "Oxford University Press (OUP)",
  volume    =  49,
  number    =  3,
  pages     = "659--668",
  month     =  "3~" # may,
  year      =  2023,
  language  = "en"
}

@ARTICLE{tang2012driver,
  title     = "Identifying controlling nodes in neuronal networks in different
               scales",
  author    = "Tang, Yang and Gao, Huijun and Zou, Wei and Kurths, Jürgen",
  journal   = "PLoS One",
  publisher = "Public Library of Science (PLoS)",
  volume    =  7,
  number    =  7,
  pages     = "e41375",
  month     =  "27~" # jul,
  year      =  2012,
  language  = "en"
}

@ARTICLE{kumar2014inputnovelty,
  title         = "Input novelty as a control metric for time varying linear
                   systems",
  author        = "Kumar, Gautam and Menolascino, Delsin and Ching, Shinung",
  journal       = "arXiv [math.OC]",
  month         =  "21~" # nov,
  year          =  2014,
  archivePrefix = "arXiv",
  primaryClass  = "math.OC"
}

@BOOK{slotine1991appnlctrl,
  title     = "Applied Nonlinear Control",
  author    = "Slotine, Jean-Jacques E and Li, Weiping",
  publisher = "Prentice-Hall",
  year      =  1991,
  language  = "en"
}

@ARTICLE{brockett1976nonlineardiff,
  title     = "Nonlinear systems and differential geometry",
  author    = "Brockett, R W",
  journal   = "Proc. IEEE Inst. Electr. Electron. Eng.",
  publisher = "Institute of Electrical and Electronics Engineers (IEEE)",
  volume    =  64,
  number    =  1,
  pages     = "61--72",
  year      =  1976,
  language  = "en"
}

@ARTICLE{haynes1970lie,
  title     = "Nonlinear controllability via lie theory",
  author    = "Haynes, G W and Hermes, H",
  journal   = "SIAM J. Control",
  publisher = "Society for Industrial \& Applied Mathematics (SIAM)",
  volume    =  8,
  number    =  4,
  pages     = "450--460",
  month     =  nov,
  year      =  1970,
  language  = "en"
}

@INPROCEEDINGS{aguilar2014netctrl,
  title     = "Necessary conditions for controllability of nonlinear networked
               control systems",
  author    = "Aguilar, Cesar O and Gharesifard, Bahman",
  booktitle = "2014 American Control Conference",
  publisher = "IEEE",
  pages     = "5379--5383",
  month     =  jun,
  year      =  2014,
  language  = "en"
}

@ARTICLE{whalen2015lie,
  title     = "Observability and controllability of nonlinear networks: The role
               of symmetry",
  author    = "Whalen, Andrew J and Brennan, Sean N and Sauer, Timothy D and
               Schiff, Steven J",
  journal   = "Phys. Rev. X.",
  publisher = "American Physical Society (APS)",
  volume    =  5,
  number    =  1,
  pages     =  011005,
  month     =  "23~" # jan,
  year      =  2015,
  language  = "en"
}

@ARTICLE{lindmark2018ctrl,
  title    = "Minimum energy control for complex networks",
  author   = "Lindmark, Gustav and Altafini, Claudio",
  journal  = "Sci. Rep.",
  volume   =  8,
  number   =  1,
  pages    =  3188,
  month    =  "16~" # feb,
  year     =  2018,
  language = "en"
}

@ARTICLE{liu2011complexnet,
  title     = "Controllability of complex networks",
  author    = "Liu, Yang-Yu and Slotine, Jean-Jacques and Barabási,
               Albert-László",
  journal   = "Nature",
  publisher = "Springer Science and Business Media LLC",
  volume    =  473,
  number    =  7346,
  pages     = "167--173",
  month     =  "12~" # may,
  year      =  2011,
  language  = "en"
}

@ARTICLE{jaffe2024elcd,
  title         = "Learning neural contracting dynamics: Extended linearization
                   and global guarantees",
  author        = "Jaffe, Sean and Davydov, Alexander and Lapsekili, Deniz and
                   Singh, Ambuj and Bullo, Francesco",
  journal       = "arXiv [cs.LG]",
  month         =  "12~" # feb,
  year          =  2024,
  archivePrefix = "arXiv",
  primaryClass  = "cs.LG"
}

@ARTICLE{brudey2015autonomicptsd,
  title     = "Autonomic and inflammatory consequences of posttraumatic stress
               disorder and the link to cardiovascular disease",
  author    = "Brudey, Chevelle and Park, Jeanie and Wiaderkiewicz, Jan and
               Kobayashi, Ihori and Mellman, Thomas A and Marvar, Paul J",
  journal   = "Am. J. Physiol. Regul. Integr. Comp. Physiol.",
  publisher = "American Physiological Society",
  volume    =  309,
  number    =  4,
  pages     = "R315--21",
  month     =  "15~" # aug,
  year      =  2015,
  language  = "en"
}

@ARTICLE{kass2023interaction,
  title    = "Identification of interacting neural populations: methods and
              statistical considerations",
  author   = "Kass, Robert E and Bong, Heejong and Olarinre, Motolani and Xin,
              Qi and Urban, Konrad N",
  journal  = "J. Neurophysiol.",
  volume   =  130,
  number   =  3,
  pages    = "475--496",
  month    =  "1~" # sep,
  year     =  2023,
  language = "en"
}

@ARTICLE{semedo2020interaction,
  title    = "Statistical methods for dissecting interactions between brain
              areas",
  author   = "Semedo, João D and Gokcen, Evren and Machens, Christian K and
              Kohn, Adam and Yu, Byron M",
  journal  = "Curr. Opin. Neurobiol.",
  volume   =  65,
  pages    = "59--69",
  month    =  dec,
  year     =  2020,
  language = "en"
}

@ARTICLE{semedo2019commsub,
  title    = "Cortical Areas Interact through a Communication Subspace",
  author   = "Semedo, João D and Zandvakili, Amin and Machens, Christian K and
              Yu, Byron M and Kohn, Adam",
  journal  = "Neuron",
  volume   =  102,
  number   =  1,
  pages    = "249--259.e4",
  month    =  "3~" # apr,
  year     =  2019,
  language = "en"
}

@ARTICLE{macdowell2023multiplex,
  title    = "Multiplexed Subspaces Route Neural Activity Across Brain-wide
              Networks",
  author   = "MacDowell, Camden J and Libby, Alexandra and Jahn, Caroline I and
              Tafazoli, Sina and Buschman, Timothy J",
  journal  = "bioRxiv",
  month    =  "12~" # feb,
  year     =  2023,
  language = "en"
}

@ARTICLE{perich2021rnn,
  title    = "Inferring brain-wide interactions using data-constrained recurrent
              neural network models",
  author   = "Perich, Matthew G and Arlt, Charlotte and Soares, Sofia and Young,
              Megan E and Mosher, Clayton P and Minxha, Juri and Carter, Eugene
              and Rutishauser, Ueli and Rudebeck, Peter H and Harvey,
              Christopher D and Rajan, Kanaka",
  journal  = "bioRxiv",
  pages    = "2020.12.18.423348",
  month    =  "11~" # mar,
  year     =  2021,
  language = "en"
}

@ARTICLE{perich2020multiregion,
  title    = "Rethinking brain-wide interactions through multi-region 'network
              of networks' models",
  author   = "Perich, Matthew G and Rajan, Kanaka",
  journal  = "Curr. Opin. Neurobiol.",
  volume   =  65,
  pages    = "146--151",
  month    =  dec,
  year     =  2020,
  language = "en"
}

@ARTICLE{andalman2019dynamics,
  title    = "Neuronal Dynamics Regulating Brain and Behavioral State
              Transitions",
  author   = "Andalman, Aaron S and Burns, Vanessa M and Lovett-Barron, Matthew
              and Broxton, Michael and Poole, Ben and Yang, Samuel J and
              Grosenick, Logan and Lerner, Talia N and Chen, Ritchie and
              Benster, Tyler and Mourrain, Philippe and Levoy, Marc and Rajan,
              Kanaka and Deisseroth, Karl",
  journal  = "Cell",
  volume   =  177,
  number   =  4,
  pages    = "970--985.e20",
  month    =  "2~" # may,
  year     =  2019,
  language = "en"
}

@ARTICLE{pinto2019dynamics,
  title    = "Task-Dependent Changes in the Large-Scale Dynamics and Necessity
              of Cortical Regions",
  author   = "Pinto, Lucas and Rajan, Kanaka and DePasquale, Brian and Thiberge,
              Stephan Y and Tank, David W and Brody, Carlos D",
  journal  = "Neuron",
  volume   =  104,
  number   =  4,
  pages    = "810--824.e9",
  month    =  "20~" # nov,
  year     =  2019,
  language = "en"
}

@ARTICLE{kleinman2021rnn,
  title   = "A mechanistic multi-area recurrent network model of decision-making",
  author  = "Kleinman, Michael and Chandrasekaran, Chandramouli and Kao,
             Jonathan",
  journal = "Adv. Neural Inf. Process. Syst.",
  volume  =  34,
  pages   = "23152--23165",
  month   =  "6~" # dec,
  year    =  2021
}

@ARTICLE{gokcen2022dlag,
  title     = "Disentangling the flow of signals between populations of neurons",
  author    = "Gokcen, Evren and Jasper, Anna I and Semedo, João D and
               Zandvakili, Amin and Kohn, Adam and Machens, Christian K and Yu,
               Byron M",
  journal   = "Nature Computational Science",
  publisher = "Nature Publishing Group",
  volume    =  2,
  number    =  8,
  pages     = "512--525",
  month     =  "18~" # aug,
  year      =  2022,
  language  = "en"
}

@INPROCEEDINGS{gokcen2023motifs,
  title     = "Uncovering motifs of concurrent signaling across multiple
               neuronal populations",
  author    = "Gokcen, Evren and Jasper, Anna Ivic and Xu, Alison and Kohn, Adam
               and Machens, Christian K and Yu, Byron M",
  booktitle = "Thirty-seventh Conference on Neural Information Processing
               Systems",
  month     =  "2~" # nov,
  year      =  2023
}

@ARTICLE{ebrahimi2022interarea,
  title    = "Emergent reliability in sensory cortical coding and inter-area
              communication",
  author   = "Ebrahimi, Sadegh and Lecoq, Jérôme and Rumyantsev, Oleg and Tasci,
              Tugce and Zhang, Yanping and Irimia, Cristina and Li, Jane and
              Ganguli, Surya and Schnitzer, Mark J",
  journal  = "Nature",
  volume   =  605,
  number   =  7911,
  pages    = "713--721",
  month    =  may,
  year     =  2022,
  language = "en"
}

@ARTICLE{rodu2018regions,
  title    = "Detecting multivariate cross-correlation between brain regions",
  author   = "Rodu, Jordan and Klein, Natalie and Brincat, Scott L and Miller,
              Earl K and Kass, Robert E",
  journal  = "J. Neurophysiol.",
  volume   =  120,
  number   =  4,
  pages    = "1962--1972",
  month    =  "1~" # oct,
  year     =  2018,
  language = "en"
}

@ARTICLE{barbosa2023gating,
  title    = "Early selection of task-relevant features through population
              gating",
  author   = "Barbosa, Joao and Proville, Rémi and Rodgers, Chris C and DeWeese,
              Michael R and Ostojic, Srdjan and Boubenec, Yves",
  journal  = "Nat. Commun.",
  volume   =  14,
  number   =  1,
  pages    =  6837,
  month    =  "27~" # oct,
  year     =  2023,
  language = "en"
}

@ARTICLE{semedo2022fffb,
  title    = "Feedforward and feedback interactions between visual cortical
              areas use different population activity patterns",
  author   = "Semedo, João D and Jasper, Anna I and Zandvakili, Amin and
              Krishna, Aravind and Aschner, Amir and Machens, Christian K and
              Kohn, Adam and Yu, Byron M",
  journal  = "Nat. Commun.",
  volume   =  13,
  number   =  1,
  pages    =  1099,
  month    =  "1~" # mar,
  year     =  2022,
  language = "en"
}

@ARTICLE{ye2015ccm,
  title    = "Distinguishing time-delayed causal interactions using convergent
              cross mapping",
  author   = "Ye, Hao and Deyle, Ethan R and Gilarranz, Luis J and Sugihara,
              George",
  journal  = "Sci. Rep.",
  volume   =  5,
  pages    =  14750,
  month    =  "5~" # oct,
  year     =  2015,
  language = "en"
}

@ARTICLE{tajima2015ccm,
  title    = "Untangling Brain-Wide Dynamics in Consciousness by Cross-Embedding",
  author   = "Tajima, Satohiro and Yanagawa, Toru and Fujii, Naotaka and
              Toyoizumi, Taro",
  journal  = "PLoS Comput. Biol.",
  volume   =  11,
  number   =  11,
  pages    = "e1004537",
  month    =  nov,
  year     =  2015,
  language = "en"
}

@INPROCEEDINGS{glaser2020switching,
  title     = "Recurrent Switching Dynamical Systems Models for Multiple
               Interacting Neural Populations",
  author    = "Glaser, Joshua I and Whiteway, Matthew and Cunningham, John P and
               Paninski, Liam and Linderman, Scott W",
  booktitle = "Advances in Neural Information Processing Systems 33 (NeurIPS
               2020)",
  pages     = "2020.10.21.349282",
  month     =  "22~" # oct,
  year      =  2020,
  language  = "en"
}

@ARTICLE{karniol-tambour2022-switching,
  title   = "Modeling communication and switching nonlinear dynamics in
             multi-region neural activity",
  author  = "Karniol-Tambour, Orren and Zoltowski, David M and Diamanti, E Mika
             and Pinto, Lucas and Tank, David W and Brody, Carlos D and Pillow,
             Jonathan W",
  journal = "bioRxiv",
  pages   = "2022.09.13.507841",
  month   =  "15~" # sep,
  year    =  2022
}

@ARTICLE{bressler2011granger,
  title    = "Wiener-Granger causality: a well established methodology",
  author   = "Bressler, Steven L and Seth, Anil K",
  journal  = "Neuroimage",
  volume   =  58,
  number   =  2,
  pages    = "323--329",
  month    =  "15~" # sep,
  year     =  2011,
  language = "en"
}

@ARTICLE{seth2015granger,
  title    = "Granger causality analysis in neuroscience and neuroimaging",
  author   = "Seth, Anil K and Barrett, Adam B and Barnett, Lionel",
  journal  = "J. Neurosci.",
  volume   =  35,
  number   =  8,
  pages    = "3293--3297",
  month    =  "25~" # feb,
  year     =  2015,
  language = "en"
}

@ARTICLE{chen2022pop,
  title     = "Population burst propagation across interacting areas of the
               brain",
  author    = "Chen, Yu and Douglas, Hannah and Medina, Bryan J and Olarinre,
               Motolani and Siegle, Joshua H and Kass, Robert E",
  journal   = "J. Neurophysiol.",
  publisher = "American Physiological Society",
  volume    =  128,
  number    =  6,
  pages     = "1578--1592",
  month     =  "1~" # dec,
  year      =  2022,
  language  = "en"
}

@ARTICLE{sugihara2012ccm,
  title    = "Detecting causality in complex ecosystems",
  author   = "Sugihara, George and May, Robert and Ye, Hao and Hsieh, Chih-Hao
              and Deyle, Ethan and Fogarty, Michael and Munch, Stephan",
  journal  = "Science",
  volume   =  338,
  number   =  6106,
  pages    = "496--500",
  month    =  "26~" # oct,
  year     =  2012,
  language = "en"
}

@ARTICLE{calderon2024taci,
  title    = "Inferring the time-varying coupling of dynamical systems with
              temporal convolutional autoencoders",
  author   = "Calderon, Josuan and Berman, Gordon J",
  journal  = "eLife",
  month    =  "12~" # nov,
  year     =  2024,
  language = "en"
}

@ARTICLE{friston2003dcm,
  title     = "Dynamic causal modelling",
  author    = "Friston, K J and Harrison, L and Penny, W",
  journal   = "Neuroimage",
  publisher = "Elsevier BV",
  volume    =  19,
  number    =  4,
  pages     = "1273--1302",
  month     =  aug,
  year      =  2003,
  language  = "en"
}

@INPROCEEDINGS{lu2023attention,
  title     = "Attention for Causal Relationship Discovery from Biological
               Neural Dynamics",
  author    = "Lu, Ziyu and Tabassum, Anika and Kulkarni, Shruti and Mi, Lu and
               Nathan Kutz, J and Shea-Brown, Eric and Lim, Seung-Hwan",
  booktitle = "NeurIPS 2023 Workshop on Causal Representation Learning",
  month     =  "12~" # nov,
  year      =  2023
}

@ARTICLE{ahamed2020continuouscomplexity,
  title     = "Capturing the continuous complexity of behaviour in
               Caenorhabditis elegans",
  author    = "Ahamed, Tosif and Costa, Antonio C and Stephens, Greg J",
  journal   = "Nat. Phys.",
  publisher = "Nature Publishing Group",
  volume    =  17,
  number    =  2,
  pages     = "275--283",
  month     =  "5~" # oct,
  year      =  2020,
  language  = "en"
}

@ARTICLE{deyle2016tracking,
  title    = "Tracking and forecasting ecosystem interactions in real time",
  author   = "Deyle, Ethan R and May, Robert M and Munch, Stephan B and
              Sugihara, George",
  journal  = "Proc. Biol. Sci.",
  volume   =  283,
  number   =  1822,
  month    =  "13~" # jan,
  year     =  2016,
  language = "en"
}

@INPROCEEDINGS{chen2018neuralode,
  title     = "Neural Ordinary Differential Equations",
  author    = "Chen, Ricky T Q and Rubanova, Yulia and Bettencourt, Jesse and
               Duvenaud, David K",
  editor    = "Bengio, S and Wallach, H and Larochelle, H and Grauman, K and
               Cesa-Bianchi, N and Garnett, R",
  booktitle = "Advances in Neural Information Processing Systems",
  publisher = "Curran Associates, Inc.",
  volume    =  31,
  year      =  2018
}

@ARTICLE{panichello2021wmattention,
  title    = "Shared mechanisms underlie the control of working memory and
              attention",
  author   = "Panichello, Matthew F and Buschman, Timothy J",
  journal  = "Nature",
  month    =  "31~" # mar,
  year     =  2021,
  language = "en"
}

@article{gomez2021torchquad,
author = {Gómez, Pablo and Hem Toftevaag, Håvard and Meoni, Gabriele},
doi = {10.21105/joss.03439},
journal = {Journal of Open Source Software},
number = {6},
title = {{torchquad: Numerical Integration in Arbitrary Dimensions with PyTorch}},
volume = {64},
year = {2021}
}

@ARTICLE{paszke2019pytorch,
  title   = "{PyTorch}: An imperative style, high-performance deep learning
             library",
  author  = "Paszke, Adam and Gross, Sam and Massa, Francisco and Lerer, Adam
             and Bradbury, James and Chanan, Gregory and Killeen, Trevor and
             Lin, Zeming and Gimelshein, N and Antiga, L and Desmaison, Alban
             and Köpf, Andreas and Yang, E and DeVito, Zach and Raison, Martin
             and Tejani, Alykhan and Chilamkurthy, Sasank and Steiner, Benoit
             and Fang, Lu and Bai, Junjie and Chintala, Soumith",
  journal = "Adv. Neural Inf. Process. Syst.",
  volume  = "abs/1912.01703",
  month   =  "3~" # dec,
  year    =  2019
}

@ARTICLE{vanderpol1926osc,
  title     = "{LXXXVIII}. On “relaxation-oscillations”",
  author    = "van der Pol, Balth",
  journal   = "The London, Edinburgh, and Dublin Philosophical Magazine and
               Journal of Science",
  publisher = "Taylor \& Francis",
  volume    =  2,
  number    =  11,
  pages     = "978--992",
  month     =  "1~" # nov,
  year      =  1926
}

@ARTICLE{lorenz1963flow,
  title     = "Deterministic Nonperiodic Flow",
  author    = "Lorenz, Edward N",
  journal   = "J. Atmos. Sci.",
  publisher = "American Meteorological Society",
  volume    =  20,
  number    =  2,
  pages     = "130--141",
  month     =  "1~" # mar,
  year      =  1963,
  language  = "en"
}

@INPROCEEDINGS{lorenz1996pred,
  title     = "Predictability: a problem partly solved",
  author    = "Lorenz, E N",
  booktitle = "ECMWF Seminar on Predictability, 4-8 September 1995",
  publisher = "European Centre for Medium-Range Weather Forecasts",
  year      =  1996,
  language  = "en"
}

@ARTICLE{gilpin2023modelscale,
  title     = "Model scale versus domain knowledge in statistical forecasting of
               chaotic systems",
  author    = "Gilpin, William",
  journal   = "Phys. Rev. Res.",
  publisher = "American Physical Society",
  volume    =  5,
  number    =  4,
  pages     =  043252,
  month     =  "15~" # dec,
  year      =  2023
}

@INPROCEEDINGS{gilpin2021chaos,
  title     = "Chaos as an interpretable benchmark for forecasting and
               data-driven modelling",
  author    = "Gilpin, William",
  editor    = "Vanschoren, J and Yeung, S",
  booktitle = "Proceedings of the Neural Information Processing Systems Track on
               Datasets and Benchmarks",
  volume    =  1,
  year      =  2021
}

@INPROCEEDINGS{tassa2012ilqr,
  title     = "Synthesis and stabilization of complex behaviors through online
               trajectory optimization",
  author    = "Tassa, Yuval and Erez, Tom and Todorov, Emanuel",
  booktitle = "2012 IEEE/RSJ International Conference on Intelligent Robots and
               Systems",
  publisher = "IEEE",
  pages     = "4906--4913",
  month     =  oct,
  year      =  2012,
  language  = "en"
}

@ARTICLE{li2004ilqr,
  title     = "Iterative Linear Quadratic Regulator Design for Nonlinear
               Biological Movement Systems",
  author    = "Li, Weiwei and Todorov, Emanuel",
  journal   = "ICINCO 2004, Proceedings of the First International Conference on
               Informatics in Control, Automation and Robotics, Setúbal,
               Portugal, August 25-28, 2004",
  publisher = "unknown",
  volume    =  1,
  pages     = "222--229",
  month     =  "1~" # jan,
  year      =  2004
}

@ARTICLE{ouala2020partial,
  title     = "Learning latent dynamics for partially observed chaotic systems",
  author    = "Ouala, S and Nguyen, D and Drumetz, L and Chapron, B and Pascual,
               A and Collard, F and Gaultier, L and Fablet, R",
  journal   = "Chaos",
  publisher = "AIP Publishing",
  volume    =  30,
  number    =  10,
  pages     =  103121,
  month     =  "20~" # oct,
  year      =  2020,
  language  = "en"
}

@BOOK{strogatz2014dyn,
  title     = "Nonlinear Dynamics and Chaos: With Applications to Physics,
               Biology, Chemistry, and Engineering",
  author    = "Strogatz, Steven H",
  publisher = "Avalon Publishing",
  month     =  "29~" # jul,
  year      =  2014,
  language  = "en"
}

@BOOK{sydel2009stab,
  title     = "Practical bifurcation and stability analysis",
  author    = "Seydel, Rudiger U",
  publisher = "Springer",
  address   = "New York, NY",
  edition   =  3,
  series    = "Interdisciplinary applied mathematics",
  month     =  "10~" # dec,
  year      =  2009,
  language  = "en"
}

@ARTICLE{kozachkov2020stable,
  title   = "Achieving stable dynamics in neural circuits",
  author  = "Kozachkov, Leo and Lundqvist, Mikael and Slotine, Jean Jacques and
             Miller, Earl K",
  journal = "PLoS Comput. Biol.",
  volume  =  16,
  number  =  8,
  pages   = "1--15",
  year    =  2020
}

@ARTICLE{kozachkov2022rnns,
  title   = "{RNNs} of {RNNs}: Recursive construction of stable assemblies of
             recurrent neural networks",
  author  = "Kozachkov, L and Ennis, M and Slotine, J",
  journal = "Adv. Neural Inf. Process. Syst.",
  year    =  2022
}

@ARTICLE{lohmiller1998contraction,
  title   = "On Contraction Analysis for Non-linear Systems",
  author  = "Lohmiller, Winfried and Slotine, Jean-Jacques E",
  journal = "Automatica",
  volume  =  34,
  number  =  6,
  pages   = "683--696",
  month   =  "1~" # jun,
  year    =  1998
}

@ARTICLE{manchester2017controlcontraction,
  title     = "Control contraction metrics: Convex and intrinsic criteria for
               nonlinear feedback design",
  author    = "Manchester, Ian R and Slotine, Jean-Jacques E",
  journal   = "IEEE Trans. Automat. Contr.",
  publisher = "Institute of Electrical and Electronics Engineers (IEEE)",
  volume    =  62,
  number    =  6,
  pages     = "3046--3053",
  month     =  jun,
  year      =  2017,
  language  = "en"
}

@ARTICLE{wensing2020contractgrad,
  title    = "Beyond convexity-Contraction and global convergence of gradient
              descent",
  author   = "Wensing, Patrick M and Slotine, Jean-Jacques",
  journal  = "PLoS One",
  volume   =  15,
  number   =  8,
  pages    = "e0236661",
  month    =  "4~" # aug,
  year     =  2020,
  language = "en"
}

@ARTICLE{dieci1997lyap,
  title     = "On the Computation of Lyapunov Exponents for Continuous Dynamical
               Systems",
  author    = "Dieci, Luca and Russell, Robert D and Van Vleck, Erik S",
  journal   = "SIAM J. Numer. Anal.",
  publisher = "Society for Industrial and Applied Mathematics",
  volume    =  34,
  number    =  1,
  pages     = "402--423",
  year      =  1997
}

@ARTICLE{christiansen1997lyap,
  title     = "Computing Lyapunov spectra with continuous Gram - Schmidt
               orthonormalization",
  author    = "Christiansen, F and Rugh, H H",
  journal   = "Nonlinearity",
  publisher = "IOP Publishing",
  volume    =  10,
  number    =  5,
  pages     =  1063,
  month     =  "1~" # sep,
  year      =  1997,
  language  = "en"
}

@INPROCEEDINGS{duchaine2008robotstab,
  title     = "Investigation of human-robot interaction stability using Lyapunov
               theory",
  author    = "Duchaine, Vincent and Gosselin, Clement M",
  booktitle = "2008 IEEE International Conference on Robotics and Automation",
  publisher = "IEEE",
  pages     = "2189--2194",
  month     =  may,
  year      =  2008,
  language  = "en"
}

@INPROCEEDINGS{hootsmans2002transjac,
  title     = "Large motion control of mobile manipulators including vehicle
               suspension characteristics",
  author    = "Hootsmans, N A M and Dubowsky, S",
  booktitle = "Proceedings. 1991 IEEE International Conference on Robotics and
               Automation",
  publisher = "IEEE Comput. Soc. Press",
  pages     = "2336--2341 vol.3",
  year      =  2002
}

@ARTICLE{tchon2015lagrangejac,
  title     = "Lagrangian Jacobian inverse for nonholonomic robotic systems",
  author    = "Tchoń, Krzysztof and Ratajczak, Adam and Góral, Ida",
  journal   = "Nonlinear Dyn.",
  publisher = "Springer Nature",
  volume    =  82,
  number    =  4,
  pages     = "1923--1932",
  month     =  dec,
  year      =  2015,
  language  = "en"
}

@ARTICLE{moosavian2007transjac,
  title     = "Modified transpose Jacobian control of robotic systems",
  author    = "Moosavian, S Ali A and Papadopoulos, Evangelos",
  journal   = "Automatica (Oxf.)",
  publisher = "Elsevier BV",
  volume    =  43,
  number    =  7,
  pages     = "1226--1233",
  month     =  "1~" # jul,
  year      =  2007,
  language  = "en"
}

@ARTICLE{lee2013genereg,
  title     = "Transcriptional regulation and its misregulation in disease",
  author    = "Lee, Tong Ihn and Young, Richard A",
  journal   = "Cell",
  publisher = "Elsevier BV",
  volume    =  152,
  number    =  6,
  pages     = "1237--1251",
  month     =  "14~" # mar,
  year      =  2013,
  language  = "en"
}

@ARTICLE{maurano2012genereg,
  title     = "Systematic localization of common disease-associated variation in
               regulatory {DNA}",
  author    = "Maurano, Matthew T and Humbert, Richard and Rynes, Eric and
               Thurman, Robert E and Haugen, Eric and Wang, Hao and Reynolds,
               Alex P and Sandstrom, Richard and Qu, Hongzhu and Brody, Jennifer
               and Shafer, Anthony and Neri, Fidencio and Lee, Kristen and
               Kutyavin, Tanya and Stehling-Sun, Sandra and Johnson, Audra K and
               Canfield, Theresa K and Giste, Erika and Diegel, Morgan and
               Bates, Daniel and Hansen, R Scott and Neph, Shane and Sabo, Peter
               J and Heimfeld, Shelly and Raubitschek, Antony and Ziegler,
               Steven and Cotsapas, Chris and Sotoodehnia, Nona and Glass, Ian
               and Sunyaev, Shamil R and Kaul, Rajinder and Stamatoyannopoulos,
               John A",
  journal   = "Science",
  publisher = "American Association for the Advancement of Science (AAAS)",
  volume    =  337,
  number    =  6099,
  pages     = "1190--1195",
  month     =  "7~" # sep,
  year      =  2012,
  language  = "en"
}

@ARTICLE{unger-avila2024genereg,
  title     = "Gene regulatory networks in disease and ageing",
  author    = "Unger Avila, Paula and Padvitski, Tsimafei and Leote, Ana
               Carolina and Chen, He and Saez-Rodriguez, Julio and Kann, Martin
               and Beyer, Andreas",
  journal   = "Nat. Rev. Nephrol.",
  publisher = "Springer Science and Business Media LLC",
  volume    =  20,
  number    =  9,
  pages     = "616--633",
  month     =  "12~" # sep,
  year      =  2024,
  language  = "en"
}

@ARTICLE{madhamshettiwar2012genereg,
  title     = "Gene regulatory network inference: evaluation and application to
               ovarian cancer allows the prioritization of drug targets",
  author    = "Madhamshettiwar, Piyush B and Maetschke, Stefan R and Davis,
               Melissa J and Reverter, Antonio and Ragan, Mark A",
  journal   = "Genome Med.",
  publisher = "Springer Nature",
  volume    =  4,
  number    =  5,
  pages     =  41,
  month     =  "1~" # may,
  year      =  2012,
  language  = "en"
}

@ARTICLE{barabasi2011genereg,
  title     = "Network medicine: a network-based approach to human disease",
  author    = "Barabási, Albert-László and Gulbahce, Natali and Loscalzo, Joseph",
  journal   = "Nat. Rev. Genet.",
  publisher = "Nature Publishing Group",
  volume    =  12,
  number    =  1,
  pages     = "56--68",
  month     =  jan,
  year      =  2011,
  language  = "en"
}

@ARTICLE{goldberger2019autonomic,
  title     = "Autonomic nervous system dysfunction: {JACC} focus seminar",
  author    = "Goldberger, Jeffrey J and Arora, Rishi and Buckley, Una and
               Shivkumar, Kalyanam",
  journal   = "J. Am. Coll. Cardiol.",
  publisher = "Elsevier BV",
  volume    =  73,
  number    =  10,
  pages     = "1189--1206",
  month     =  "19~" # mar,
  year      =  2019,
  language  = "en"
}

@ARTICLE{xiong2019autonomic,
  title     = "Autonomic dysfunction in neurological disorders",
  author    = "Xiong, Li and Leung, Thomas W H",
  journal   = "Aging (Albany NY)",
  publisher = "Impact Journals, LLC",
  volume    =  11,
  number    =  7,
  pages     = "1903--1904",
  month     =  "9~" # apr,
  year      =  2019,
  language  = "en"
}

@ARTICLE{alvares2016autonomic,
  title     = "Autonomic nervous system dysfunction in psychiatric disorders and
               the impact of psychotropic medications: a systematic review and
               meta-analysis",
  author    = "Alvares, Gail A and Quintana, Daniel S and Hickie, Ian B and
               Guastella, Adam J",
  journal   = "J. Psychiatry Neurosci.",
  publisher = "CMA Joule Inc.",
  volume    =  41,
  number    =  2,
  pages     = "89--104",
  month     =  mar,
  year      =  2016,
  language  = "en"
}

@INPROCEEDINGS{gilpin2020deeprecon,
  title     = "Deep reconstruction of strange attractors from time series",
  author    = "Gilpin, William",
  booktitle = "Proceedings of the 34th International Conference on Neural
               Information Processing Systems",
  publisher = "Curran Associates Inc.",
  address   = "Red Hook, NY, USA",
  number    = "Article 18",
  pages     = "204--216",
  series    = "NIPS'20",
  month     =  "6~" # dec,
  year      =  2020
}

@ARTICLE{young2023recon,
  title     = "Deep learning delay coordinate dynamics for chaotic attractors
               from partial observable data",
  author    = "Young, Charles D and Graham, Michael D",
  journal   = "Phys. Rev. E.",
  publisher = "American Physical Society",
  volume    =  107,
  number    = "3-1",
  pages     =  034215,
  month     =  "30~" # mar,
  year      =  2023,
  language  = "en"
}

@ARTICLE{ozalp2023recon,
  title     = "Reconstruction, forecasting, and stability of chaotic dynamics
               from partial data",
  author    = "Özalp, Elise and Margazoglou, Georgios and Magri, Luca",
  journal   = "Chaos",
  publisher = "AIP Publishing",
  volume    =  33,
  number    =  9,
  pages     =  093107,
  month     =  "1~" # sep,
  year      =  2023,
  language  = "en"
}

@ARTICLE{lu2022recon,
  title     = "Discovering sparse interpretable dynamics from partial
               observations",
  author    = "Lu, Peter Y and Ariño Bernad, Joan and Soljačić, Marin",
  journal   = "Commun. Phys.",
  publisher = "Springer Science and Business Media LLC",
  volume    =  5,
  number    =  1,
  pages     = "1--7",
  month     =  "12~" # aug,
  year      =  2022,
  language  = "en"
}

@article{colgin_frequency_2009,
	title = {Frequency of gamma oscillations routes flow of information in the hippocampus},
	volume = {462},
	issn = {1476-4687},
	url = {https://doi.org/10.1038/nature08573},
	doi = {10.1038/nature08573},
	number = {7271},
	journal = {Nature},
	author = {Colgin, Laura Lee and Denninger, Tobias and Fyhn, Marianne and Hafting, Torkel and Bonnevie, Tora and Jensen, Ole and Moser, May-Britt and Moser, Edvard I.},
	month = nov,
	year = {2009},
	pages = {353--357},
}

@article{muldoon2016,
    doi = {10.1371/journal.pcbi.1005076},
    author = {Muldoon, Sarah Feldt AND Pasqualetti, Fabio AND Gu, Shi AND Cieslak, Matthew AND Grafton, Scott T. AND Vettel, Jean M. AND Bassett, Danielle S.},
    journal = {PLOS Computational Biology},
    publisher = {Public Library of Science},
    title = {Stimulation-Based Control of Dynamic Brain Networks},
    year = {2016},
    month = {09},
    volume = {12},
    url = {https://doi.org/10.1371/journal.pcbi.1005076},
    pages = {1-23},
    number = {9},

}

@ARTICLE{bassett2017networkneuroscience,
  title    = "Network neuroscience",
  author   = "Bassett, Danielle S and Sporns, Olaf",
  journal  = "Nat. Neurosci.",
  volume   =  20,
  number   =  3,
  pages    = "353--364",
  month    =  "23~" # feb,
  year     =  2017,
  language = "en"
}

@ARTICLE{ni2024distributeddynamic,
  title     = "Distributed and dynamical communication: a mechanism for flexible
               cortico-cortical interactions and its functional roles in visual
               attention",
  author    = "Ni, Shencong and Harris, Brendan and Gong, Pulin",
  journal   = "Commun. Biol.",
  publisher = "Springer Science and Business Media LLC",
  volume    =  7,
  number    =  1,
  pages     =  550,
  month     =  "8~" # may,
  year      =  2024,
  language  = "en"
}

@ARTICLE{antzoulatos2014learn,
  title     = "Increases in functional connectivity between prefrontal cortex
               and striatum during category learning",
  author    = "Antzoulatos, Evan G and Miller, Earl K",
  journal   = "Neuron",
  publisher = "Elsevier BV",
  volume    =  83,
  number    =  1,
  pages     = "216--225",
  month     =  "2~" # jul,
  year      =  2014,
  language  = "en"
}

@ARTICLE{hoffman2019jacpen,
  title         = "Robust learning with Jacobian regularization",
  author        = "Hoffman, Judy and Roberts, Daniel A and Yaida, Sho",
  journal       = "arXiv [stat.ML]",
  month         =  "7~" # aug,
  year          =  2019,
  archivePrefix = "arXiv",
  primaryClass  = "stat.ML"
}

@ARTICLE{schneider2025timeseries-att,
  title         = "Time-series attribution maps with regularized contrastive
                   learning",
  author        = "Schneider, Steffen and Laiz, Rodrigo González and Filippova,
                   Anastasiia and Frey, Markus and Mathis, Mackenzie Weygandt",
  journal       = "arXiv [stat.ML]",
  month         =  "17~" # feb,
  year          =  2025,
  archivePrefix = "arXiv",
  primaryClass  = "stat.ML"
}

@ARTICLE{wikner2024reservoirstab,
  title     = "Stabilizing machine learning prediction of dynamics: Novel
               noise-inspired regularization tested with reservoir computing",
  author    = "Wikner, Alexander and Harvey, Joseph and Girvan, Michelle and
               Hunt, Brian R and Pomerance, Andrew and Antonsen, Thomas and Ott,
               Edward",
  journal   = "Neural Netw.",
  publisher = "Elsevier BV",
  volume    =  170,
  pages     = "94--110",
  month     =  "1~" # feb,
  year      =  2024,
  language  = "en"
}

@ARTICLE{pathak2017esn,
  title    = "Using machine learning to replicate chaotic attractors and
              calculate Lyapunov exponents from data",
  author   = "Pathak, Jaideep and Lu, Zhixin and Hunt, Brian R and Girvan,
              Michelle and Ott, Edward",
  journal  = "Chaos",
  volume   =  27,
  number   =  12,
  pages    =  121102,
  month    =  dec,
  year     =  2017,
  language = "en"
}

@ARTICLE{lu2017esn,
  title     = "Reservoir observers: Model-free inference of unmeasured variables
               in chaotic systems",
  author    = "Lu, Zhixin and Pathak, Jaideep and Hunt, Brian and Girvan,
               Michelle and Brockett, Roger and Ott, Edward",
  journal   = "Chaos",
  publisher = "AIP Publishing",
  volume    =  27,
  number    =  4,
  pages     =  041102,
  month     =  "5~" # apr,
  year      =  2017,
  language  = "en"
}

@ARTICLE{banerjee2019esn,
  title     = "Using machine learning to assess short term causal dependence and
               infer network links",
  author    = "Banerjee, Amitava and Pathak, Jaideep and Roy, Rajarshi and
               Restrepo, Juan G and Ott, Edward",
  journal   = "Chaos",
  publisher = "AIP Publishing",
  volume    =  29,
  number    =  12,
  pages     =  121104,
  month     =  "26~" # dec,
  year      =  2019,
  language  = "en"
}

@ARTICLE{rigotti2013mixed,
  title     = "The importance of mixed selectivity in complex cognitive tasks",
  author    = "Rigotti, Mattia and Barak, Omri and Warden, Melissa R and Wang,
               Xiao-Jing and Daw, Nathaniel D and Miller, Earl K and Fusi,
               Stefano",
  journal   = "Nature",
  publisher = "Springer Science and Business Media LLC",
  volume    =  497,
  number    =  7451,
  pages     = "585--590",
  month     =  "30~" # may,
  year      =  2013,
  language  = "en"
}

@ARTICLE{pathak2018esn,
  title    = "Model-Free Prediction of Large Spatiotemporally Chaotic Systems
              from Data: A Reservoir Computing Approach",
  author   = "Pathak, Jaideep and Hunt, Brian and Girvan, Michelle and Lu,
              Zhixin and Ott, Edward",
  journal  = "Phys. Rev. Lett.",
  volume   =  120,
  number   =  2,
  pages    =  024102,
  month    =  "12~" # jan,
  year     =  2018,
  language = "en"
}

%%%%%%%%%%%%%%%%%%%%%%%%%%%%%%%%%%%%%%%%%%%%%%%%%%%%%%%%%%%%

\appendix

\clearpage
\beginsupplement
\section{JacobianODE technical details}
\label{supp:jacobianODE-technical-detail}

\subsection{Parameterizing the time derivative via the Jacobian} \label{supp:jacparam}
Recall that, for the time derivative function $\mathbf{f}$, we have that the path integral is independent of the choice of path: \begin{equation}
\mathbf{f}(\mathbf{x}(t_f)) - \mathbf{f}(\mathbf{x}(t_i)) \ \ = \ \ \int_{\mathcal{C}}\mathbf{J}\ ds \ \ = \ \ \int_{t_i}^{t_f} \mathbf{J}(\mathbf{c}(r))\mathbf{c}'(r) \ dr,
\end{equation}
where $\mathcal{C}$ is a piecewise smooth curve in $\mathbb{R}^n$ and $\mathbf{c}: [t_i, t_f] \to \mathcal{C}$ is a parameterization of $\mathcal{C}$ with $\mathbf{c}(t_i) = \mathbf{x}(t_i)$  and $\mathbf{c}(t_f) = \mathbf{x}(t_f)$ (also see equation \ref{eq:pathint}). Path integration notably provides only differences between $\mathbf{f}$ at distinct points. Thus, knowing the value of $\mathbf{f}$ at one point is needed for trajectory reconstruction, which we do not assume. To address this, we note it is possible to represent $\mathbf{f}$ solely through the Jacobian at a point $\mathbf{x}(t)$ as \begin{equation}
\mathbf{f}(\mathbf{x}(t)) = \frac{G(t_0, t; \mathbf{J}, \mathbf{c}) + \mathbf{x}(t) - \mathbf{x}(t_0)}{t - t_0},
\label{eq:f_from_jac}
\end{equation}
where
\begin{equation}
G(t_0, t; \mathbf{J}, \mathbf{c}) = \int_{t_0}^{t} \int_{s}^{t} \mathbf{J}(\mathbf{c}_{s,t}(r))\mathbf{c}_{s,t}'(r)dr ds
\end{equation} and we have abbreviated $\mathbf{c}_{s,t}(r) = \mathbf{c}(r;s, t, \mathbf{x}(s), \mathbf{x}(t))$, a piecewise smooth curve on $[s, t]$ parameterized by $r$ and beginning and ending at $\mathbf{x}(s)$ and $\mathbf{x}(t)$ respectively. Intuitively, we can circumvent the need to know $\mathbf{f}$ by recognizing that integrating $\mathbf{f}$ between time points will produce the difference in the system states at those time points, which are known to us. With this formalism, we avoid the need to represent $\mathbf{f}$ directly, thereby enabling all gradients to backpropagate through the Jacobian network. The proof of this formalism is presented below.

\begin{proposition}[Jacobian-parameterized ODEs]
Let $\dot{\mathbf{x}}(t) = \mathbf{f}(\mathbf{x}(t))$ and let $\mathbf{J}_{\mathbf{f}}(\mathbf{x}(t)) = \mathbf{J}(\mathbf{x}(t)) =\frac{\partial}{\partial \mathbf{x}}\mathbf{f}(\mathbf{x}(t))$. Then given times $t_0, t$ we can express $\mathbf{f}$ parameterized by the Jacobian as\begin{equation*}
\mathbf{f}(\mathbf{x}(t)) = \frac{G(t_0, t; \mathbf{J}, \mathbf{c}) + \mathbf{x}(t) - \mathbf{x}(t_0)}{t - t_0},
\end{equation*}
where
\begin{equation*}
G(t_0, t; \mathbf{J}, \mathbf{c}) = \int_{t_0}^{t} \int_{s}^{t} \mathbf{J}(\mathbf{c}_{s,t}(r))\mathbf{c}_{s,t}'(r)dr ds
\end{equation*}

and we have abbreviated $\mathbf{c}_{s,t}(r) = \mathbf{c}(r;s, t, \mathbf{x}(s), \mathbf{x}(t))$, a piecewise smooth curve on $[s, t]$ parameterized by $r$ and beginning and ending at $\mathbf{x}(s)$ and $\mathbf{x}(t)$ respectively.
\end{proposition}
\begin{proof}
For given times $t$, $t_0$, and $s$, from the fundamental theorem of calculus, we have that

\[
\mathbf{x}(t) - \mathbf{x}(t_0) = \int_{t_0}^t \mathbf{f}(\mathbf{x}(s)) ds
\]
and

\[\mathbf{f}(\mathbf{x}(t)) - \mathbf{f}(\mathbf{x}(s)) = \int_{s}^{t} \mathbf{J}(\mathbf{c}_{s,t}(r))\mathbf{c}_{s,t}'(r)dr\]

Letting 
\[
H(s, t) = \int_{s}^{t} \mathbf{J}(\mathbf{c}_{s,t}(r))\mathbf{c}_{s,t}'(r)dr = \mathbf{f}(\mathbf{x}(t)) - \mathbf{f}(\mathbf{x}(s))
\]
We then have that

\begin{align*}
\int_{t_0}^{t} H(s, t) ds &= \int_{t_0}^{t} \mathbf{f}(\mathbf{x}(t)) - \mathbf{f}(\mathbf{x}(s)) ds \\
&=  \int_{t_0}^{t} \mathbf{f}(\mathbf{x}(t)) ds - \int_{t_0}^{t} \mathbf{f}(\mathbf{x}(s)) ds \\
&= (t-t_0)\mathbf{f}(\mathbf{x}(t)) - (\mathbf{x}(t) - \mathbf{x}(t_0))\\
\end{align*}
Letting now
\[
G(t_0, t; \mathbf{J}, \mathbf{c}) = \int_{t_0}^{t} H(s, t) ds =\int_{t_0}^{t} \int_{s}^{t} \mathbf{J}(\mathbf{c}_{s,t}(r))\mathbf{c}_{s,t}'(r)dr ds
\]
We can see that
\[
G(t_0, t; \mathbf{J}, \mathbf{c}) = (t-t_0)\mathbf{f}(\mathbf{x}(t)) - (\mathbf{x}(t) - \mathbf{x}(t_0))
\]
and thus 

$$\mathbf{f}(\mathbf{x}(t)) = \frac{G(t_0, t; \mathbf{J}, \mathbf{c}) + \mathbf{x}(t) - \mathbf{x}(t_0)}{t - t_0}$$

\end{proof}

\subsection{Path integrating to generate predictions}
\label{supp:path-pred-gen}
Given an initial observed trajectory $\mathbf{x}(t_0 + k\Delta t), k=0,\dots,b$ of length at least two ($b \geq 1$), we compute an estimate of $\hat{\mathbf{f}}(\mathbf{x}(t_0 + b\Delta t))$ of $\mathbf{f}(\mathbf{x}(t_0 + b\Delta t))$  by replacing $\mathbf{J}$ with $\hat{\mathbf{J}}^{\theta}$ in equation \ref{eq:f_from_jac}. Specifically, we compute an estimate of $\hat{\mathbf{f}}(\mathbf{x}(t_b))$ of $\mathbf{f}(\mathbf{x}(t_b))$ as

\begin{equation}
\hat{\mathbf{f}}(\mathbf{x}(t_b)) = \frac{G(t_0, t_b; \hat{\mathbf{J}}^{\theta}, \mathbf{c}) + \mathbf{x}(t_b) - \mathbf{x}(t_0)}{t_b- t_0},
\end{equation}

In practice, we compute this estimate by constructing a cubic spline on the initial observed trajectory using 15 points (i.e., $b = 14$). Given that the computation of $G$ involves a double integral - one over states and over time - using a spline is computationally advantageous. This is because the path integral (and all intermediate steps) can be quickly computed along the full spline, with the result of each intermediate step along the path then being summed to approximate the time integral. The integral is computed by interpolating 4 points for every gap between observed points, resulting in a discretization of 58 points along the spline. Integrals are computed using the trapezoid method from \texttt{torchquad} \cite{gomez2021torchquad}. 

Once we have constructed our estimate of $\hat{\mathbf{f}}(\mathbf{x}(t_b))$ we can estimate $\mathbf{f}$ at any other point $\mathbf{x}(t)$ as
\[
\hat{\mathbf{f}}(\mathbf{x}(t)) =
\begin{cases} 
 H(t_b, t) + \hat{\mathbf{f}}(\mathbf{x}(t_b)), & \text{if } t_b< t \\
  \hat{\mathbf{f}}(\mathbf{x}(t_b)) - H(t, t_b), & \text{if } t_b> t \\
  \mathbf{f}(\mathbf{x}(t_b)) & \text{if } t_b = t
\end{cases}\]

where by convention we integrate forwards in time and $H$ is the path integral defined above, with $\hat{\mathbf{J}}$ in place of $\mathbf{J}$. In practice, for the integration path $\mathbf{c}(r;s,t,\mathbf{x}(s),\mathbf{x}(t))$, we construct a line from $\mathbf{x}(s)$ to $\mathbf{x}(t)$ as

\[
\mathbf{c}(r;s,t,\mathbf{x}(s),\mathbf{x}(t)) = \left(1 - \frac{r - s}{t - s}\right) \mathbf{x}(s) + \frac{r - s}{t -s}\mathbf{x}(t)
\]
to maintain the interpretability of having $r$ in the range $[s, t]$ however it can be easily seen that setting $r' = \frac{r - s}{t - s}$ we recover the familiar line

\[
\mathbf{c}(r') = (1 - r')\mathbf{x}(s) +r'\mathbf{x}(t)
\]
with $r'$ taking values on $[0,1]$. Line integrals are computed with 20 discretization steps using the trapezoid method from \texttt{torchquad} \cite{gomez2021torchquad}.

Using $\hat{\mathbf{f}}(\mathbf{x}(t))$, we can then generate predictions as
\begin{equation}
\hat{\mathbf{x}}(t +\Delta t) = \mathbf{x}(t) + \int_{t}^{t + \Delta t} \hat{\mathbf{f}}(\mathbf{x}(\tau)) d\tau
\label{eq:pred-gen}
\end{equation}
where the integral can be computed by a standard ODE solver (we used the RK4 method from \texttt{torchdiffeq} with default values for the relative and absolute tolerance).

We can then compute the trajectory reconstruction loss as
\begin{equation}
\mathcal{L}_{\text{traj}}(\theta;\mathbf{x}) = \sum_{k=b+1}^{T-1} d\left(\mathbf{x}(t_0 + k\Delta t), \,  \hat{\mathbf{x}}(t_0 + k\Delta t)\right)
\end{equation} where $d$ is a distance measure between trajectories (e.g. mean squared error).

\subsection{Loop closure loss}
\label{supp:loop-closure}
While the space of Jacobian functions $\hat{\mathbf{J}}^{\theta}$ that solve the dynamics problem may be quite large, we are interested only in functions $\hat{\mathbf{J}}^{\theta}$ that \textit{both} solve the dynamics problem and generate matrices with rows that are conservative vector fields. To effectively restrict our optimization to this space, we employ a self-supervised loss building on the loss introduced by \citet{iyer2024velocities}. This loss imposes a conservative constraint on the rows of the Jacobian matrix. Specifically, for any piecewise smooth curve $\mathcal{C}_{\text{loop}}$ starting and ending at the same state $\mathbf{x}_0$, we have that $\left \Vert 
\int_{\mathcal{C}_{\text{loop}}} \mathbf{J} ds \right \Vert_2 = 0$.

Thus, given $n_{\text{loops}}$ sets of any $L$ (not necessarily sequential)  points $\{\mathbf{x}^{(l)}(t_1),\mathbf{x}^{(l)}(t_2), ..., \mathbf{x}^{(l)}(t_L)\}$ we can form a loop $\mathcal{C}^{(^{(l)})}_{\text{loop}}$ consisting of the sequence lines from $\mathbf{x}^{(l)}(t_i)$ to $\mathbf{x}^{(l)}(t_{i+1})$, $i=1,...,L-1$, followed by a line from $\mathbf{x}^{(l)}(t_L)$ to $\mathbf{x}^{(l)}(t_1)$. We  then define the following regularization loss to enforce this conservation constraint:
\begin{equation}
\mathcal{L}_{\text{loop}}(\theta;\mathbf{x}) = \frac{1}{n_{\text{loops}}}\sum_{l=1}^{n_{\text{loops}}}\frac{1}{n}\left \Vert \int_{\mathcal{C}^{(l)}_{\text{loop}}} \hat{\mathbf{J}}^{\theta}ds\right\Vert^2_2
\end{equation}
We note that while other choices for loops would have been possible (for instance, forward and backward passes along observed trajectories, or arbitrary circles beginning and ending at the same point), the approach presented here has several advantages (as discussed in Section \ref{sec:lossfuncs}). Namely, it balances computational tractability with uniform sampling of the directions in the tangent space. Enforcing that all the points that comprise the loop are on the data manifold ensures that the loops are integrated in directions that are as informative as possible for estimating the Jacobian of the given dynamical system.

\section{Control-theoretic analysis details}
\label{supp:control-supp}

\subsection{Gramian computation}
\label{supp:gramian-comp}

We begin with equation \ref{eq:area-interaction} from Section \ref{sec:wmtask-control-analysis}, in which we separate the locally-linearized dynamics in the tangent space into separate pairwise inter-subsystem control interactions. These pairwise control interactions are in general of the form

\[
\delta\dot{\mathbf{x}}^A(t) = \mathbf{A}(t)\delta\mathbf{x}^A(t) + \mathbf{B}(t) \delta \mathbf{x}^B(t)
\]

where $\mathbf{A}$ is the within-subsystem Jacobian, $\mathbf{B}$ is the across subsystem Jacobian and $\mathbf{x}^A$, $\mathbf{x}^B$ are the states of subsystems A and B, respectively. For a given control system, the time-varying reachability Gramian on the interval $[t_0,t_1]$ is defined as

\[
\mathbf{W}_r(t_0, t_1) \triangleq \int_{t_0}^{t_1}\mathbf{\Phi}(t_1, \tau)\mathbf{B} (\tau)\mathbf{B}^T(\tau)\mathbf{\Phi}^T(t_1, \tau) d\tau
\]

where $\mathbf{\Phi}$ denotes the state-transition matrix of the intrinsic dynamics of subsystem A without any input (i.e., $\delta\mathbf{x}^A(t) = \mathbf{\Phi}(t, t_0)\delta\mathbf{x}^A(t_0)$) \cite{antsaklis2006linear}.  Differentiating with respect to $t$, we obtain $\delta \dot{\mathbf{x}}^A(t) = \frac{\partial}{\partial t} \mathbf{\Phi}(t, t_0)\delta\mathbf{x}^A(t_0)$. Noting also that, in the absence of input from subsystem B, we have
\[
\delta \dot{\mathbf{x}}^A(t) = \mathbf{A}(t)\delta\mathbf{x}^A(t) = \mathbf{A}(t)\mathbf{\Phi}(t, t_0)\delta\mathbf{x}^A(t_0)
\]
Thus setting the equations equal to each other and canceling $\delta \mathbf{x}^A(t_0)$ from both sides we obtain
\[
\frac{\partial}{\partial t} \mathbf{\Phi}(t, t_0) = \mathbf{A}(t)\mathbf{\Phi}(t, t_0)
\]
Note that $\mathbf{\Phi}(t_0, t_0) = \mathbf{I}$. Now, letting $\mathbf{\Gamma}(t, \tau) =\mathbf{\Phi}(t, \tau)\mathbf{B} (\tau)\mathbf{B}^T(\tau)\mathbf{\Phi}^T(t, \tau) $ and differentiating the Gramian expression with respect to the second argument (and using the Leibniz integral rule) yields

\[
\begin{aligned}
\frac{\partial}{\partial t} \mathbf{W}_r(t_0, t) &= \frac{\partial}{\partial t}\int_{t_0}^{t}\mathbf{\Gamma}(t, \tau) d\tau \\
&= \mathbf{\Gamma}(t, t) \frac{\partial}{\partial t} (t) - \mathbf{\Gamma}(t, t_0) \frac{\partial}{\partial t} (t_0) + \int_{t_0}^{t} \frac{\partial}{\partial t}\mathbf{\Gamma}(t, \tau) d\tau \\
&= \mathbf{I} \mathbf{B}(t)\mathbf{B}^T (t) \mathbf{I} (1) - 0 + \int_{t_0}^t\frac{\partial}{\partial t}\mathbf{\Gamma}(t, \tau)  d\tau
\end{aligned}
\]

and observing

\[
\begin{aligned}
\frac{\partial}{\partial t}\mathbf{\Gamma}(t, \tau)  &=
\left(\frac{\partial}{\partial t} \mathbf{\Phi}(t, t_0) \right)\mathbf{B} (\tau)\mathbf{B}^T(\tau)\mathbf{\Phi}^T(t, \tau)  + \mathbf{\Phi}(t, t_0) \mathbf{B} (\tau)\mathbf{B}^T(\tau)\left(\frac{\partial}{\partial t}\mathbf{\Phi}(t, \tau)\right)^T \\
& = \mathbf{A}(t)\mathbf{\Phi}(t, t_0) \mathbf{B} (\tau)\mathbf{B}^T(\tau)\mathbf{\Phi}^T(t, \tau) + \mathbf{\Phi}(t, t_0) \mathbf{B} (\tau)\mathbf{B}^T(\tau)\mathbf{\Phi}^T(t, \tau)\mathbf{A}^T(t) \\
&= \mathbf{A}(t)\mathbf{\Gamma}(t, \tau) + \mathbf{\Gamma}(t, \tau) \mathbf{A}^T(t)
\end{aligned}
\]
we can continue

\[
\begin{aligned}
\frac{\partial}{\partial t} \mathbf{W}_r(t_0, t) &= \mathbf{B}(t)\mathbf{B}^T (t) + \mathbf{A}(t)\int_{t_0}^{t}\mathbf{\Gamma}(t, \tau) d\tau +\left( \int_{t_0}^{t}\mathbf{\Gamma}(t, \tau) d\tau \right)\mathbf{A}^T(t) \\
&= \mathbf{B}(t)\mathbf{B}^T (t) + \mathbf{A}(t)\mathbf{W}_r(t_0, t) + \mathbf{W}_r(t_0, t)\mathbf{A}^T(t)
\end{aligned}
\]

which illustrates that the reachability Gramian can be solved for using an ODE integrator (with initial condition $\mathbf{W}_r(t_0, t_0) = \mathbf{0}$) \cite{tyner2010geojac}. In practice, to compute the reachability Gramians using the trained JacobianODE models, we fit a cubic spline $\mathbf{c}(t)$ to the reference trajectory $\mathbf{x}(t)$ and compute $\mathbf{J}(t) = \mathbf{J}(\mathbf{c}(t))$. We can then parse the Jacobian matrix into its component submatrices and compute the Gramians accordingly.

The reachability Gramian is a symmetric positive semidefinite matrix \cite{antsaklis2006linear}. The optimal cost of driving the system from state $\delta\mathbf{x}_0$ to state $\delta\mathbf{x}_1$ on time interval $[t_0, t_1]$ can be computed as
\[
\left(\delta\mathbf{x}_1 - \mathbf{\Phi}(t_1, t_0)\delta\mathbf{x}_0\right)^T \mathbf{W}^{-1}_r(t_0, t_1)\left(\delta\mathbf{x}_1 - \mathbf{\Phi}(t_1, t_0)\delta\mathbf{x}_0\right)
\]
Thus each eigenvalue of $\mathbf{W}_r$ reflects the ease of control along the corresponding eigenvector \cite{antsaklis2006linear, lindmark2018ctrl}. Eigenvectors with larger corresponding eigenvalues will have smaller inverse eigenvalues, and thus scale the cost down along those directions when computing the cost as above.

\subsection{Extension to non-autonomous dynamics}
While we deal with autonomous systems in this work, we note that this construction can easily be extended to include non-autonomous dynamical systems, of the form $\dot{\mathbf{x}}(t) = \mathbf{f}(\mathbf{x}, t)$ by constructing an augmented state $\tilde{\mathbf{x}} \in \mathbb{R}^{n + 1}$ which is simply the concatenation of $\mathbf{x}$ with $t$. This yields the autonomous dynamics $\dot{\tilde{\mathbf{x}}} = \tilde{\mathbf{f}}(\tilde{\mathbf{x}})$, where $\tilde{\mathbf{f}}: \mathbb{R}^{n+1} \to \mathbb{R}^{n+1}$ is the concatenation of $\mathbf{f}$ with a function that maps all states to $1$.  

\subsection{Extension to more than two subsystems}
\label{supp:extend-multiple}

Consider a nonlinear dynamical system in $\mathbb{R}^n$, defined by $\dot{\mathbf{x}}(t) = \mathbf{f}(\mathbf{x}(t))$. Suppose now that the system is composed of $K$ subsystems, where the time evolution of subsystem $k$ is given by $\mathbf{x}^{(k)}(t)\in \mathbb{R}^{n_k}$. $\mathbf{x}(t)$ is then comprised of a concatenation of the $\mathbf{x}^{(k)}(t)$, with $\sum_{k=1}^K n_k = n$. Given a particular reference trajectory, $\mathbf{x}(t)$, with $\mathbf{J}$ as the Jacobian of $\mathbf{f}$, then the tangent space dynamics around the reference trajectory for subsystem $\alpha \in \{1,...,K\}$ are given by \[
\delta \dot{\mathbf{x}}^{(\alpha)}(t) = \sum_{k=1}^{K} \mathbf{J}^{k\to\alpha}(\mathbf{x}(t))\delta\mathbf{x}^{(k)}(t)
\] where $\mathbf{J}^{k\to\alpha}(\mathbf{x}(t)) \in \mathbb{R}^{n_{\alpha}\times n_k}$ is the submatrix of $\mathbf{J}(\mathbf{x}(t))$ in which the columns correspond to subsystem $k$ and the rows correspond to subsystem $\alpha$. Now, for a given subsystem $\beta \in \{1, ..., K\}$ with $\beta \neq \alpha$, we wish to analyze the ease with which $\beta$ can control $\alpha$ locally around the reference trajectory, without intervention from other subsystems. Discounting the interventions from other subsystems equates to setting $\delta\mathbf{x}^{(k)}(t) = 0$ for $k \neq \alpha,\beta$, leaving the expression \[
\delta \dot{\mathbf{x}}^{(\alpha)}(t) =\mathbf{J}^{\alpha\to\alpha}(\mathbf{x}(t))\delta\mathbf{x}^{(\alpha)}(t) + \mathbf{J}^{\beta\to\alpha}(\mathbf{x}(t))\delta\mathbf{x}^{(\beta)}(t)
\] which quantifies the influence $\beta$ can exert over $\alpha$ in the absence of perturbations from any other subsystem. Using this representation, the reachability Gramian can be computed as described above. Performing this procedure for all pairs of subsystems $\alpha,\beta \in \{1,...,K\}$ thus characterizes all pairwise control relationships between subsystems along the reference trajectory.

\section{Supplementary results}
\label{supp:supp-results}

\subsection{Control and communication capture different phenomena}
\label{supp:control-v-comm}

\begin{figure}[!htbp] % 'h' means "here" (try placing it near the text)
    \vspace{-2mm}
    \centering
    \includegraphics[width=1.0\linewidth]{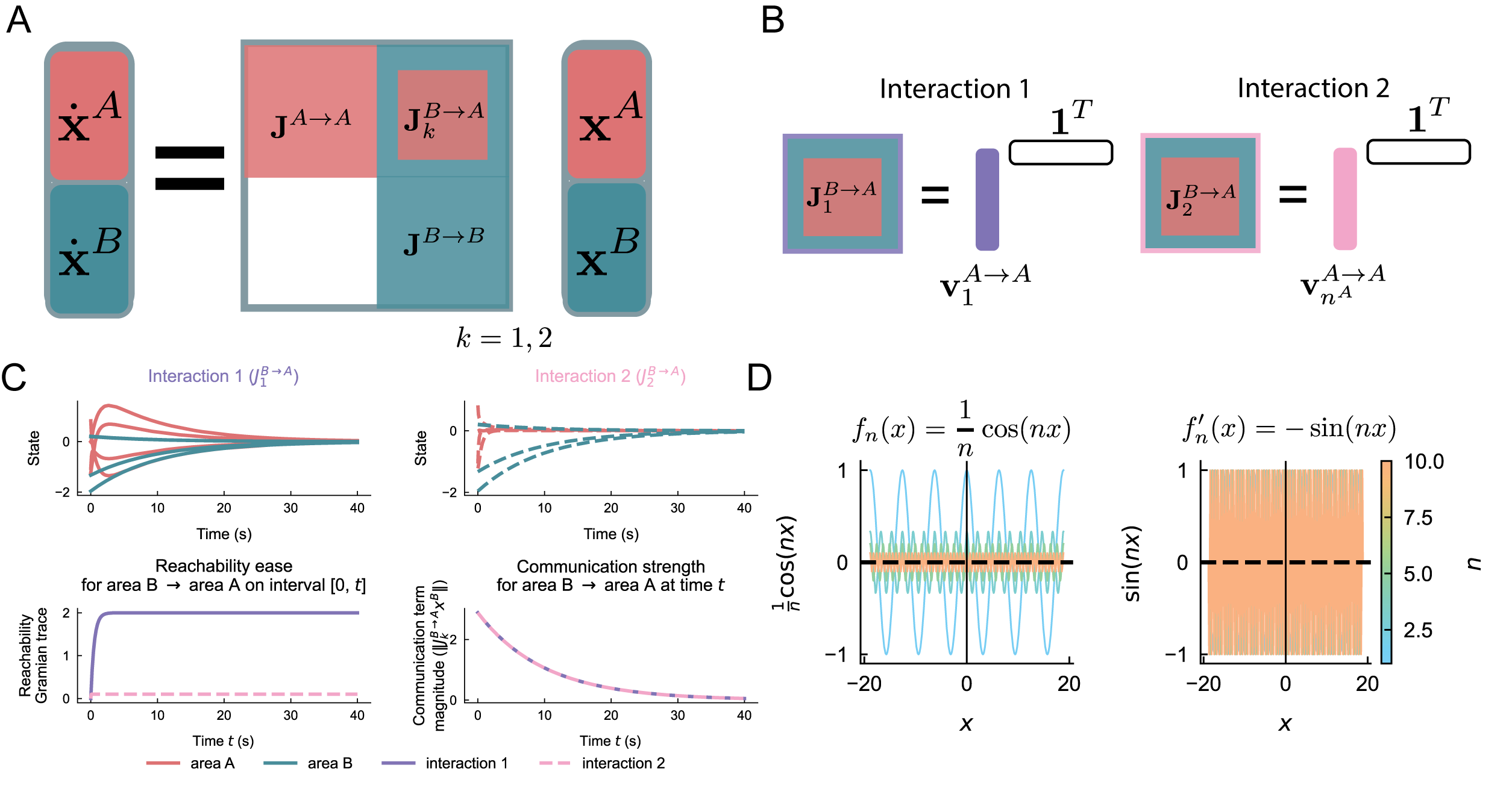} % Adjust width as needed
    \caption{\textbf{Communication versus control in linear systems and the challenge of Jacobian estimation.} (A) Setup of two linearly connected brain areas, A and B. (B) Interaction matrices projecting signals from area B onto either the most stable (Interaction 1) or least stable (Interaction 2) eigenvectors of area A. (C) Although both interactions communicate identical signal magnitudes, Interaction 2 provides significantly enhanced reachability due to alignment with unstable modes of area A dynamics. (D) An illustrative example demonstrating that accurate approximation of a function (left panel) does not guarantee accurate approximation of its derivative (right panel), emphasizing the necessity of directly estimating the Jacobian.}
    \label{suppfig:control-v-comm}
    \vspace{-1mm}
\end{figure}

In the Introduction (Section \ref{sec:introduction}) we note that measuring communication between two systems is different than measuring control. To illustrate this, consider two brain areas, A and B, whose dynamics are internally linear. The areas are connected only through a linear feedforward interaction term from area B to area A (Figure \ref{suppfig:control-v-comm}A). Concretely, we consider the dynamics:

\begin{equation*}
\begin{aligned}
\dot{\mathbf{x}}^A(t) &= \mathbf{J}^{A \to A} \mathbf{x}^A (t)+ \mathbf{J}^{B \to A} \mathbf{x}^B(t) \\
\dot{\mathbf{x}}^B(t) &= \mathbf{J}^{B \to B} \mathbf{x}^B(t)
\end{aligned}
\end{equation*}

Here, the $\mathbf{J}$ matrices are time-invariant. When considering communication between brain areas, one might aim to find the subspaces in which communication between B and A occurs, as well as the messages passed \cite{semedo2019commsub, karniol-tambour2022-switching}. In this construction, the subspace in which area B communicates with area A is explicitly given by $\mathbf{J}^{B \to A}$, and thus the message, or input, from area B to area A at any time $t$ is simply given by $\mathbf{J}^{B \to A}\mathbf{x}(t)$.  We pick the dimensions of each region to be $n_A = n_B = 4$ and let the eigenvectors of the (negative definite matrix) $\mathbf{J}^{A \to A}$ be given by $\mathbf{v}_i, i = 1, 2, 3, 4$. The eigenvectors are numbered in order of decreasing real part of their corresponding eigenvalue. Now, suppose we construct the interaction matrix $\mathbf{J}^{B \to A}$ in two different ways: If we let (1) $\mathbf{J}_1^{B \to A} = \mathbf{v_1}\mathbf{1}^T$ (Figure \ref{suppfig:control-v-comm}B, left) , then the signal from B is projected onto the most stable mode of region A, whereas if we let (2) $\mathbf{J}_2^{B \to A} = \mathbf{v_4}\mathbf{1}^T$ (Figure \ref{suppfig:control-v-comm}B, right), the signal is projected onto the least stable mode. While interaction 1 and interaction 2 communicated messages with identical magnitudes, interaction 2 led to much lower cost reachability control  (Figure \ref{suppfig:control-v-comm}C). This illustrates that control depends not only on the directions along which the areas can communicate, but also on how \textit{aligned} the communication is with the target area's dynamics.

\subsection{Derivative estimation is not implied by function estimation}
\label{supp:deriv-v-func}

An alternative approach to directly estimating the Jacobians would be to learn an approximation $\hat{\mathbf{f}}$ of the function $\mathbf{f}$, and then approximate the Jacobian via automatic differentiation (i.e., an estimate $\hat{\mathbf{J}}= \frac{\partial}{\partial x} \hat{\mathbf{f}}$, as with the NeuralODEs). While this can be effective in certain scenarios, it is not generally the case that approximating a function well will yield a good approximation of its derivative. To illustrate this, we recall an example from \citet{latremoliere2022jacest}, in which functions $f_n = \frac{1}{n} \cos(nx)$ are used to approximate the function $f(x) = 0$ (Figure \ref{suppfig:control-v-comm}D). While these approximations improve with increasing $n$ (i.e., $\lim_{n \to \infty} f_n = f$), this is not the case for the derivative ($\lim_{n \to \infty} f'_n  = -\sin(nx) \neq f'$).  This demonstrates the necessity of learning $\mathbf{J}$ directly (rather than first approximating $\mathbf{f}$), which we also demonstrate empirically. In the context of machine learning, this setting could be interpreted as overfitting \cite{latremoliere2022jacest}. As long as function estimates match at the specific points in the training set, how the function fluctuates between these points is not constrained to match the true function. For this reason, we included the Frobenius norm Jacobian regularization in our implementation of the NeuralODEs (Appendix \ref{supp:neuralode-details}).

\vspace{-3mm}

\subsection{Full benchmark dynamical systems results}
\label{supp:dyn-sys-supp-results}

\begin{figure}[!htbp] % 'h' means "here" (try placing it near the text)
    \vspace{-2mm}
    \centering
    \includegraphics[width=1.0\linewidth]{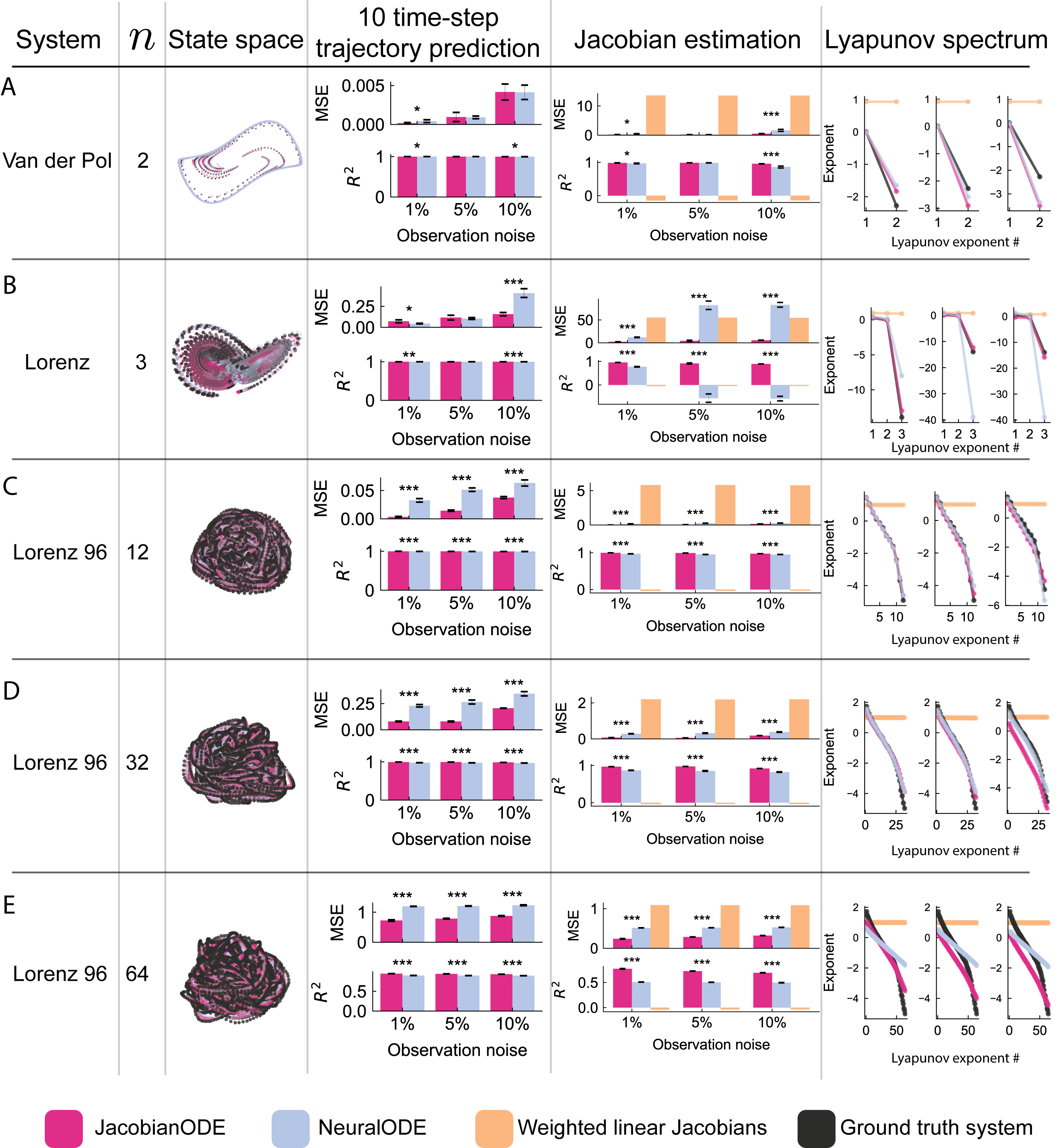} % Adjust width as needed
    \caption{\textbf{Full dynamical systems prediction results.} State space representations (using spectral embedding for systems that have more than 3 dimensions), 10 time-step trajectory predictions, Jacobian estimation, and Lyapunov spectrum estimates for each of (A) the Van der Pol oscillator, (B) the Lorenz system, and the Lorenz 96 system with (C) 12, (D) 32, and (E) 64 dimensions.}
    \label{suppfig:full-dyn-sys}
    \vspace{-1mm}
\end{figure}

We here present the full results for all considered example dynamical systems. 10 time-step trajectory predictions along with Jacobian estimation and estimated Lyapunov spectra are displayed in Figure \ref{suppfig:full-dyn-sys}, along with MSE and $R^2$ in Table \ref{tab:traj_metrics} (trajectory prediction) and Table \ref{tab:jac_metrics} (Jacobian estimation). Jacobian estimation errors using the 2-norm are presented in Table \ref{tab:2_norm_metrics}.

\begin{table}[ht]
  \centering
  \caption{Trajectory prediction metrics (MSE and $R^2$) for each system and training noise. Errors are mean $\pm$ standard deviation across  five random initializations of the model architectures.}
  \label{tab:traj_metrics}
  \scriptsize
  \setlength{\tabcolsep}{6pt}
  \begin{adjustbox}{width=\textwidth}
    \begin{tabular}{llc@{\hspace{0.5em}}cc@{\hspace{0.5em}}cc}
      \toprule
      & & \multicolumn{2}{c}{MSE} & \multicolumn{2}{c}{$R^2$} \\
      \cmidrule(lr){3-4} \cmidrule(lr){5-6}
      Project & Training noise & JacobianODE & NeuralODE & JacobianODE & NeuralODE \\
      \midrule
      VanDerPol (2 dim) & 1\% & \textbf{0.00017 ± 0.00007} & 0.0004 ± 0.0002 & \textbf{0.99987 ± 0.00005} & 0.9996 ± 0.0002 \\
       & 5\% & 0.0010 ± 0.0006 & \textbf{0.0009 ± 0.0002} & \textbf{0.9995 ± 0.0003} & 0.99949 ± 0.00009 \\
       & 10\% & 0.004 ± 0.001 & \textbf{0.0041 ± 0.0009} & \textbf{0.9979 ± 0.0005} & 0.9969 ± 0.0003 \\
      \midrule
      Lorenz (3 dim) & 1\% & 0.07 ± 0.02 & \textbf{0.045 ± 0.004} & 0.99979 ± 0.00004 & \textbf{0.99986 ± 0.00001} \\
       & 5\% & 0.12 ± 0.03 & \textbf{0.10 ± 0.01} & 0.9996 ± 0.0001 & \textbf{0.99958 ± 0.00004} \\
       & 10\% & \textbf{0.16 ± 0.02} & 0.41 ± 0.05 & \textbf{0.99928 ± 0.00008} & 0.9981 ± 0.0003 \\
      \midrule
      Lorenz 96 (12 dim) & 1\% & \textbf{0.0035 ± 0.0007} & 0.033 ± 0.003 & \textbf{0.99969 ± 0.00007} & 0.9969 ± 0.0003 \\
       & 5\% & \textbf{0.014 ± 0.001} & 0.051 ± 0.003 & \textbf{0.9985 ± 0.0001} & 0.9955 ± 0.0003 \\
       & 10\% & \textbf{0.038 ± 0.002} & 0.063 ± 0.005 & \textbf{0.9960 ± 0.0002} & 0.9940 ± 0.0007 \\
      \midrule
      Lorenz 96 (32 dim) & 1\% & \textbf{0.079 ± 0.004} & 0.23 ± 0.01 & \textbf{0.9939 ± 0.0003} & 0.977 ± 0.002 \\
       & 5\% & \textbf{0.079 ± 0.004} & 0.26 ± 0.02 & \textbf{0.9939 ± 0.0003} & 0.974 ± 0.002 \\
       & 10\% & \textbf{0.205 ± 0.001} & 0.34 ± 0.02 & \textbf{0.9837 ± 0.0001} & 0.968 ± 0.002 \\
      \midrule
      Lorenz 96 (64 dim) & 1\% & \textbf{0.72 ± 0.03} & 1.194 ± 0.005 & \textbf{0.946 ± 0.002} & 0.8997 ± 0.0007 \\
       & 5\% & \textbf{0.79 ± 0.01} & 1.205 ± 0.007 & \textbf{0.9400 ± 0.0010} & 0.8986 ± 0.0010 \\
       & 10\% & \textbf{0.87 ± 0.01} & 1.23 ± 0.01 & \textbf{0.9332 ± 0.0009} & 0.896 ± 0.001 \\
      \midrule
      Task-trained RNN & 1\% & \textbf{0.0042 ± 0.0006} & 0.00558 ± 0.00002 & \textbf{0.9981 ± 0.0003} & 0.99762 ± 0.00001 \\
       & 5\% & \textbf{0.0043 ± 0.0006} & 0.00596 ± 0.00005 & \textbf{0.9981 ± 0.0002} & 0.99747 ± 0.00002 \\
       & 10\% & 0.0089 ± 0.0001 & \textbf{0.00655 ± 0.00003} & 0.99591 ± 0.00004 & \textbf{0.99721 ± 0.00001} \\
      \midrule
      \bottomrule
    \end{tabular}
  \end{adjustbox}
\end{table}
\begin{table}[ht]
  \centering
  \caption{Jacobian estimation metrics (MSE and $R^2$) for each system and training noise. Errors are mean $\pm$ standard deviation across  five random initializations of the model architectures.}
  \label{tab:jac_metrics}
  \scriptsize
  \setlength{\tabcolsep}{6pt}
  \begin{adjustbox}{width=\textwidth}
    \begin{tabular}{llc@{\hspace{0.5em}}cc@{\hspace{0.5em}}cc}
      \toprule
      & & \multicolumn{2}{c}{MSE} & \multicolumn{2}{c}{$R^2$} \\
      \cmidrule(lr){3-4} \cmidrule(lr){5-6}
      Project & Training noise & JacobianODE & NeuralODE & JacobianODE & NeuralODE \\
      \midrule
      VanDerPol (2 dim) & 1\% & \textbf{0.18 ± 0.06} & 0.4 ± 0.1 & \textbf{0.985 ± 0.005} & 0.97 ± 0.01 \\ 
       & 5\% & \textbf{0.16 ± 0.02} & 0.17 ± 0.03 & \textbf{0.987 ± 0.002} & 0.985 ± 0.002 \\ 
       & 10\% & \textbf{0.50 ± 0.04} & 1.6 ± 0.3 & \textbf{0.958 ± 0.003} & 0.86 ± 0.03 \\ 
      \midrule
      Lorenz (3 dim) & 1\% & \textbf{2.4 ± 0.1} & 12.0 ± 0.7 & \textbf{0.954 ± 0.002} & 0.77 ± 0.01 \\ 
       & 5\% & \textbf{4.6 ± 1.2} & 82.0 ± 9.2 & \textbf{0.91 ± 0.02} & -0.6 ± 0.2 \\ 
       & 10\% & \textbf{5.7 ± 0.2} & 83.0 ± 5.2 & \textbf{0.891 ± 0.004} & -0.58 ± 0.10 \\ 
      \midrule
      Lorenz 96 (12 dim) & 1\% & \textbf{0.012 ± 0.003} & 0.18 ± 0.01 & \textbf{0.9979 ± 0.0006} & 0.969 ± 0.002 \\ 
       & 5\% & \textbf{0.053 ± 0.006} & 0.27 ± 0.02 & \textbf{0.991 ± 0.001} & 0.953 ± 0.003 \\ 
       & 10\% & \textbf{0.153 ± 0.008} & 0.27 ± 0.01 & \textbf{0.973 ± 0.001} & 0.951 ± 0.002 \\ 
      \midrule
      Lorenz 96 (32 dim) & 1\% & \textbf{0.075 ± 0.004} & 0.28 ± 0.01 & \textbf{0.965 ± 0.002} & 0.866 ± 0.006 \\ 
       & 5\% & \textbf{0.060 ± 0.003} & 0.32 ± 0.02 & \textbf{0.972 ± 0.001} & 0.850 ± 0.009 \\ 
       & 10\% & \textbf{0.180 ± 0.003} & 0.38 ± 0.02 & \textbf{0.915 ± 0.001} & 0.820 ± 0.008 \\ 
      \midrule
      Lorenz 96 (64 dim) & 1\% & \textbf{0.238 ± 0.007} & 0.515 ± 0.002 & \textbf{0.773 ± 0.007} & 0.509 ± 0.002 \\ 
       & 5\% & \textbf{0.286 ± 0.003} & 0.519 ± 0.003 & \textbf{0.727 ± 0.003} & 0.505 ± 0.003 \\ 
       & 10\% & \textbf{0.321 ± 0.004} & 0.527 ± 0.005 & \textbf{0.694 ± 0.003} & 0.497 ± 0.005 \\ 
      \midrule
      Task-trained RNN & 1\% & \textbf{2.2 ± 0.2} & 5.2777 ± 0.0009 & \textbf{0.59 ± 0.03} & -0.0008 ± 0.0002 \\ 
       & 5\% & \textbf{1.70 ± 0.08} & 5.2900 ± 0.0001 & \textbf{0.68 ± 0.01} & -0.00315 ± 0.00002 \\ 
       & 10\% & \textbf{1.99 ± 0.01} & 5.2889 ± 0.0001 & \textbf{0.623 ± 0.002} & -0.00295 ± 0.00002 \\ 
      \midrule
      \bottomrule
    \end{tabular}
  \end{adjustbox}
\end{table}
\begin{table}[ht]
  \centering
  \caption{Mean 2-norm error on Jacobian estimation, $\langle\|\mathbf{J} - \hat{\mathbf{J}}\|_2 \rangle$, for each system and noise level. Errors are reported as mean $\pm$ standard deviation, with mean and standard deviation computed over five random initializations of the model architectures.}
  \label{tab:2_norm_metrics}
  \scriptsize
  \setlength{\tabcolsep}{6pt}
  \begin{adjustbox}{width=\textwidth}
    \begin{tabular}{llccc}
      \toprule
      Project & Training noise & JacobianODE & NeuralODE & Weighted Linear \\
      \midrule
      VanDerPol (2 dim) & 1\% & \textbf{0.7 ± 0.1} & 1.0 ± 0.3 & 6.05 \\
       & 5\% & \textbf{0.71 ± 0.05} & \textbf{0.71 ± 0.08} & 6.03 \\
       & 10\% & \textbf{1.31 ± 0.05} & 2.2 ± 0.2 & 6.02 \\
      \midrule
      Lorenz (3 dim) & 1\% & \textbf{3.2 ± 0.2} & 8.6 ± 0.3 & 17.06 \\
       & 5\% & \textbf{4.9 ± 0.9} & 25.9 ± 1.5 & 17.02 \\
       & 10\% & \textbf{6.0 ± 0.1} & 26.4 ± 0.9 & 16.95 \\
      \midrule
      Lorenz 96 (12 dim) & 1\% & \textbf{0.8 ± 0.1} & 3.5 ± 0.2 & 14.94 \\
       & 5\% & \textbf{1.75 ± 0.09} & 4.1 ± 0.1 & 14.93 \\
       & 10\% & \textbf{2.91 ± 0.09} & 4.2 ± 0.1 & 14.92 \\
      \midrule
      Lorenz 96 (32 dim) & 1\% & \textbf{3.75 ± 0.08} & 7.8 ± 0.1 & 16.29 \\
       & 5\% & \textbf{3.10 ± 0.09} & 8.1 ± 0.2 & 16.26 \\
       & 10\% & \textbf{4.89 ± 0.05} & 8.6 ± 0.1 & 16.28 \\
      \midrule
      Lorenz 96 (64 dim) & 1\% & \textbf{8.4 ± 0.1} & 12.63 ± 0.02 & 16.84 \\
       & 5\% & \textbf{9.04 ± 0.07} & 12.66 ± 0.03 & 16.81 \\
       & 10\% & \textbf{9.42 ± 0.06} & 12.72 ± 0.05 & 16.82 \\
      \midrule
      Task-trained RNN & 1\% & \textbf{30.4 ± 0.4} & 38.565 ± 0.006 & 39.29 \\
       & 5\% & \textbf{29.4 ± 0.9} & 38.5611 ± 0.0004 & 38.91 \\
       & 10\% & \textbf{36.0 ± 0.1} & 38.5600 ± 0.0004 & 38.70 \\
      \midrule
      \bottomrule
    \end{tabular}
  \end{adjustbox}
\end{table}

\subsection{Ablation studies}
\label{supp:ablation-studies}

To determine the value of the different components of the JacobianODE learning framework, we performed several ablation studies. We chose to evaluate on the Lorenz system and the task-trained RNNs, as these together provide two common settings of chaos and stability, as well as low- and high-dimensional dynamics.

\insettitle{Ablating the Jacobian-parameterized ODEs} The JacobianODE framework constructs an estimate of the initial time derivative $\mathbf{f}$ via a double integral of the Jacobian as described in Section \ref{sec:param-ode-jac}. To briefly recall, Jacobian path integration is described as 

\begin{equation}
\mathbf{f}(\mathbf{x}(t_f)) - \mathbf{f}(\mathbf{x}(t_i)) \ \ = \ \ \int_{\mathcal{C}}\mathbf{J}\ ds \ \ = \ \ \int_{t_i}^{t_f} \mathbf{J}(\mathbf{c}(r))\mathbf{c}'(r) \ dr,
\label{jaceqnsupp1}
\end{equation}

When we do not have access to the initial derivative and base point $\mathbf{f}(\mathbf{x}(t_i)), \mathbf{x}(t_i)$, we use the following formulation

 \begin{equation}
\mathbf{f}(\mathbf{x}(t)) = \frac{G(t_0, t; \mathbf{J}, \mathbf{c}) + \mathbf{x}(t) - \mathbf{x}(t_0)}{t - t_0},
\end{equation}

where
\begin{equation}
G(t_0, t; \mathbf{J}, \mathbf{c}) = \int_{t_0}^{t} \int_{s}^{t} \mathbf{J}(\mathbf{c}_{s,t}(r))\mathbf{c}_{s,t}'(r)dr ds
\end{equation} 

Alternatively, one could use Equation \ref{jaceqnsupp1} and learn $\mathbf{f}(\mathbf{x}(t_i))$ and $\mathbf{x}(t_i)$ as learnable parameters ($\mathbf{x}(t_i)$ is needed as the first point of the path, i.e. $\mathbf{c}(t_i) = \mathbf{x}(t_i)$ inside the integral) \cite{beik-mohammadi2024-ncds}. Here, the path integral between $\mathbf{x}(t_i)$ and $\mathbf{x}(t_f)$ is a linear interpolation, as before. These ablated models were trained on 10 time-step prediction, as with the original JacobianODEs. For these models, we completed a full sweep over the loop closure loss weight $\lambda_{\text{loop}}$ in order to determine the best hyperparameter.

\insettitle{Ablating teacher forcing} We trained models without any teacher-forcing. That is, models were able to generate only one-step predictions, without any recursive predictions. Again we did a full hyperparameter sweep to pick $\lambda_{\text{loop}}$.

\insettitle{Ablating loop closure loss} We ablated the loop closure loss in two ways. The first was to set $\lambda_{\text{loop}}=0$ to illustrate what would happen if there were no constraints placed on the learned Jacobians. The second was to instead use the Jacobian Frobenius norm regularization that was used for the NeuralODEs (details are in Appendix \ref{supp:neuralode-details}). We did a full sweep to pick $\lambda_{\text{jac}}$, the Frobenius norm regularization weight.

\begin{table}[ht]
  \centering
  \caption{Mean Frobenius norm error $\langle\|\mathbf{J} - \hat{\mathbf{J}}\|_F \rangle$ for different model ablations on the Lorenz system with 10\% observation noise. Errors are reported as mean $\pm$ standard deviation, with statistics computed over 8 test trajectories, each consisting of 1200 points.}
  \label{tab:ablation_metrics_lorenz}
  \scriptsize
  \setlength{\tabcolsep}{6pt}
  \begin{adjustbox}{width=0.8\textwidth}
    \begin{tabular}{lc}
      \toprule
      Model variant & Frobenius norm error \\
      \midrule
      JacobianODE (original) & \textbf{6.46 ± 1.50} \\
      With learned base derivative point & 7.87 ± 1.14 \\
      No teacher forcing& 11.59 ± 1.30 \\
      No loop closure & 58.22 ± 2.00 \\
      With Jacobian penalty instead of loop closure & 9.59 ± 2.45 \\
      \bottomrule
    \end{tabular}
  \end{adjustbox}
\end{table}

\insettitle{Ablation results} The performance of the ablated models on Jacobian estimation in the Lorenz system are presented in Table \ref{tab:ablation_metrics_lorenz}. The original JacobianODE outperforms all ablated models, indicating that all components of the JacobianODE training framework improve the model's performance in this setting. Ablating the Jacobian-parameterized initial derivative estimate resulted in a slight decrease in the estimation loss. This is potentially because the network could offload some of the responsibility for generating correct trajectory predictions onto the estimated base point $\mathbf{x}(t_i)$ and derivative estimate $\hat{\mathbf{f}}(\mathbf{x}(t_i))$, slightly reducing the necessity of estimating correct Jacobians. Ablating the teacher forcing annealing predictably led to worse Jacobian estimation, as the network no longer has to consider how errors will propagate along the trajectory. The most dramatic increase in error was with the ablation of the loop closure loss. Without this important regularization, the learned Jacobians reproduced the dynamics but were not constrained to be conservative, resulting in poor Jacobian estimation. The inclusion of the Frobenius penalty on the Jacobians mitigated this, although it did not encourage accurate Jacobian estimation to the same degree as the loop closure loss. 

\begin{table}[ht]
  \centering
  \caption{Mean Frobenius norm error $\langle\|\mathbf{J} - \hat{\mathbf{J}}\|_F \rangle$ for different model ablations on the task-trained RNN with 10\% observation noise. Errors are reported as mean $\pm$ standard deviation, with statistics computed over 409 test trajectories, each consisting of the 49 points from the second delay and response epochs.}
  \label{tab:ablation_metrics_wmtask}
  \scriptsize
  \setlength{\tabcolsep}{6pt}
  \begin{adjustbox}{width=0.8\textwidth}
    \begin{tabular}{lc}
      \toprule
      Model variant & Frobenius norm error \\
      \midrule
      JacobianODE (original) & 180.13 ± 1.55 \\
      With learned base derivative point & 186.28 ± 1.17 \\
      No teacher forcing& 187.72 ± 1.63 \\
      No loop closure & 313.31 ± 29.88 \\
      With Jacobian penalty instead of loop closure & \textbf{163.69 ± 4.22} \\
      \bottomrule
    \end{tabular}
  \end{adjustbox}
\end{table}
We then tested the ablated the models on Jacobian estimation in the task-trained RNN, with results presented in Table \ref{tab:ablation_metrics_wmtask}. Again, ablating the Jacobian-parameterized derivative estimates, teacher forcing, and loop closure resulted in worse Jacobian estimation. Interestingly, in this setting, the inclusion of a penalty on the Frobenius norm of the Jacobians outperformed the use of the loop closure loss. This could potentially be because the loop closure loss is more difficult to drive to zero in high dimensional systems, or because the loop closure loss is more important in chaotic systems like the Lorenz system considered above. Future work should consider in what contexts each kind of regularization is most beneficial to JacobianODE models.

\subsection{NeuralODEs achieve improved performance at the cost of increased inference time}
\label{supp:computation-scaling}

In the main paper, we implemented both the JacobianODEs and the NeuralODEs as four-layer MLPs, with the four layers having sizes of 256, 1024, 2048, and 2048 respectively. This was done for the fairest architectural comparison between the models, to ensure that both models had the same representational capacity when generating their respective outputs. However, there are many architectural changes that we could make to this setup that impact performance. We hypothesized based on the discussion in Appendix \ref{supp:deriv-v-func} that increasing the hidden layer size of the NeuralODEs would improve Jacobian estimation, as larger models have been known to learn smoother representations. Furthermore, we wondered whether including residual blocks in place of the standard MLP implementation would improve Jacobian estimation.

To test this, we implemented the NeuralODEs as four-layer residual networks and tested three different sizes of hidden layer: 1024, 2048, and 4096. Results are in Table \ref{tab:fro_errors_by_noise}. For nearly all models, these changes yielded only marginal improvements over the original NeuralODE model. Only the model with 4096-dimensional hidden layers under 10\% training noise achieves a performance near the \textit{original} JacobianODEs. However, the performance limit is still below that of the JacobianODEs, even with a large increase in the representational capacity of the model. It is furthermore of note that the NeuralODEs were able to significantly improve performance only in the high noise setting. This suggests that high noise is necessary for the model to be forced to learn the response of the system to perturbation. In contrast, JacobianODEs perform similarly across all noise levels, indicating a stronger inductive bias to learn the response of the system to perturbation.
\begin{table}[ht]
  \centering
  \caption{Mean Frobenius norm error $\langle\|\mathbf{J} - \hat{\mathbf{J}}\|_F \rangle$ for different model types on the task-trained RNN data across observation noise levels. Errors are reported as mean $\pm$ standard deviation, with statistics computed over 409 test trajectories.}
  \label{tab:fro_errors_by_noise}
  \scriptsize
  \setlength{\tabcolsep}{6pt}
  \begin{adjustbox}{width=0.8\textwidth}
    \begin{tabular}{lccc}
      \toprule
      & \multicolumn{1}{c}{Training noise} & \multicolumn{1}{c}{Training noise} & \multicolumn{1}{c}{Learnable
parameters} \\
      Model type & 5\% & 10\% & \\
      \midrule
      JacobianODE (original) & \textbf{174.1 ± 2.4} & \textbf{180.1 ± 1.5} & 4.02e+07 \\
      NeuralODE (original) & 294.0 ± 2.4 & 293.9 ± 2.4 & 6.85e+06 \\
      NeuralODE (1024 dim. hidden layers) & 291.6 ± 2.3 & 289.8 ± 2.3 & 3.41e+06 \\
      NeuralODE (2048 dim. hidden layers) & 288.5 ± 2.3 & 275.0 ± 2.7 & 1.31e+07 \\
      NeuralODE (4096 dim. hidden layers) & 275.8 ± 2.4 & 184.2 ± 5.6 & 5.14e+07 \\
      \bottomrule
    \end{tabular}
  \end{adjustbox}
\end{table}

\begin{figure}[!htbp] % 'h' means "here" (try placing it near the text)
    \vspace{-2mm}
    \centering
    \includegraphics[width=1.0\linewidth]{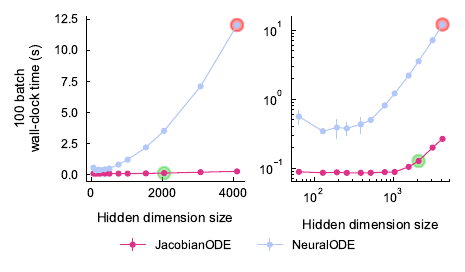} % Adjust width as needed
    \caption{\textbf{JacobianODEs achieve highly efficient Jacobian inference.} Jacobian inference times computed over ten repetitions of 100 batches (error bars indicate mean $\pm$ standard deviation). Red circles indicate the inference time corresponding to the largest hidden layer dimension of the highest performing NeuralODE model in Table \ref{tab:fro_errors_by_noise}. Green circles indicate the inference time corresponding to the largest hidden layer dimension of the highest performing JacobianODE model in Table \ref{tab:fro_errors_by_noise}. Each plot illustrates the same data but with different $x$ and $y$ scaling.}
    \label{suppfig:computation-scaling}
    \vspace{-1mm}
\end{figure}

While the increased hidden layer size seems to induce smoother hidden representations (as expected), it comes with increasing computational cost. Recall that the NeuralODEmodels directly parameterize the dynamics as $\dot{\mathbf{x}} = \hat{\mathbf{f}}^\theta(\mathbf{x})$. The Jacobian at state $\mathbf{x}$ is computed by directly backpropagating through the Neural ODE $\hat{\mathbf{f}}^\theta$, thereby significantly increasing compute. On the other hand, the JacobianODEs only require a forward pass.  We evaluated the Jacobian inference time of JacobianODE and NeuralODE models with four hidden layers of the same size on an H100 GPU. Each model was timed on inferring the Jacobians of 100 batches of the 128-dimensional task-trained RNN, with each batch containing 16 sequences of length 25. Timings were repeated ten times for each model. Results are plotted in Figure \ref{suppfig:computation-scaling}. 

The JacobianODE achieves much faster inference times than the NeuralODE -- approximately two orders of magnitude faster at large hidden dimension sizes. Furthermore, as shown in Table \ref{tab:fro_errors_by_noise}, the JacobianODE with a maximum hidden layer size of 2048 outperforms the NeuralODE with a maximum hidden layer size of 4096 on Jacobian estimation, and does so with orders of magnitude faster inference (Figure \ref{suppfig:computation-scaling}, green and red circles, Table \ref{tab:fro_errors_by_noise}). This suggests that while both architectures appear to have inference times that scale approximately exponentially, the JacobianODE achieves more favorable scaling across every hidden layer size we tested. Our analysis therefore illustrates that it is possible to improve the NeuralODEs' Jacobian estimation with larger models, but the inference time scaling renders these models ill-equipped for important settings such as real-time control and the analysis of high-volume neural data.

\subsection{Comparison to echo state networks}
Echo state networks (ESNs, or reservoir networks) are tools for time-series forecasting and analysis. ESNs are typically implemented as RNNs with fixed, sparse, hidden connectivity. The output weights of the ESNs can then be trained via linear regression to reproduce specific temporal patterns. ESNs have been shown to be able to forecast the Lorenz 63 system for a particular choice of hyperparameters \cite{pathak2017esn}. We note that \citet{pathak2017esn} implements the same method as \citet{pathak2018esn} .  We implemented the ESNs exact hyperparameters reported in Pathak et. al. 2017 for the Lorenz system, which involved a reservoir of 300 nodes. Our implementation of the Lorenz system was simulated with a sampling interval of \~0.015 s, very similar to the 0.02 s used in the paper. Thus, for consistency of comparison with JacobianODEs, we fit ESN models to data sampled every 0.015 s. 

To enable Jacobian estimation, we changed one small implementation detail from \citet{pathak2017esn}. Specifically, we trained the network output weights to predict the third variable $z$ using only the network state $\mathbf{r}(t)$, as opposed to the vector $\tilde{\mathbf{r}}(t)$ for which the node state is squared ($\tilde{r}_i(t) = r_i^2(t)$) for half of the nodes $i$. This is because squaring the reservoir state complicates Jacobian computation, and was stated by \citet{pathak2017esn} and \citet{lu2017esn} to only be necessary for symmetry breaking if only the $z$ variable of the Lorenz system were provided. We trained the ESNs on the same volume of data as the JacobianODEs and NeuralODEs, with 1\% observation noise. We swept the regularization parameter $\beta$ to ensure a good fit. We furthermore generated predictions using a 150 time-step lead-up to ensure the ESNs converged to a good internal representation. The ESNs predicted the Lorenz system well, achieving an MSE of 0.03 on the 10-step prediction task from our paper (on test data). This was very similar to the reported 10-step prediction MSEs of 0.07 and 0.04 for the JacobianODEs and NeuralODEs, respectively.

In \citet{pathak2017esn}, the Lyapunov exponents of the system were computed using the Jacobian of the \textit{interior} dynamics of the ESN, as the recursion necessary to compute the Jacobian can be formulated uniquely only in regard to this state representation. Thus, each of these Jacobians would be a $300 \times 300$ matrix. While this suffices to compute the Lyapunov exponents, it does not suffice for our purposes of obtaining Jacobian estimates of the dynamics of the \textit{original} system. For this purpose, it has been shown that it is possible to obtain an estimate of the Jacobian of the original dynamical system (i.e., not the echo state space) using ESNs \cite{banerjee2019esn}. However, as \citet{banerjee2019esn} noted, it is necessary to invert the non-square ESN output matrix in order to obtain these Jacobians, thus they are non-uniquely determined, and are used in the paper "only for causality estimation purposes". Nevertheless, we computed ESN-estimated Jacobians on the 1\% observation noise test data. We used a large (7,200 time-step) lead-up and ensured coherence to the trajectory by feeding in the true state as input. The Frobenius norm error over five random seeds was $85.8 \pm 3.1$, much larger than the reported errors of $3.3 \pm 0.2$ and $8.7 \pm 0.3$ for the JacobianODEs and NeuralODEs respectively. Thus, similar to the NeuralODE, while ESNs can achieve comparable prediction on chaotic systems, they do so without enabling an accurate reconstruction of the Jacobian of the original dynamical system.

\subsection{Analyzing the dimensionality of Jacobian manifolds}
\label{supp:dimensionality}
RNN activity is often confined to low-dimensional manifolds, potentially simplifying Jacobian estimation. To address the extent to which this influences our results, we performed the following analyses.

\textbf{Jacobian rank}: We first note that both the RNN hidden weight matrix and nearly all Jacobians had full rank (128; \textasciitilde 0.1\% had rank 127). To assess the \textit{effective} rank, we computed the participation ratio (PR), a singular value-based measure of intrinsic dimensionality. The RNN weight matrix had a PR of 123, and the Jacobians had a mean PR of \textasciitilde 106 (std 0.7).

\textbf{Intrinsic-data dimensionality}: Another possibility is that the data itself may be low-dimensional. Since the Jacobian is a function of the trajectory data, the Jacobian manifold’s intrinsic dimension is at most that of the trajectory manifold. To estimate the intrinsic dimensions, we pooled time points from a large sample of trajectories. For RNN state vectors, we computed PR directly; for Jacobians, we flattened them into $n^2$-dimensional vectors before computing PR. We found the RNN trajectory PR to be \textasciitilde 13.5, and the Jacobian PR to be \textasciitilde 6. Given the 128D extrinsic space, this confirms both RNN states and Jacobians lie on lower-dimensional manifolds.

Notably, prior work estimated the dimensionality of prefrontal cortex delay activity at 24 using \textasciitilde 4000 neurons, falling to \textasciitilde 6 when subsampled to 100 neurons \cite{rigotti2013mixed}. These values were interpreted as high-dimensional. Thus, the dimensionalities we observe are consistent with biological systems performing complex tasks.

Furthermore, in the 64-dimensional Lorenz 96 system, the trajectory PR was \textasciitilde 47.5, and the Jacobians had a PR of \textasciitilde 50, which we believe to be high-dimensional enough to demonstrate the effectiveness of our method on complex systems.

We note that in larger systems, approximating the Jacobian as low rank may be more efficient. Rather than computing each of the outputs directly, we can constrain the Jacobian to be low-rank by designing the output as a low-rank combination of vectors (e.g., it learns to output a low-rank singular value decomposition).

% \subsection{Time series visualizations}
% *****

%as the RNN derivative includes a negative residual term (appendix \ref{supp:task-trained-rnn-details}).

%\begin{itemize}
%\item NeuralODE f-estimates
%\item NeuralODE on noisy vs noiseless data
%\end{itemize}

\section{Experimental details}
\label{supp:experimental-details}

\subsection{Dynamical systems data}
\label{supp:dyn-sys-data}
We used the following dynamical systems for the testing and validation of Jacobian estimation.

\insettitle{Van der Pol oscillator} We implement the classic Van der Pol oscillator \cite{vanderpol1926osc}. The system is governed by the following equations

\[
\begin{aligned}
\dot{x} \ \ &=\ \ y \\
\dot{y} \ \ &= \ \ \mu(1 - x^2)y - x
\end{aligned}
\]
We pick $\mu = 2$ in our implementation.

\insettitle{Lorenz} We implement the Lorenz system introduced by \citet{lorenz1963flow} as

\[
\begin{aligned}
\dot{x} \ \ &= \ \sigma(y - x) \\
\dot{y} \ \ &= x(\rho - z) - y \\
\dot{z} \ \ &= xy - \beta z
\end{aligned}
\]

with the typical choices for parameters ($\sigma=10, \rho=28, \beta=8/3$).

\insettitle{Lorenz 96} We implement the Lorenz 96 system introduced by \citet{lorenz1996pred} and defined by
\[
\dot{x_i} = (x_{(i+1) \pmod{N}} - x_{(i-2) \pmod{N}})x_{(i-1)\pmod{N}} - x_{i \pmod{N}} +F
\]
with $F=8$ and $N\in\{12, 32, 64\}.$

\insettitle{Simulation} We use the \texttt{dysts} package to simulate all dynamical systems \cite{gilpin2021chaos, gilpin2023modelscale}. The characteristic timescale of their Fourier spectrum $\tau$ is selected and the systems are sampled with respect to $\tau$. For all systems, the training data consisted of 26 trajectories of 12 periods, sampled at 100 time steps per $\tau$. The validation data consisted of 6 trajectories of 12 periods sampled at 100 time steps per $\tau$. The test data consisted of 8 trajectories of 12 periods sampled at 100 time steps per $\tau$. Trajectories were initialized using a random normal distribution with standard deviation 0.2. The simulation algorithm used was Radau. Batches were constructed by moving a sliding window along the signal. The sequence length was selected such that the generated predictions would generate 10 novel time points (i.e., 11 time steps for the NeuralODE, and 25 time steps for the JacobianODE, due to the 15 time steps used to estimate the initial time derivative $\mathbf{f}$). 

\insettitle{Noise} We define $P\%$ observation noise with $P = 100p$ in the following way. Let $A_{\text{signal}} = \mathbb{E}\left[ \left\Vert x(t)\right\Vert_2 ^2\right]$ be the expected squared norm of the signal with $\mathbf{x}(t) \in\mathbb{R}^n$. Then consider a noise signal $\mathbf{\eta}(t) \in \mathbb{R}^n$ where each component $\eta_i(t) \sim \mathcal{N}(0, \frac{1}{\sqrt{n}}p\sqrt{A_{\text{signal}}})$. Then
\[
\mathbb{E}\left[\left\Vert \mathbf{\eta}(t)\right\Vert_2^2\right] = \mathbb{E}\left[\sum_{i=1}^n\mathbf{\eta}_i^2(t)\right] = \sum_{i=1}^n \mathbb{E}\left[\mathbf{\eta}_i^2(t)\right] = \sum_{i=1}^n \frac{1}{n}p^2A_{\text{signal}} = p^2A_{\text{signal}}
\]
and thus the noise percent is
\[
\sqrt{\frac{\mathbb{E}\left[\left\Vert \mathbf{\eta}(t)\right\Vert_2^2\right]}{\mathbb{E}\left[ \left\Vert x(t)\right\Vert_2 ^2\right]}} = \sqrt{\frac{p^2 A_{\text{signal}}}{A_{\text{signal}}}} = p
\]

\subsection{Task-trained RNN}
\label{supp:task-trained-rnn-details}

The task used to train the RNN was exactly as defined in Section \ref{sec:wmtask-control-analysis}. The hidden dimensionality of the RNN was 128, and the input dimensionality was 10, where the first four dimensions represented the one-hot encoded "upper" color, the second four dimensions represented the one-hot encoded "lower" color, and the last two dimensions represented the one-hot encoded cue. The RNN used for the task had hidden dynamics defined by

\[
\begin{aligned}
\tau\dot{\mathbf{h}}(t) &= -\mathbf{h} + \mathbf{W}_{hh}\sigma( \mathbf{h}(t))+ \mathbf{W}_{hi}\mathbf{u}(t) + \mathbf{b} \\
\mathbf{o}(t) &= \mathbf{W}_{oh} \mathbf{h}(t)
\end{aligned}\]

with $\tau = 50$ ms, which for the purposes of training was discretized with Euler integration with a time step of $\Delta t = 20$ ms. $\mathbf{W}_{hh}$ is the 128x128 dimensional matrix that defines the internal dynamics, $\mathbf{W}_{hi}$ is the 128 x 10 dimensional matrix that maps the input into the hidden state,  $\mathbf{W}_{oh}$ is the 4x128 dimensional output matrix that maps the hidden state to a four-dimensional output $\mathbf{o}(t)$, and $\mathbf{b}$ is a static bias term. $\sigma$ was taken to be an exponential linear unit activation with $\alpha = 1$. The RNN hidden state was split into two "areas" each with 64 dimensions. The input matrix $\mathbf{W}_{hi}$ was masked during training so that inputs could only flow into the first 64 dimensions -- the "visual" area. The same procedure was performed for the output matrix $\mathbf{W}_{oh}$, except the mask was such that outputs could stem only from the second group of 64 dimensions -- the 
"cognitive" area. The within-area subblocks of the matrix $\mathbf{W}_{hh}$ were first initialized such that the real part of the eigenvalues were randomly distributed on the interval $[-0.1, 0]$ and the imaginary part of the eigenvalues were randomly distributed on the interval $[0, 2\pi]$. The eigenvectors were random orthonormal matrices. We then computed the matrix exponential of this matrix. The across-area weights were first initialized to be random normal, then divided by the 2-norm of the resulting matrix and multiplied by 0.05. The input and output matrices were initialized as random normal and then scaled by the 2-norm of the resulting matrix. The static bias $\mathbf{b}$ was initialized at $\mathbf{0}$. After initialization, all weights could be altered unimpeded (except for the masks). Notably, the inputs $\mathbf{u}(t)$ were present only during the stimulus and cue presentation epochs -- otherwise the network evolved autonomously. The loss was computed via cross entropy loss on the RNN outputs during the response period (the final 250 ms of the trial).

For the training data, we generated 4096 random trials, and used 80\% for training and the remainder for validation. The batch size used was 32. Training was performed for 40 epochs. The learning rate was 0.0005. For use with the Jacobian estimation models, data was batched and used for training exactly as was done with the other dynamical systems data (see Appendix \ref{supp:dyn-sys-data}). Observation noise was also computed in the same way.

\subsection{NeuralODE details}
\label{supp:neuralode-details}

NeuralODE models directly estimate the time derivative $\mathbf{f}$ with a neural-network parameterized function $\hat{\mathbf{f}}^\theta$. Then the Jacobians can be computed as $\hat{\mathbf{J}} = \frac{\partial}{\partial\mathbf{x}}\hat{\mathbf{f}}^{\theta}$.

The NeuralODEs were implemented as described in Section \ref{sec:jac-est} and ODE integration was done exactly as for the JacobianODE using the \texttt{torchdiffeq} package with the RK4 method \cite{chen2018neuralode}. To regularize the NeuralODE we implemented a Frobenius norm penalty on the estimated Jacobians, i.e.
\[
\mathcal{L_{\text{jac}}} = \lambda_{\text{jac}} \langle \left\Vert \hat{\mathbf{J}}(\mathbf{x}(t))\right\Vert_F \rangle
\]
where $\hat{\mathbf{J}}$ is the estimated Jacobian computed via automatic differentation and $\lambda_{\text{jac}}$ is a hyperparameter that controls the relative weighting of the Jacobian penalty \cite{hoffman2019jacpen, wikner2024reservoirstab, schneider2025timeseries-att}. As mentioned in the main text, this penalty prevents the model from learning unnecessarily large eigenvalues and encourages better Jacobian estimation.

\subsection{Weighted linear Jacobian details}
\label{supp:weighted-linear-jacobian}
We implemented a baseline Jacobian estimation method using weighted linear regression models as described in \citet{deyle2016tracking}. Given a reference point $\mathbf{x}(t^*)$ at which the (discrete) Jacobian will be computed, all other points are weighted according to
\[
w_k = \exp\frac{-\theta \left\Vert \mathbf{x}(t_k) - \mathbf{x}(t^*)\right\Vert}{\bar{d}}
\]
where
\[
\bar{d} = \sum_{i=1}^T \left\Vert \mathbf{x}(t_i) - \mathbf{x}(t^*) \right\Vert
\]
is the average distance from $\mathbf{x}(t^*)$ to all other points. We then perform a linear regression using the weighted points (and a bias term), the result of which is an estimate of the discrete Jacobian at $\mathbf{x}(t^*)$, which can be converted to continuous time by subtracting the identity matrix and dividing by the sampling time step (i.e., $\hat{\mathbf{J}} = \frac{\hat{\mathbf{J}}_{\text{discrete}} - \mathbf{I}}{\Delta t}$, where $\hat{\mathbf{J}}_{\text{discrete}}$ is the discrete Jacobian). The parameter $\theta$ tunes how strongly the regression is weighted towards local points. To pick $\theta$, we sweep over values range from 0 to 10, and pick the value that yields the best one-step prediction according to
\[
\mathbf{x}(t^* + 2\Delta t) = \mathbf{x}(t^* + \Delta t) +e^{\hat{\mathbf{J}}\Delta t}(\mathbf{x}(t ^* + \Delta t) - \mathbf{x}(t^*))
\]
where $\hat{\mathbf{J}}$ is the estimated Jacobian. This form of prediction has been previously been used to learn Jacobians in machine learning settings \cite{latremoliere2022jacest}. To test the method, we pick $\theta$ based on data with observation noise at a particular noise level, then add in the denoised data to the data pool in order to compute regressions and estimate Jacobians at the true points.

\subsection{Model details}
\label{supp:model-details}
All models were implemented as four-layer MLPS, with the four layers having sizes of 256, 1024, 2048, and 2048 respectively. All models used a sigmoid linear unit activation. JacobianODE models output to the dimension $n^2$ which was then reshaped into the matrix of the appropriate dimension. NeuralODEs output to the dimension $n$.

\subsection{Lyapunov spectrum computation}
To compute the Lyapunov spectrum, we employ a QR based algorithm \cite{dieci1997lyap, christiansen1997lyap}. We discretize the Jacobians using the matrix exponential (i.e., $\hat{\mathbf{J}}_{\text{discrete}} = e^{\hat{\mathbf{J}\Delta t}}$) and then propagate a bundle of small vectors through the Jacobians, using QR to ensure the perturbations remain bounded.

\subsection{Iterative Linear Quadratic Regulator (ILQR)}

We implement the standard algorithm for ILQR, the details of which can be found in \citet{li2004ilqr} and \citet{tassa2012ilqr}. In brief, the ILQR algorithm linearizes the system dynamics around a nominal trajectory using the Jacobian, and then iteratively optimizes the control sequence using forward and backward passes to minimize the total control cost. The state cost matrix $\mathbf{Q}$ was a diagonal matrix with 1.0 along the diagonal. The final state cost matrix $\mathbf{Q}_f$ was a diagonal matrix with 1.0 along the diagonal. The control cost matrix $\mathbf{R}$ was a diagonal matrix with 0.01 along the diagonal.  The control matrix was a 128 $\times$ 128 matrix in which the 64 $\times$ 64 block corresponding to the first 64 neurons (the "visual" area) was the 64-dimensional identity matrix. The control algorithm was seeded with only the initial state of the test trajectory with 5\% noise. The control sequence was initialized random normal with standard deviation 0.001 and mean 0. The ILQR algorithm was run for a max of 100 iterations. The regularization was initialized at 1.0, with a minimum of  $1 \times 10^{-6}$ and a maximum of $1 \times 10^{10}$. $\Delta_0$ was set to 2, as in \citet{tassa2012ilqr}. If the backward pass failed 20 times in a row, the optimization was stopped. The list of values for the line search parameter $\alpha$ was $1.1^{-k^2}$ for $k \in {0,...,9}$ (see \citet{tassa2012ilqr}). The linear model used for the linear baseline was computed via linear regression.

\subsection{Training details}
\label{supp:training-details}

All models were implemented in PyTorch. The batch size used was 16. Gradients were accumulated for 4 batches. Training epochs were limited to 500 shuffled batches. Validation epochs were limited to 100 randomly chosen batches. Testing used all testing data. Training was run for a maximum of 1000 epochs, 3 hours, or until the early stopping was activated (see Appendix \ref{supp:early-stopping}), whichever came first.

\subsubsection{Generalized Teacher Forcing}
\label{supp:teacher-forcing}
The Jacobian can be best learned when training predictions are generated recursively (i.e., replacing $\mathbf{x}(t)$ by $\hat{\mathbf{x}}(t)$). However, in chaotic systems, and/or systems with measurement noise (as considered here), this could lead to catastrophic divergence of the predicted trajectory from the true trajectory during training. We therefore employ Generalized Teacher Forcing when training all models \cite{hess2023generalizedtf}. Generalized Teacher Forcing prevents catastrophic divergence by forcing the generated predictions along the line from the prediction to the true state. Specifically, for a given predicted state $\hat{\mathbf{x}}(t)$ and true state $\mathbf{x}(t)$, the teacher forced state is
\[
\tilde{\mathbf{x}}(t) = (1 - \alpha) \hat{\mathbf{x}}(t) +  \alpha\mathbf{x}(t)
\]
with $\alpha \in [0,1]$. This effectively forces the predictions along a line from the prediction to the true state, by an amount with proportion $\alpha$. $\alpha = 1$ corresponds to fully-forced prediction (i.e., one-step prediction) and $\alpha=0$ corresponds to completely unforced prediction (i.e., autonomous prediction). \citet{hess2023generalizedtf} suggested that a good estimate of $\alpha$ is
\begin{equation}
\alpha = \max\left(\max_p \left [ 1 - \frac{1}{\Vert\mathcal{G}(\mathbf{J}_{T:2}^{(p)})\Vert}\right], 0\right)
\label{eq:alpha-gtf}
\end{equation}
where $\mathbf{J}_{T:2}^{(p)}$ are the Jacobians of the modeled dynamics computed at data-constrained states, $p$ indicates the batch or sequence index, and

\[
\Vert\mathcal{G}(\mathbf{J}_{T:2}^{(p)})\Vert = \left\Vert \left(\prod_{k=0}^{T-2} \mathbf{J}_{T-k}\right)^{\frac{1}{T-1}}\right\Vert
\]
effectively computes the discrete maximum Lyapunov exponent. In our implementation, we compute $\Vert\mathcal{G}(\mathbf{J}_{T:2}^{(p)})\Vert $ using a QR-decomposition-based Lyapunov computation algorithm \cite{dieci1997lyap}. As the Jacobian of the dynamics is necessary to compute this quantity, the JacobianODEs enjoy an advantage over other models in that the Jacobians are directly output by the model, and do not have to be computed via differentiating the model itself.

We furthemore employ a slightly modified version of the suggested annealing process in \citet{hess2023generalizedtf}, which sets $\alpha_0 = 1$ and updates $\alpha_n$ as 
\[
\alpha_n = \gamma \alpha_{n-1} + (1 - \gamma)\alpha
\]
where $\alpha$ is computed according to equation \ref{eq:alpha-gtf}. Following the suggested hyperparameters, we set $\gamma = 0.999$ and update $\alpha_n$ every 5 batches. Once the teacher forced state $\tilde{\mathbf{x}}(t)$ is computed, it can simply replace $\mathbf{x}(t)$ in equation \ref{eq:pred-gen} to generate predictions. 

\subsubsection{Loop closure loss}
\label{supp:loop-training}
We implemented a loop closure loss as discussed in Section \ref{sec:lossfuncs} and Appendix \ref{supp:loop-closure}. For each loop, we used 20 randomly chosen points from the batch. For each batch, we constructed the same number of loops as there were batches. Path integrals were discretized in 20 steps and computed using the trapezoid method from \texttt{torchquad} \cite{gomez2021torchquad}. 

\subsubsection{Validation loss}
All models were validated on 10 time-step prediction task with teacher forcing parameter $\alpha=0$  (i.e., autonomous prediction).

\subsubsection{Learning rate scheduling}
For all models, the learning rate was annealed in accordance with teacher forcing annealing. Given an initial and final learning rates $\eta_i$ and $\eta_f$ we compute the effective learning rate as
\[
\eta = \eta_f + \sigma(\alpha_n) (\eta_i - \eta_f)
\]
where $\alpha_n$ is the current value of the teacher forcing parameter and 
\[
\sigma(\alpha_n) = \frac{\alpha_n}{\alpha_n + (1 - \alpha_n)e^{-k\alpha_n}}
\]
$\sigma(\alpha_n)$ is a scaling function with $\sigma(1) = 1$ and $\sigma(0) =0$ and for which the shape of the scaling is controlled by the parameter $k$. For positive values of $k$, the scaling is super-linear, and for negative values of $k$ it is sub-linear. We use $k=1$, ensuring that the learning rate does not decrease too quickly at the start of learning. We set $\eta_i = 10^{-4}$ and $\eta_f = 10^{-6}$ for all models.

\subsubsection{Optimizer and weight decay}
All models were trained with PyTorch's AdamW optimizer with the learning rate as descried above, and weight decay parameter $10^{-4}$. All other parameters were default ($\beta_1 = 0.9, \beta_2 = 0.999, \epsilon = 10^{-8}$). We also used gradient norm clipping, with a clipping value of 1.0.

\subsubsection{Early stopping}
\label{supp:early-stopping}
For all models, we implemented an early stopping scheme that halted the training if the validation loss improved by less than 1\% for two epochs in a row.

\subsubsection{Added noise during learning}

For the models trained on the task-trained RNN dynamics, we added 5\% Gaussian i.i.d. noise (defined relative to the norm of the training data with observation noise already added). Noise was sampled for each batch and added prior to the trajectory generation step of the learning process. Additional noise was not added for the loop closure computation.

\subsubsection{Hyperparameter selection}
\label{supp:hyperparameter-selection}
For the JacobianODEs, the primary hyperparameter to select is the loop closure loss $\lambda_{\text{loop}}$. To select this hyperparameter, we trained JacobianODE models with $\lambda_{\text{loop}} \in [0, 10^{-6}, 10^{-5}, 10^{-4}, 10^{-3}, 10^{-2}, 10^{-1}, 1, 10]$. For each run, the epoch with the lowest trajectory validation loss ($\mathcal{L}_{\text{traj}}$) is kept. Then, for this model, we compute the one-step prediction error on validation data, the validation loop closure loss ($\mathcal{L}_{\text{loop}}$), and the percentage of Jacobian eigenvalues on all validation data that have a decay rate faster than the sampling rate $\frac{1}{\Delta t}$. We exclude any models that meet any of the following criteria:
\begin{enumerate}
\item \textbf{One-step prediction error greater than the persistence baseline}. The persistence baseline is computed as the mean error between each time step $k\Delta t$ and the subsequent time step $(k+1)\Delta t$ across the dataset, and constitutes a sanity check for whether a model is capturing meaningful information about the dynamics.
\item \textbf{Loop closure loss greater than $\sqrt{n}$}, where $n$ is the system dimension (see Appendix \ref{supp:lc-derivation} for the derivation of this bound). As discussed in the main text, we are interested in Jacobians that not only solve the trajectory prediction problem, but that also are constructed so that the rows of the matrix are approximately conservative vector fields.
\item \textbf{More than 0.1\% of the Jacobian eigenvalues have a decay rate faster than the sampling rate $\frac{1}{\Delta t}$}. Since large negative eigenvalues do not impact trajectory prediction, the models may erroneously learn Jacobians with large negative eigenvalues. If the decay rate of these eigenvalues is faster than the sampling rate, we can infer that the eigenvalues are not aligned with the observed data.
\end{enumerate}

If none of the models that meet criterion (2) meet criterion (1), we discount criterion (2), as this suggests that a loop closure loss below $\sqrt{n}$ bound is too strict to obtain good prediction on this system. Additionally, if none of the models that meet criterion (3) meet criterion (1), we discount criterion (1), as this suggests that noise is very high in the data, which leads to both high one-step prediction error, and large negative eigenvalues to compensate for the perturbations introduced by the noise. Of the remaining models, we select the one with the lowest trajectory validation loss $\mathcal{L}_{\text{traj}}$.

For the NeuralODEs, we needed to select the hyperparameter $\lambda_{\text{jac}}$, which regularized the mean frobenius norm of the Jacobians computed through automatic differentiation.  Again, to select this hyperparameter, we trained JacobianODE models with $\lambda_{\text{jac}} \in [0, 10^{-6}, 10^{-5}, 10^{-4}, 10^{-3}, 10^{-2}, 10^{-1}, 1, 10]$.  We followed exactly the above procedure with the exception of criterion (2), which was deemed unnecessary, as computing the Jacobians implicitly via a gradient of the model (using automatic differentiation) ensures that the rows of the matrix are conservative.

All other hyperparameters (model size, learning rate, length of initial trajectory, number of discretization steps, etc.) were fixed for all systems. Given the wide range of systems and behaviors and dimensionalities that the JacobianODEs are capable of capturing, this indicates that the method is robust given a reasonable choice of these hyperparameters. 

\subsubsection{Model hyperparameters and training details}

Below are presented the details of all Jacobian estimation models considered in the main paper.

\begin{table}[ht]
  \centering
  \caption{Hyperparameters used and model details for each system and noise level. Training time is reported in seconds.}
  \label{tab:hyperparameters}
  \scriptsize
  \setlength{\tabcolsep}{4pt}
  \begin{adjustbox}{width=\textwidth}
    \begin{tabular}{llllllllllllll}
      \toprule
      System & Noise & Model & Loop closure weight & Jacobian penalty & Training time (s) & Final epoch & Learning rate & Min learning rate & Weight decay & Learnable parameters \\
      \midrule
      VanDerPol (2 dim) & 1\% & JacobianODE & 0.0010 & 0 & 882.139 & 15 & 0.0001 & 1.00e-06 & 0.0001 & 6.568e+06 \\
       & 1\% & NeuralODE & 0 & 1.00e-06 & 949.217 & 21 & 0.0001 & 1.00e-06 & 0.0001 & 6.564e+06 \\
       & 5\% & JacobianODE & 0.0001 & 0 & 692.980 & 11 & 0.0001 & 1.00e-06 & 0.0001 & 6.568e+06 \\
       & 5\% & NeuralODE & 0 & 0 & 249.451 & 6 & 0.0001 & 1.00e-06 & 0.0001 & 6.564e+06 \\
       & 10\% & JacobianODE & 0.010 & 0 & 678.794 & 10 & 0.0001 & 1.00e-06 & 0.0001 & 6.568e+06 \\
       & 10\% & NeuralODE & 0 & 0.0010 & 713.221 & 16 & 0.0001 & 1.00e-06 & 0.0001 & 6.564e+06 \\
      \midrule
      Lorenz (3 dim) & 1\% & JacobianODE & 0.0010 & 0 & 972.268 & 16 & 0.0001 & 1.00e-06 & 0.0001 & 6.578e+06 \\
       & 1\% & NeuralODE & 0 & 0.0010 & 889.819 & 18 & 0.0001 & 1.00e-06 & 0.0001 & 6.566e+06 \\
       & 5\% & JacobianODE & 0.010 & 0 & 1.147e+03 & 18 & 0.0001 & 1.00e-06 & 0.0001 & 6.578e+06 \\
       & 5\% & NeuralODE & 0 & 0.0010 & 680.651 & 15 & 0.0001 & 1.00e-06 & 0.0001 & 6.566e+06 \\
       & 10\% & JacobianODE & 0.010 & 0 & 388.346 & 6 & 0.0001 & 1.00e-06 & 0.0001 & 6.578e+06 \\
       & 10\% & NeuralODE & 0 & 0.010 & 421.488 & 8 & 0.0001 & 1.00e-06 & 0.0001 & 6.566e+06 \\
      \midrule
      Lorenz 96 (12 dim) & 1\% & JacobianODE & 0.0010 & 0 & 1.817e+03 & 31 & 0.0001 & 1.00e-06 & 0.0001 & 6.857e+06 \\
       & 1\% & NeuralODE & 0 & 1.00e-06 & 1.892e+03 & 32 & 0.0001 & 1.00e-06 & 0.0001 & 6.587e+06 \\
       & 5\% & JacobianODE & 0.0010 & 0 & 716.408 & 12 & 0.0001 & 1.00e-06 & 0.0001 & 6.857e+06 \\
       & 5\% & NeuralODE & 0 & 0.0010 & 1.147e+03 & 19 & 0.0001 & 1.00e-06 & 0.0001 & 6.587e+06 \\
       & 10\% & JacobianODE & 0.010 & 0 & 1.213e+03 & 18 & 0.0001 & 1.00e-06 & 0.0001 & 6.857e+06 \\
       & 10\% & NeuralODE & 0 & 0.0001 & 851.803 & 14 & 0.0001 & 1.00e-06 & 0.0001 & 6.587e+06 \\
      \midrule
      Lorenz 96 (32 dim) & 1\% & JacobianODE & 0.010 & 0 & 5.160e+03 & 85 & 0.0001 & 1.00e-06 & 0.0001 & 8.665e+06 \\
       & 1\% & NeuralODE & 0 & 0.0010 & 4.204e+03 & 43 & 0.0001 & 1.00e-06 & 0.0001 & 6.633e+06 \\
       & 5\% & JacobianODE & 0.0010 & 0 & 1.341e+03 & 22 & 0.0001 & 1.00e-06 & 0.0001 & 8.665e+06 \\
       & 5\% & NeuralODE & 0 & 0.0010 & 3.855e+03 & 39 & 0.0001 & 1.00e-06 & 0.0001 & 6.633e+06 \\
       & 10\% & JacobianODE & 0.0001 & 0 & 868.262 & 14 & 0.0001 & 1.00e-06 & 0.0001 & 8.665e+06 \\
       & 10\% & NeuralODE & 0 & 0.0010 & 2.695e+03 & 28 & 0.0001 & 1.00e-06 & 0.0001 & 6.633e+06 \\
      \midrule
      Lorenz 96 (64 dim) & 1\% & JacobianODE & 0.0010 & 0 & 2.749e+03 & 40 & 0.0001 & 1.00e-06 & 0.0001 & 1.497e+07 \\
       & 1\% & NeuralODE & 0 & 0.010 & 4.801e+03 & 30 & 0.0001 & 1.00e-06 & 0.0001 & 6.706e+06 \\
       & 5\% & JacobianODE & 0.0001 & 0 & 1.748e+03 & 27 & 0.0001 & 1.00e-06 & 0.0001 & 1.497e+07 \\
       & 5\% & NeuralODE & 0 & 0.010 & 4.879e+03 & 30 & 0.0001 & 1.00e-06 & 0.0001 & 6.706e+06 \\
       & 10\% & JacobianODE & 0.0010 & 0 & 1.701e+03 & 25 & 0.0001 & 1.00e-06 & 0.0001 & 1.497e+07 \\
       & 10\% & NeuralODE & 0 & 0.010 & 4.237e+03 & 26 & 0.0001 & 1.00e-06 & 0.0001 & 6.706e+06 \\
      \midrule
      Task-trained RNN & 1\% & JacobianODE & 0.0001 & 0 & 2.496e+03 & 32 & 0.0001 & 1.00e-06 & 0.0001 & 4.016e+07 \\
       & 1\% & NeuralODE & 0 & 0 & 9.777e+03 & 38 & 0.0001 & 1.00e-06 & 0.0001 & 6.854e+06 \\
       & 5\% & JacobianODE & 0.0001 & 0 & 2.352e+03 & 29 & 0.0001 & 1.00e-06 & 0.0001 & 4.016e+07 \\
       & 5\% & NeuralODE & 0 & 1.00e-05 & 7.770e+03 & 27 & 0.0001 & 1.00e-06 & 0.0001 & 6.854e+06 \\
       & 10\% & JacobianODE & 0.010 & 0 & 2.172e+03 & 27 & 0.0001 & 1.00e-06 & 0.0001 & 4.016e+07 \\
       & 10\% & NeuralODE & 0 & 1.00e-05 & 5.260e+03 & 18 & 0.0001 & 1.00e-06 & 0.0001 & 6.854e+06 \\
      \midrule
      \bottomrule
    \end{tabular}
  \end{adjustbox}
\end{table}

\subsection{Information about computing resources and efficiency}
\label{supp:compute-efficiency}

All models were able to be trained on a single H100 GPU, with 80 GB of memory.

\insettitle{Jacobian inference times} Jacobian inference times for the JacobianODE and NeuralODE models are discussed in Appendix \ref{supp:computation-scaling}. As discussed, models were implemented with four hidden layers of the same size, and tested on 100 batches of the 128-dimensional task-trained RNN data, with each batch consisting of 16 sequences of length 25. Timings were repeated ten times for each model (see Figure \ref{suppfig:computation-scaling} for details).

\insettitle{Training time} Total training times for each of the chosen models are presented in Table \ref{tab:hyperparameters}. Furthermore, we include the training time (including backward pass) for 100 batches (with 16 sequences per batch), using 10 time-step prediction, in Table \ref{tab:trajectory_times}. 

\begin{table}[ht]
  \centering
  \caption{Trajectory training time (seconds) for each system and noise level. Training used 100 batches, with 16 sequences per batch, as well as 10 time-step prediction.}
  \label{tab:trajectory_times}
  \scriptsize
  \setlength{\tabcolsep}{4pt}
  \begin{adjustbox}{width=1.0\textwidth}
    \begin{tabular}{lcccccc}
      \toprule
      Model & Lorenz (3 dim) & VanDerPol (2 dim) & Lorenz 96 (12 dim) & Lorenz 96 (32 dim) & Lorenz 96 (64 dim) & Task-trained RNN \\
      \midrule
      \multicolumn{7}{c}{1\% noise} \\
      \midrule
      JacobianODE & 15.772 & 13.737 & 11.604 & 16.469 & 16.214 & 23.855 \\
      NeuralODE & 8.215 & 8.403 & 11.852 & 12.075 & 18.192 & 14.955 \\
      \midrule
      \multicolumn{7}{c}{5\% noise} \\
      \midrule
      JacobianODE & 12.422 & 16.772 & 10.655 & 13.121 & 12.949 & 23.197 \\
      NeuralODE & 6.191 & 5.841 & 13.501 & 8.021 & 21.307 & 30.209 \\
      \midrule
      \multicolumn{7}{c}{10\% noise} \\
      \midrule
      JacobianODE & 12.590 & 15.719 & 18.447 & 10.506 & 18.900 & 22.482 \\
      NeuralODE & 11.083 & 9.875 & 15.269 & 8.008 & 17.826 & 33.074 \\
      \midrule
      \bottomrule
    \end{tabular}
  \end{adjustbox}
\end{table}

\subsection{Statistical details}
All statistics were computed using \texttt{scipy}. For the comparison between JacobianODE and NeuralODE trajectory and Jacobian predictions, as well as the comparison of Gramian traces and minimum eigenvalues, we used a two-sample t-test. For the comparison of ILQR control accuracies and errors, we used a Wilcoxon signed-rank test.

\subsection{Derivation of loop closure loss bound}
\label{supp:lc-derivation}

We consider the loop closure loss as defined in \ref{sec:lossfuncs}. We are interested in estimating a bound on the error
\[
\mathbb{E}\left[\frac{1}{n}\left \Vert \int_{\mathcal{C}^{(l)}_{\text{loop}}} \mathbf{J}ds\right\Vert^2_2 \right]
\]

where $n$ is the system dimension. While in theory this quantity should be 0, in practice due to numerical estimation error, it will not be. First recall, that

\[
\int_{\mathcal{C}^{(l)}_{\text{loop}}} \mathbf{J}ds = \sum_{i =1}^L \int_{\mathbf{x}(t_{i \pmod{L}})}^{\mathbf{x}(t_{(i + 1) \pmod{L}})}  \mathbf{J}(\mathbf{c}(r))\mathbf{c}'(r)dr
\]

where $L$ is the number of loop points and $\mathbf{c}$ is a line from $\mathbf{x}(t_{i \pmod{L}})$ to $\mathbf{x}(t_{(i + 1) \pmod{L}}) $. We assume that

\[
\mathbb{E}\left[\left\Vert\int_{\mathbf{x}(t_{i \pmod{L}})}^{\mathbf{x}(t_{(i + 1) \pmod{L}})}  \mathbf{J}(\mathbf{c}(r))\mathbf{c}'(r)dr\right\Vert_2^2 \right] = \mathbb{E}\left[\left\Vert\int_{\mathbf{x}(t_{j \pmod{L}})}^{\mathbf{x}(t_{(j + 1) \pmod{L}})}  \mathbf{J}(\mathbf{c}(r))\mathbf{c}'(r)dr\right\Vert_2^2 \right]
\]
$\forall i,j\in[1,..,L]$, which is justified as the numerical error accrued will likely be similar along different lines for the same system. Thus

\[
\begin{aligned}
\mathbb{E}\left[\frac{1}{n}\left \Vert \int_{\mathcal{C}^{(l)}_{\text{loop}}} \mathbf{J}ds\right\Vert^2_2 \right] &= \mathbb{E}\left[\frac{1}{n}\left \Vert\sum_{i =1}^L \int_{\mathbf{x}(t_{i \pmod{L}})}^{\mathbf{x}(t_{(i + 1) \pmod{L}})}  \mathbf{J}(\mathbf{c}(r))\mathbf{c}'(r)dr\right\Vert_2^2\right] \\
&\leq \frac{1}{n}  \mathbb{E}\left[\sum_{i =1}^L\left \Vert \int_{\mathbf{x}(t_{i \pmod{L}})}^{\mathbf{x}(t_{(i + 1) \pmod{L}})}  \mathbf{J}(\mathbf{c}(r))\mathbf{c}'(r)dr\right\Vert_2^2\right] \\
&= \frac{1}{n}\sum_{i =1}^L \mathbb{E}\left[ \left \Vert \int_{\mathbf{x}(t_{i \pmod{L}})}^{\mathbf{x}(t_{(i + 1) \pmod{L}})}  \mathbf{J}(\mathbf{c}(r))\mathbf{c}'(r)dr\right\Vert_2^2\right] \\
&= \frac{1}{n}\sum_{i =1}^L \mathbb{E}\left[ \left \Vert \int_{\mathbf{x}(t_0)}^{\mathbf{x}(t_1)}  \mathbf{J}(\mathbf{c}(r))\mathbf{c}'(r)dr\right\Vert_2^2\right] \\
&= \frac{L}{n}\mathbb{E}\left[ \left \Vert \int_{\mathbf{x}(t_0)}^{\mathbf{x}(t_1)}  \mathbf{J}(\mathbf{c}(r))\mathbf{c}'(r)dr\right\Vert_2^2\right]
\end{aligned}
\]
where we have used the fact that the errors are assumed to be equivalent for each of the line segments comprising the overall loop path. Recall now that for the trapezoid integration rule, the error $E$ in integrating $\int_a^b f(x) dx$ can be computed as $E=-\frac{(b-a)^3}{12M^2}f''(u)$ for some $u \in [a, b]$. Thus the squared error is bounded as
\[
E^2 \leq \left\Vert\max_{u \in [a, b]}  \frac{(b-a)^3}{12M^2}f''(u) \right\Vert_2^2
\]
In our case, when path integrating lines, we are effectively integrating from $a=0$ to $b=1$. Furthermore, let $\mathbf{g}(r) = \mathbf{J}(\mathbf{c}(r))\mathbf{c}'(r)$, the integrand of our line integrations. We assume that $\mathbb{E}\left[ \left\Vert \max_{r}\mathbf{g}''(r)\right\Vert_2^2\right] \propto n \cdot \sqrt{n}$ where the first $n$ comes from the fact that the norm involves summing over the $n$ components of the vector $\mathbf{g}''(r)$ and the second $\sqrt{n}$ involves an assumption that loop closure loss will be more difficult to compute accurately in higher dimensions, though this will be more pronounced as dimensionality initially starts increasing. Thus $\mathbb{E}\left[ \left\Vert \max_{r}\mathbf{g}''(r)\right\Vert_2^2\right] = k n \cdot \sqrt{n}$ for some $k \in \mathbb{R}^+$. Now, continuing on,

\[
\begin{aligned}
\mathbb{E}\left[\frac{1}{n}\left \Vert \int_{\mathcal{C}^{(l)}_{\text{loop}}} \mathbf{J}ds\right\Vert^2_2 \right] &\leq \frac{L}{n}\mathbb{E}\left[ \left \Vert \int_{\mathbf{x}(t_0)}^{\mathbf{x}(t_1)}  \mathbf{J}(\mathbf{c}(r))\mathbf{c}'(r)dr\right\Vert_2^2\right] \\
&\leq \frac{L}{n}\mathbb{E}\left[ \left \Vert   \frac{1^3}{12M^2}  \max_r \mathbf{g}''(r)\right\Vert_2^2\right] \\
&= \frac{L}{12^2nM^4}\mathbb{E}\left[ \left \Vert  \max_r \mathbf{g}''(r)\right\Vert_2^2\right] \\
&= \frac{Lkn\cdot\sqrt{n}}{12^2nM^4} \\
&= \frac{Lk\sqrt{n}}{12^2 M^4}
\end{aligned}
\]

Finally, assuming that the number of discretization steps $M$ was chosen to be large enough such that $M^4 \geq \frac{1}{12^2} Lk$ we finally obtain
\[
\mathbb{E}\left[\frac{1}{n}\left \Vert \int_{\mathcal{C}^{(l)}_{\text{loop}}} \mathbf{J}ds\right\Vert^2_2 \right] \leq \sqrt{n}
\]

% \subsection{Data and code availability}

% The data for all analyzed systems can be downloaded \href{https://drive.google.com/file/d/1SNDWxlpIPB3OtOfR8pvr9pFBYxqixr8M/view?usp=sharing}{here}. Code to implement all models, training, and control theoretic analyses are included with the Supplementary Material submission.

%\section{Notes}
%\begin{itemize}

% \item %space of Jacobians can be no greater than $n$, reduced space, self-supervised loss encourages the restriction

%\end{itemize}

%%%%%%%%%%%%%%%%%%%%%%%%%%%%%%%%%%%%%%%%%%%%%%%%%%%%%%%%%%%%

\clearpage
\section*{NeurIPS Paper Checklist}

\begin{enumerate}

\item {\bf Claims}
    \item[] Question: Do the main claims made in the abstract and introduction accurately reflect the paper's contributions and scope?
    \item[] Answer: \answerYes{} % Replace by \answerYes{}, \answerNo{}, or \answerNA{}.
    \item[] Justification: %\justificationTODO{}
    The abstract and introduction claim the the paper contributes robust data-driven Jacobian estimation, a framework to characterize control between interacting subsystems, inference of control dynamics in trained recurrent neural networks, and accurate control of the trained network. All claims are substantiated in detail through a description of the presented method and analytical framework, as well as comprehensive testing. See Sections \ref{sec:jacobianode}, \ref{sec:jac-est}, and \ref{sec:wmtask-control-analysis}.

\item {\bf Limitations}
    \item[] Question: Does the paper discuss the limitations of the work performed by the authors?
    \item[] Answer: \answerYes{} % Replace by \answerYes{}, \answerNo{}, or \answerNA{}.
    \item[] Justification: The primary limitation of the present approach is that it was not tested on partially observed system dynamics, a limitation which is discussed in the Discussion (Section \ref{sec:discussion}). The method was tested on a wide range of datasets (Section \ref{sec:jac-est}) and computational efficiency is discussed in Appendix \ref{supp:compute-efficiency}. %\justificationTODO{}

\item {\bf Theory assumptions and proofs}
    \item[] Question: For each theoretical result, does the paper provide the full set of assumptions and a complete (and correct) proof?
    \item[] Answer: \answerYes{}
    %\answerTODO{} % Replace by \answerYes{}, \answerNo{}, or \answerNA{}.
    \item[] Justification: A complete and correct proof of the presented theoretical result, that the time derivative $\mathbf{f}$ can be represented in terms of its Jacobian, is provided in Appendix \ref{supp:jacparam}.%\justificationTODO{}

    \item {\bf Experimental result reproducibility}
    \item[] Question: Does the paper fully disclose all the information needed to reproduce the main experimental results of the paper to the extent that it affects the main claims and/or conclusions of the paper (regardless of whether the code and data are provided or not)?
    \item[] Answer: \answerYes{} % Replace by \answerYes{}, \answerNo{}, or \answerNA{}.
    \item[] Justification:
    The paper provides full information about the presented methods and experiments, with sufficient detail that all results could be reproduced.
    %\justificationTODO{}

\item {\bf Open access to data and code}
    \item[] Question: Does the paper provide open access to the data and code, with sufficient instructions to faithfully reproduce the main experimental results, as described in supplemental material?
    \item[] Answer: \answerYes{} % Replace by \answerYes{}, \answerNo{}, or \answerNA{}.
    \item[] Justification:
    As part of the submission, we provide an anonymized link to download the analyzed data, as well as all code implementing the models, baselines, control analyses, and data generation. If the paper is accepted, the full code will be open sourced.
    %\justificationTODO{}

\item {\bf Experimental setting/details}
    \item[] Question: Does the paper specify all the training and test details (e.g., data splits, hyperparameters, how they were chosen, type of optimizer, etc.) necessary to understand the results?
    \item[] Answer: \answerYes{} % Replace by \answerYes{}, \answerNo{}, or \answerNA{}.
    \item[] Justification:
    The paper specifies all training and test details, including data splits, hyperparameter selection, and training details such as optimizers.
    %\justificationTODO{}

\item {\bf Experiment statistical significance}
    \item[] Question: Does the paper report error bars suitably and correctly defined or other appropriate information about the statistical significance of the experiments?
    \item[] Answer: \answerYes{} 
    % Replace by \answerYes{}, \answerNo{}, or \answerNA{}.
    \item[] Justification: The paper includes error bars and statistical information. Additionally, the factors of variability that the errors are capturing are clearly stated.

\item {\bf Experiments compute resources}
    \item[] Question: For each experiment, does the paper provide sufficient information on the computer resources (type of compute workers, memory, time of execution) needed to reproduce the experiments?
    \item[] Answer: \answerYes{} % Replace by \answerYes{}, \answerNo{}, or \answerNA{}.
    \item[] Justification: We provide information about compute resources needed to reproduce the experiments in Appendix \ref{supp:compute-efficiency}.%\justificationTODO{}
    
\item {\bf Code of ethics}
    \item[] Question: Does the research conducted in the paper conform, in every respect, with the NeurIPS Code of Ethics \url{https://neurips.cc/public/EthicsGuidelines}?
    \item[] Answer: \answerYes{} % Replace by \answerYes{}, \answerNo{}, or \answerNA{}.
    \item[] Justification: Ethical considerations have been taken into account and the research conforms fully to the NeurIPS code of ethics. %\justificationTODO{}

\item {\bf Broader impacts}
    \item[] Question: Does the paper discuss both potential positive societal impacts and negative societal impacts of the work performed?
    \item[] Answer: \answerYes{} % Replace by \answerYes{}, \answerNo{}, or \answerNA{}.
    \item[] Justification: Yes, the paper discusses societal impacts, including potential applications to modeling biological dysfunction in disease. There are no expected negative impacts. %\justificationTODO{}
    
\item {\bf Safeguards}
    \item[] Question: Does the paper describe safeguards that have been put in place for responsible release of data or models that have a high risk for misuse (e.g., pretrained language models, image generators, or scraped datasets)?
    \item[] Answer: \answerNA{} % Replace by \answerYes{}, \answerNo{}, or \answerNA{}.
    \item[] Justification: This paper contributes an analytical framework for characterizing nonlinear control in interacting subsystems, and poses no risk of misuse. We do not make use of any language models or image generators, and do not use any scraped datasets. We therefore do not describe safeguards.%\justificationTODO{}

\item {\bf Licenses for existing assets}
    \item[] Question: Are the creators or original owners of assets (e.g., code, data, models), used in the paper, properly credited and are the license and terms of use explicitly mentioned and properly respected?
    \item[] Answer: \answerYes{} % Replace by \answerYes{}, \answerNo{}, or \answerNA{}.
    \item[] Justification: Yes, we implement existing methods in our paper and cite the original creators appropriately.%\justificationTODO{}

\item {\bf New assets}
    \item[] Question: Are new assets introduced in the paper well documented and is the documentation provided alongside the assets?
    \item[] Answer: \answerYes{} % Replace by \answerYes{}, \answerNo{}, or \answerNA{}.
    \item[] Justification: We provide well-documented code for generating the models and implementing the datasets used in this paper. %\justificationTODO{}

\item {\bf Crowdsourcing and research with human subjects}
    \item[] Question: For crowdsourcing experiments and research with human subjects, does the paper include the full text of instructions given to participants and screenshots, if applicable, as well as details about compensation (if any)? 
    \item[] Answer: \answerNA{} % Replace by \answerYes{}, \answerNo{}, or \answerNA{}.
    \item[] Justification: This paper does not involve crowdsourcing nor research with human subjects.%\justificationTODO{}

\item {\bf Institutional review board (IRB) approvals or equivalent for research with human subjects}
    \item[] Question: Does the paper describe potential risks incurred by study participants, whether such risks were disclosed to the subjects, and whether Institutional Review Board (IRB) approvals (or an equivalent approval/review based on the requirements of your country or institution) were obtained?
    \item[] Answer: \answerNA{} % Replace by \answerYes{}, \answerNo{}, or \answerNA{}.
    \item[] Justification: This paper does not involve crowdsourcing nor research with human subjects. %\justificationTODO{}

\item {\bf Declaration of LLM usage}
    \item[] Question: Does the paper describe the usage of LLMs if it is an important, original, or non-standard component of the core methods in this research? Note that if the LLM is used only for writing, editing, or formatting purposes and does not impact the core methodology, scientific rigorousness, or originality of the research, declaration is not required.
    %this research? 
    \item[] Answer: \answerNA{} % Replace by \answerYes{}, \answerNo{}, or \answerNA{}.
    \item[] Justification:
    The core method development in this research does not involve LLMs as any important, original, or non-standard components.%\justificationTODO{}

\end{enumerate}

\end{document}